%% file: paper.tex
\pdfoutput=1

\documentclass[format=acmsmall,letterpaper,nonacm]{acmart}
\PassOptionsToPackage{hyphens}{url}
\usepackage[hyphens]{url}
\usepackage{hyperref}
\hypersetup{pdfstartview=FitV,
  breaklinks=true, colorlinks=true, linkcolor=black, citecolor=black, urlcolor=blue
}
\usepackage{xspace}
\usepackage{enumitem}
\usepackage{dcolumn}
\usepackage{multirow}

\usepackage{subcaption}
\usepackage{microtype}
\usepackage{color}
\usepackage{balance}
\usepackage{makecell}

\usepackage{tikz}
\usepackage{placeins}
\usepackage{booktabs}

\usetikzlibrary{automata, positioning}
\tikzset{node distance=2cm}

\graphicspath{{figures/}}

\newcommand{\eg}{e.g.,\ }
\newcommand{\ie}{i.e.,\ }
\newcommand{\etal}{et al.\ }

\newcommand{\pt}[1]{\item \textbf{#1:}}
\newcommand{\maxmind}{MaxMind\xspace}

\newcommand{\squishlist}{
 \begin{list}{${\bullet}$}
  { \setlength{\itemsep}{0pt}
     \setlength{\parsep}{3pt}
     \setlength{\topsep}{3pt}
     \setlength{\partopsep}{0pt}
     \setlength{\leftmargin}{1.5em}
     \setlength{\labelwidth}{1em}
     \setlength{\labelsep}{0.5em} } }
\newcommand{\squishend}{
  \end{list}  }

\begin{document}

\author{Matthieu Gouel}
\affiliation{%
  \institution{Sorbonne Universit\'{e}}
}
\email{matthieu.gouel@sorbonne-universite.fr}

\author{Kevin Vermeulen}
\affiliation{%
  \institution{Columbia University}
}
\email{kevin.vermeulen@columbia.edu}


\author{Olivier Fourmaux}
\affiliation{%
  \institution{Sorbonne Universit\'{e}}
}
\email{olivier.fourmaux@sorbonne-universite.fr}

\author{Timur Friedman}
\affiliation{%
  \institution{Sorbonne Universit\'{e}}
}
\email{timur.friedman@sorbonne-universite.fr}

\author{Robert Beverly}
\orcid{0000-0002-5005-7350}
\affiliation{%
  \institution{Naval Postgraduate School}
}
\email{rbeverly@nps.edu}

\title[]{Longitudinal Study of an IP Geolocation Database}

\begin{abstract}
IP geolocation -- the process of mapping network identifiers to
physical locations -- has myriad applications.  We examine a
large collection of snapshots from a popular 
geolocation database and take a first look at its
\emph{longitudinal} properties.  
We define metrics of
IP geo-persistence, prevalence, coverage, and movement, and analyze 10 years
of geolocation data at different location
granularities.
Across different classes of IP addresses, we find that
significant location differences can exist even between successive instances of
the database -- a previously
underappreciated source of potential error when using geolocation
data: 47\% of end users IP addresses move by more than 40 km in
2019.
To assess the sensitivity of research results to the instance of the
geo database, we reproduce prior 
research that depended on geolocation lookups.  In this case study, which
analyzes geolocation database performance on routers, we
demonstrate impact of these temporal effects: median distance from ground 
truth shifted from 167 km to 40 km when using a two months 
apart snapshot.
Based on our findings, we make
recommendations for best practices when using geolocation databases
in order to best encourage reproducibility and sound measurement. 
\end{abstract}




\maketitle

\input{intro}
\input{survey}
\input{methodology}

\input{data}
\input{evaluation}

\input{usecases}
\input{related-work}
\input{conclusion}
\input{appendix}

\section*{Acknowledgments}
We thank the anonymous reviewers for their constructive input and our
shepherd, Roman Kolcun for guidance. 
Matthieu Gouel, Olivier Fourmaux, and Timur Friedman
are associated with Sorbonne Université, CNRS, Laboratoire
d’informatique de Paris 6, LIP6, F-75005 Paris, France. Matthieu Gouel and Timur Friedman are associated with the Laboratory of Information, Networking and Communication Sciences, LINCS, F-75013 Paris, France.  Matthieu Gouel, Olivier Fourmaux, and Timur Friedman
were supported in part by a university research grant from the
French Ministry of Defense. 
Robert Beverly was supported in part by
NSF grant CNS-1855614.  Views and conclusions are those of the authors
and should not be interpreted as representing the official policies or
position of the U.S.\ government or the NSF.

\newpage
\clearpage
\balance
\urlstyle{same}
\bibliographystyle{ACM-Reference-Format}
\bibliography{geo}

\end{document}

%% file: intro.tex
%
%

\section{Introduction}
\label{sec:intro}

Determining the physical location of Internet hosts is important for a
range of applications including, but not limited to, advertising, content and
language customization, security and forensics, and policy 
enforcement~\cite{katz_2006,huffaker2011,wang_2011}.
However, the Internet
architecture includes no explicit notion of physical location and
hosts may be unable or unwilling to share their location.  As a
result,
the
process of third-party IP geolocation -- mapping an IP address to a
physical location -- emerged as a research topic~\cite{10.1145/383059.383073} 
more than two decades ago and has
since matured into commercial service offerings,
\eg~\cite{edgescape,maxmind,hexasoft}.

IP addresses represent network attachment points, thus IP geolocation
is often inferential.
Commercial geolocation providers compete, so the methodologies for
creating their databases are proprietary.  
State-of-the-art techniques
include combining latency constraints~\cite{gueye2006constraint}, 
topology~\cite{katz_2006}, registries~\cite{10.1145/383059.383073}, public
data~\cite{eriksson_2010}, and privileged feeds~\cite{geofeeds}.

This work takes a fresh look at IP geolocation data from
a \emph{temporal} perspective.
Specifically, we examine the longitudinal stability of locations in
an IP
geolocation database, the characteristics of location changes when they
do occur, and the extent to which a particular instance of a geolocation
database impacts conclusions that depend on locations.  To wit,
network and systems researchers frequently utilize available IP
geolocation database snapshots.  However, the date of the snapshot may
only loosely align with the time of the lookup operation, or the
lookups may span multiple snapshots, \eg a long-running measurement
campaign.
We show that snapshots of the same geolocation database separated even
closely in time can have a  
non-trivial effect on research results and findings.  

For example, across database snapshots in three month window,
we find up to 22\% of
IP addresses move more than 40km,
while coverage (the simple presence or absence of an address in the
database) varies by as much as 18\%.
Despite this temporal sensitivity, the
\emph{date} of the geolocation database snapshot is rarely reported
in the academic literature -- an omission that we show confounds 
scientific reproducibility.

As a first step to understanding the temporal characteristics of IP
geolocation databases, we define metrics of persistence, prevalence, coverage,
and distance.  We then analyze
10 years of data from the most popular, publicly available, and frequently used
database: \maxmind~\cite{maxmind}.
Our contributions include:

\begin{itemize}
  \item A survey of how recent systems and networking literature 
        utilizes and depends on IP geolocation data.
  \item Defining metrics to understand the Internet-wide 
        stability and behavior of IP geolocation databases.
  \item The first longitudinal study of a widely used IP geolocation database
        where we find significant short-term dynamics.
  \item A case study of prior research that depended on geolocation,
        showing that the results fundamentally
        differ based on the instance of the geolocation
        database used.
  \item Recommendations for the sound use of IP geolocation data in
        research. 
\end{itemize}

%% file: survey.tex
\section{Motivation}
\label{sec:survey}

To better understand IP geolocation as
used
in the network and systems research community, we surveyed the
academic literature.
We performed full-text queries, over all time, on four popular digital libraries for
three common geolocation databases, \maxmind~\cite{maxmind},
NetAcuity~\cite{net-acuity}, and IP2Loc~\cite{ip2loc}.
Table~\ref{tab:geolocation-databases} shows the number of papers in
each library. \maxmind is clearly the most popular by an order
of magnitude.  Therefore the analysis in the reminder of 
this work focuses on \maxmind.

\subsection{MaxMind}
Founded in 2002, \maxmind is a commercial entity specializing 
in IP geolocation and
related services.  \maxmind offers two IP geolocation databases, one
that is free (GeoLite) and one that requires a license
(GeoIP).  
The academic literature predominantly uses the free database, GeoLite.

GeoLite
is available as a complete database
``snapshot.''  Snapshots are currently updated weekly and available
for public download.  
GeoLite snapshots contain variable length IP prefixes, each with an
associated geolocation.  The geolocation may include country, city,
latitude/longitude, and accuracy (in
km); however many prefixes only provide
a geolocation at the country granularity.
This work studies the IPv4 GeoLite databases.  Henceforth, we
refer to GeoLite (and its successor, GeoLite2) informally as ``\maxmind'' for
simplicity.

\subsection{Survey Methodology}

We characterized the use of \maxmind 
across nine systems, security and
networking conferences during the five year period from 2016-2020.
To find papers in the literature using \maxmind, as well understand
how it is used, 
we adopt a semi-automatic method:
first, for a given conference venue, we obtain the complete
proceedings 
and perform a case-insensitive search for the string ``maxmind.''  
We manually inspect each 
paper found to contain ``maxmind''
to determine whether the work utilizes the database or is simply
referencing \maxmind.  
For example, in~\cite{kotronis2017shortcuts}, 
``maxmind'' appears only as a citation to the sentence
``Current IP-based geolocation services do not provide city-level 
accuracy$\ldots$''  Only those papers that used \maxmind's database
for their research are included in our survey.

Keeping in mind the variety of research questions and geolocation
requirements inherent in the various papers, we sought to distinguish
what was being geolocated and at what granularity.
We manually extract from each paper the granularity
required (country, city, or AS) and the type of IP addresses 
geolocated (all, end users, end host infrastructure, and router).
The ``end user'' category contains IP addresses belonging to 
residential users (e.g.,~\cite{padmanabhan2019residential}), or, 
more broadly, end users 
issuing web traffic (e.g.,~\cite{papadopoulos2017if}). 
The ``end host infrastructure'' category includes addresses belonging to Internet 
infrastructure, typically web~\cite{deng2017internet}, 
proxies~\cite{weinberg2018catch}, or DNS~\cite{pearce2017global} servers.
``Routers'' include the IP addresses of network router interfaces.
Finally, the ``all'' category contains papers that geolocate all types
of addresses such as~\cite{lee2016identifying,winter2019geo}.
Note that these sets are mutually exclusive, but a paper can use \maxmind
on several types of IP addresses.  For instance,
\cite{antonakakis2017understanding} studies the Mirai botnet where 
the infected IP addresses can
belong to both end users and end host infrastructure.


\begin{table}[t]
\caption{Count of papers referencing geolocation databases in
four digital libraries.}
\vspace{-3mm}
\label{tab:geolocation-databases}
\footnotesize{
\begin{tabular}{| r | c c c c|}
\cline{2-5}
  \multicolumn{1}{c|}{} & ACM & IEEE & arXiv.org & Springer \\
  \hline
  \maxmind & 171 & 373 & 96 & 162 \\
  \hline  
  NetAcuity & 10 &  10 & 8& 7\\ 
  \hline
  IP2Loc & 3 & 3 & 0 & 0\\ 
  \hline
\end{tabular}
}
\vspace{-3mm}
\end{table}

\begin{table*}[t]
\caption{Literature survey of MaxMind use in academic venues
(2016-2020). ``Affected'' column specifies if \maxmind was used
in methodology (Y) or for validation (V).}
\vspace{-4mm}
\resizebox{\textwidth}{!}{
\label{tab:survey}
\footnotesize{
\begin{tabular}{|l|l|l| lll | llll | ll | ll | lll |}
\hline
                &               &                 & \multicolumn{7}{c|}{\textbf{\maxmind}}   & \multicolumn{2}{c|}{\textbf{Affected}} & \multicolumn{2}{c|}{\textbf{\makecell{Snapshot \\ date \\ specified}}} &  \multicolumn{3}{c|}{\textbf{\makecell{Free (F)\\Paid(P)\\(N/A))}}}  \\
\cline{4-17}
                &               &                 &            \multicolumn{3}{c|}{\textbf{Granularity}} & \multicolumn{4}{c|}{\textbf{IP type}} &    \multirow{2}{*}{Y} & \multirow{2}{*}{V} & \multirow{2}{*}{Y} & \multirow{2}{*}{N} & \multirow{2}{*}{F} & \multirow{2}{*}{P} & \multirow{2}{*}{N/A} \\
\cline{4-10}                
 \textbf{Conference} & \textbf{Area} & \textbf{Papers} &  AS  & Country & City & All &End user & \makecell[l]{End host \\infrastructure}& Router & & & &      & & & \\
\hline
IMC         & Meas.\ &  16 & 1 & 13 & 3 & 2 & 5 & 8 & 1 & 12 & 4 & 1 & 15 & 8 & 3 & 6\\
PAM             & Meas.\ & 6 & 0 & 2 & 4 & 1 & 1 & 2 & 0 & 5 & 1 & 0 & 6 &3 & 1 & 2\\
TMA             & Meas.\ &  4 & 1 & 0 & 3 & 3 & 0 & 1 & 0 & 3 & 1 & 2 & 2 & 1 & 2 & 1\\
\hline
USENIX Sec      & Security      & 10 & 0 & 7 & 3 & 0 & 4 & 7 & 0 & 10 & 0 & 2 & 8 & 1 & 4 & 5 \\
CCS         & Security      &  6 & 2 & 1 & 3 & 0 & 1 & 2 & 3 & 6 & 0 & 0 & 6 & 2 & 1 & 1\\
\hline
SIGCOMM     & Systems       & 3 & 0 & 1 & 2 & 0 & 3 & 1 & 0 & 2 & 1 & 0 & 3 & 0 & 3 & 0\\
NSDI& Systems & 1 & 0 & 0 & 1 & 0 & 0 & 0 & 1 & 1 & 0 & 0 & 1  & 0 & 0 & 1\\
CoNEXT      & Systems       & 2 & 1 & 1 & 0 & 1 & 0 & 1 & 0 & 2 & 0 & 0 & 2 &1 & 0 & 1\\
\hline
WWW      & Web & 10 & 0 & 7 & 3 & 0 & 4 & 7 & 0 & 10 & 0 & 2 & 8 & 2 & 2 & 6\\
\hline 
\hline
Total & All & 58 & 7 & 30 & 22 & 8 & 20 & 28 & 6 & 51 & 7 & 6 & 52 & 18 & 18 & 23\\
\hline
\end{tabular}
}
}
\end{table*}

\subsection{\maxmind in the literature}
\label{sec:survey-results}

Table~\ref{tab:survey} summarizes our findings.  We follow the
rhetorical structure of Scheitle \etal~\cite{10.1145/3278532.3278574}
to classify the impact of \maxmind on the paper's results.  
\begin{list}{-}{}
\item Affected ``Y'' are papers that use \maxmind in their methodology
to obtain a result. For example,
Papadopoulos
\etal~\cite{papadopoulos2017if}
use \maxmind 
to build a
classifier to infer how much advertisers pay to reach 
users.
\item Affected ``V'' are papers that do not use \maxmind to 
obtain results, but rather to \emph{compare} their results.
For example, Weinberg \etal~\cite{weinberg2018catch} 
compare their inferred proxy locations to \maxmind's locations.
\end{list}
The ``Date'' column indicates whether or not the paper
explicitly provides the \maxmind snapshot date.
The last column indicates which \maxmind version is used, either free, paid or if the
info was not available.

From a macro perspective, \maxmind is both used at country (53\%) 
and city (37\%)
granularity. Second, it is mostly used to geolocate end users (35\%) and end host 
infrastructure (49\%) rather than 
routers (9\%). 
Then, the majority of papers (86\%) use
\maxmind  
to obtain results, and few (11\%) provide the snapshot date.
Finally, we see that free and paid version of \maxmind are equally
used by the community.
Note that the totals do not sum to the number of papers as, \eg
a paper may use \maxmind for both AS and country information~\cite{lee2016identifying}, 
or use both the free and paid version~\cite{gharaibeh_2017}.


\paragraph{\textbf{Lesson}}
MaxMind is the most popular
geolocation database to support other research.  Further, the results
of many papers may be sensitive to geolocation variation, 
especially given the lack of snapshot dates, large windows of
measurement or data, and no explicit alignment between data 
collected and the geolocation snapshot.


%% file: methodology.tex

\section{Metrics}
\label{sec:methodology}

In this report, we will:
(1) characterize the impact of selecting one geolocation database
snapshot rather than a different snapshot in time; and
(2) survey the dynamics of snapshots, to gain a deeper
understanding of how they change.
This section defines the metrics used for these purposes:
impact in \S\ref{sec:impact}
and dynamics in \S\ref{sec:methodology:dynamics}. 
We assume that snapshots contain IP prefixes and their 
associated locations; where necessary we trivially assume the
expansion of prefixes to the set of individual addresses within those
prefixes. 
We further make the simplifying assumption the snapshots are
uniformly distributed over time with an arbitrary inter-snapshot
interval (we achieve this in practice via sampling in
\S\ref{sec:time-unit}).

\subsection{Impact: Comparing Two Snapshots}\label{sec:impact}
This section provides metrics to answer the practical question:
what is the impact of choosing one snapshot rather than 
another
from the same time interval?
We define metrics on the basis of two concepts for comparing two
geodatabase snapshots: 
\textit{coverage difference} and \textit{distance distribution}.
For these definitions:
\begin{list}{-}{}
\item Let $A$ be a set of IPv4 addresses. 
\item Let $L$ be the set of all possible locations present in a
geodatabase (either latitude-longitude pairs, cities, or countries).
\item Let $M$ be a snapshot of this database, defined as a set of pairs $(x,l)$ that map 
addresses to locations.
\item Let $A_M \subset A$ be the set of addresses that 
appear in snapshot $M$.
Each address $x \in A_M$ appears in only one pair in $M$. 
\end{list}
\subsubsection{Coverage difference}
Intuitively, coverage difference means the portion of IP addresses
that appear in one snapshot or another, but not both.
Two identical snapshots have a coverage difference of zero
and for snapshots that have no IP address in common, the value is one.

Formally, coverage difference is an extension of the concept of `coverage', which
for snapshot $M$ with respect to the set of IPv4 addresses $A$ is:
\begin{equation}
 \mathrm{coverage}(M) = \frac{|A_M|}{|A|}
 \label{eq:coverage}
\end{equation}
The 
coverage difference between two snapshots $M_i$ and $M_j$ on $A$
is the Jaccard
distance between $A_{M_i}$ and $A_{M_j}$:
\begin{equation}
\mathrm{covdiff}(M_i, M_j) 
= \frac{|(A_{M_i} \bigcup  A_{M_j}) - (A_{M_i} \bigcap  A_{M_j})|}
{|A_{M_i} \bigcup  A_{M_j}|}
\label{eq:delta-coverage}
\end{equation}

\subsubsection{Distance distribution}\label{sec:methodology:impact:distance}
As an address $x$ can appear in one location in one geodatabase snapshot
and another location in a second snapshot, we let 
$\mathrm{dist}(x, M_i, M_j)$ be the Haversine distance~\cite{cajori1993history}
between the locations 
of $x$ in $M_i$ and $M_j$, using latitude-longitude values for each location.
The distance distribution between the
two snapshots is the set of distances, one for each address that appears in
both snapshots: 
\begin{equation}
\mathrm{D}_\mathrm{dist}(a, M_i, M_j) 
=  \lbrace 
\mathrm{dist}(a, M_i, M_j) 
\mid 
  x \in A_{M_i} \bigcap A_{M_j}
   \rbrace 
   \label{eq:delta-distance-distribution}
\end{equation}


\noindent To compare two distance distributions,
we define the following metric:
\begin{equation}
\mathrm{distdiff(M_i,M_j) = (mean}(\lbrace 
\mathrm{log_{10}(dist}(a, M_i, M_j)) 
\mid 
  x \in A_{M_i} \bigcap A_{M_j}
   \rbrace )
\label{eq:distance-metric-average}
\end{equation}
By taking the mean of the log of the distances, we diminish the possibility
for outliers (\eg distances potentially up to 20,000 km) to disproportionately
outweigh lower but nonetheless meaningful distances, \eg on the order of 100 km.
Note that while the median is a more robust statistic,
typically more than 50\% of the address have zero distance, \ie did not move
between snapshots (\S~\ref{sec:evaluation:impact}).  Thus, the mean
provides a meaningful non-zero measure.
Other metrics that we define on the distance distributions are the quantiles
of a distribution, along with the maximum value.


\subsection{Dynamics: How do Snapshots Evolve}\label{sec:methodology:dynamics}
As will be shown in Sec.~\ref{sec:evaluation:impact},
the choice of one snapshot rather than another can have a large impact,
even within a short time interval. 
We want to gain a deeper understanding of the underlying dynamics of a
geolocation system
and therefore define the following concepts:
the \textit{coverage difference} between snapshots,
the \textit{prevalence} and \textit{persistence} of locations, 
and the \textit{distance distribution} between snapshots.
Whereas the concepts in the previous section concerned pairs of
snapshots, in this section we consider sequences of many snapshots
generated over time.

\subsubsection{Coverage difference}\label{sec:methodology:coverage}
The coverage difference for a set of $n$ geodatabase snapshots $M_1,...,M_n$
is defined to be a value between zero and one as determined by the generalized 
Jaccard distance:
\begin{equation}
\mathrm{covdif}(M_1,...,M_n) = \frac{|\bigcup M_i - \bigcap M_i |}{|\bigcup M_i |}
\label{eq:jaccard-distance-coverage}
\end{equation}
If, for example, the coverage difference is 0.30 over the course of a year, 
this means that for 30\% of the IP addresses, each of those 
addresses does not appear in at least one snapshot for that year.
The lower the coverage difference, informally, the more
`stable' we might consider the
set of IP addresses as viewed by the geolocation system.

\subsubsection{Prevalence and persistence}\label{sec:methodology:prev-pers}
For location dynamics, we adopt prevalence and persistence metrics that are
similar to those originally proposed by Paxson~\cite{paxson} to capture notions
of the `dominance' and `persistence' of a route in the Internet.

Table~\ref{tab:pp_example} serves as a basis for illustrating how we have applied these
notions.  It shows eight time periods\footnote{Of arbitrary
granularity, for instance eight weeks or
eight months.} and the corresponding locations of 
two IP
addresses, $x$ and $y$. There are three possible locations: $\alpha$, $\beta$, $\gamma$
(which might be city, country, or latitude-longitude
pair).
\begin{table}[!h]
  \caption{Example for prevalence and persistence}
  \label{tab:pp_example}
  \begin{tabular}{@{}lcccccccc@{}}
    \toprule
    Time period     &  0       &  1         &  2       &  3         &  4         &  5         &  6         &  7       \\ \midrule
    Location of $x$ & $\alpha$ & $\alpha$   & $\alpha$  & $\gamma$   & $\beta$   & $\beta$    & $\beta$    & $\beta$  \\
    Location of $y$ & $\alpha$ & $\alpha$ & $\beta$ & $\beta$ & $\gamma$ & $\beta$ & $\alpha$ & $\alpha$ \\
    \bottomrule
  \end{tabular}
\end{table}
\paragraph{Prevalence}
The prevalence of an address $a$ for a location $l$,
$\mathrm{prev}(a,l)$ is the portion of time periods at which $a$ is
mapped to $l$. Our survey looks at the distribution of prevalence
values per address, as illustrated in
Table~\ref{tab:prevalence_per_addr}. These values are frequencies that
add up to 1 for each address. For example, address $x$ is mapped to
location $\beta$ for 4 out of the 8 time periods, so
$\mathrm{prev}(x,\beta) = 4/8$. 


Our survey also looks at the distribution of prevalence values per
location. The same values are simply transposed, as shown in
Table~\ref{tab:prevalence_per_loc}, which is a transposition of
Table~\ref{tab:prevalence_per_addr}. Distributed in this way, the
values do not necessarily add up to 1.

Seen in this way, the values give some insight into the likelihood
that a mapping to a given location will remain in that location across
time periods. A mapping to location $\gamma$ is a rare event, for
instance, as reflected in prevalence values of $1/8$ and $1/8$ that are
low on average. A mapping to location $\alpha$ can be said to be more
likely to be consistent across snapshots, with its prevalence values
of $3/8$ and $4/8$ that are higher on average.

\begin{table}
 \centering
 \caption{Prevalence values}
 \subfloat[per address]{
  \begin{tabular}{@{}lccc@{}}
    \toprule
    $l$                   &  $\alpha$ &  $\beta$ &  $\gamma$ \\ \midrule
    $\mathrm{prev}(x,l)$  &  3/8      &  4/8     &  1/8      \\
    $\mathrm{prev}(y,l)$  &  4/8      &  3/8       &  1/8   \\
    \bottomrule
  \end{tabular}
  \label{tab:prevalence_per_addr}
 }
 \quad \quad
 \subfloat[per location]{
  \begin{tabular}{@{}lcc@{}}
    \toprule
    $a$                          &  $x$   &  $y$  \\ \midrule
    $\mathrm{prev}(a,\alpha)$    &  3/8   &  4/8  \\
    $\mathrm{prev}(a,\beta)$     &  4/8   &  3/8   \\
    $\mathrm{prev}(a,\gamma)$    &  1/8   &  1/8    \\
    \bottomrule
  \end{tabular}
  \label{tab:prevalence_per_loc}
 }
\end{table}




Formally,
\begin{list}{-}{}
\item Let $T$ be the set of time periods.
\item Let $\mathcal{L}_x$ be the set of pairs $(t,l)$, one for each
  time period $t \in T$, where the snapshot for $t$
  designates $x \in A$ as being at location $l \in L$. 
\end{list}
where $A$ and $L$ have the same definition as in Section~\ref{sec:impact}.
Then the prevalence of an address $x$ at a given location $\alpha$ is: 
$$
 \mathrm{prev}(x,\alpha) = \frac{|\{(t,l) \in \mathcal{L}_x : l=\alpha\}|}{|\mathcal{L}_x|}
$$ 
and the maximum prevalence of an address $x$ is:
\begin{equation}
max(\lbrace \mathrm{prev}(x,l) : l \in L \rbrace)
 \label{eq:max-prevalence-ip}
\end{equation}

\noindent To define the distribution of prevalence values for a $l \in L$, we need to 
define $A_l$, the subset of IP addresses with at least one location equal to $l$:
$$
A_l=\{x \in A : (\exists \ (t,l_t) \in \mathcal{L}_x : l_t = l)\}
$$
and  the distribution of prevalence values is then:
\begin{equation}
D_{\mathrm{prev}}(l) = \lbrace \mathrm{prev}(x, l) : x \in A_l \rbrace
\label{eq:prevalence-location}
\end{equation}
and the mean value of this distribution is then:
\begin{equation}
\mathrm{mean}(D_{\mathrm{prev}}(l))
\label{eq:prevalence-location-average}
\end{equation}
This last metric permits sorting locations by their prevalence.

\paragraph{Persistence}
The persistence of an address $a$ for a location $l$, $\mathrm{pers}(a, l)$
is the probability for $a$, to stay in $l$ between two consecutive time periods.
Our survey looks at the distribution of persistence values per address, as
illustrated in Table~\ref{tab:persistence_per_addr}.
These values go from 0 to 1, but do not necessarily add up to 1 for each 
address. 
For example, when address $x$ is located in $\beta$, it always stays 
in $\beta$ in the subsequent time period, so $\mathrm{pers}(x, \beta)=1$.

The mean persistence for an address $a$, $\mathrm{mean}(\mathrm{pers}(a))$, 
is the generalization of $\mathrm{pers}(a, l)$, but for any location $l$.
In our example, $\mathrm{mean}(\mathrm{pers}(x))=5/7$ and 
$\mathrm{mean}(\mathrm{pers}(y))=3/7$.

Like for prevalence, our survey also looks at persistence values per location, as
shown in Table~\ref{tab:persistence_per_loc}. Seen in this way, it gives some
insight into the likelihood that a mapping to a given location will remain in that
location in the subsequent time period. For instance, it is 
very unlikely that a mapping 
to $\gamma$ will stay in $\gamma$ in the subsequent time period, 
as the persistence values are 0.

\begin{table}
 \centering
 \caption{Persistence values}
 \subfloat[per address]{
  \begin{tabular}{@{}lccc@{}}
    \toprule
    $l$                   &  $\alpha$ &  $\beta$ &  $\gamma$ \\ \midrule
    $\mathrm{pers}(x,l)$  &  2/3      &  1     &  0      \\
    $\mathrm{pers}(y,l)$  &  1/2      &  1/3  &  0   \\
    \bottomrule
  \end{tabular}
  \label{tab:persistence_per_addr}
 }
 \quad \quad
 \subfloat[per location]{
  \begin{tabular}{@{}lcc@{}}
    \toprule
    $a$                          &  $x$   &  $y$  \\ \midrule
    $\mathrm{prev}(a,\alpha)$    &  2/3   &  1/2  \\
    $\mathrm{prev}(a,\beta)$     &  1   &  1/3   \\
    $\mathrm{prev}(a,\gamma)$    &  0   &  0    \\
    \bottomrule
  \end{tabular}
  \label{tab:persistence_per_loc}
 }
\end{table}

Formally, 
by keeping the same notations for $\mathcal{L}_x$ and 
$A_l$ defined in the prevalence paragraph,  
the persistence of an address $x$ at a given location $\alpha$ is:
\begin{equation}
\mathrm{pers}(x, \alpha) = \frac{|\{ \{ (t,l_t), (t+1,l_{t+1}) \} \in \mathcal{L}_x : l_t=l_{t+1}=\alpha \} |}
{|\{ \{ (t,l_t), (t+1,l_{t+1}) \} \in \mathcal{L}_x : l_t=\alpha \} |}
 \label{eq:persistence-location-ip}
\end{equation}
The mean persistence of an address $x$ is:
\begin{equation}
\mathrm{mean(}\mathrm{pers}(x)) = \frac{|\{ \{ (t,l_t), (t+1,l_{t+1}) \} \in \mathcal{L}_x : l_t=l_{t+1}\} |}
{|\{ \{ (t,l_t), (t+1,l_{t+1}) \} \in \mathcal{L}_x \} |}
 \label{eq:avg-persistence-ip}
\end{equation}

\noindent The distribution of persistence values for a location $l \in L$ is:
\begin{equation}
D_{\mathrm{pers}}(l) = \lbrace \mathrm{pers}(x, l) : x \in A_l \rbrace
\label{eq:persistence-location}
\end{equation}

and the mean value of this distribution is:
\begin{equation}
\mathrm{mean}(D_{\mathrm{pers}}(l))
\label{eq:persistence-location-average}
\end{equation}
which permits sorting  locations by their persistence.
\subsubsection{Distance}\label{sec:methodology:distance}
Prevalence and persistence characterize the dynamics from a discrete 
point of view; in Table~\ref{tab:pp_example},
$x$ would have the same values of maximum prevalence and 
mean persistence whatever the values of 
$(\alpha,\beta, \gamma)$. We are now interested in studying the distances
between these locations. 

Our survey looks at the distribution of distance values per address.
We define the maximum distance of an address $x$ as being the maximum distance
between two of its locations.
Formally, the distribution of distances of $x$ is:
$$
D(x) = \lbrace \mathrm{dist}(l_{t_i}, l_{t_j}) : (t_i, l_{t_i}), (t_j, l_{t_j}) \in \mathcal{L}_x \} \rbrace
$$
and the maximum distance of $x$ is then:
\begin{equation}
\mathrm{max}(D(x))
\label{eq:max-distance-metric}
\end{equation}

Our survey also looks at the distribution of maximum distance per 
geographic entity (either a continent, country). 
For instance, for the addresses that geolocate
within a particular country, we wish to understand the maximum
distance they move within that country. It allows us to understand if IP addresses
tend to have a higher/lower maximum distance depending on which countries
they belong. 
Notice the distinction here between the terms location and geographic entity, 
as the locations refer to latitude/longitude pairs in this paragraph.

To define the distribution of maximum distance for a geographic entity $c$, 
we first define 
the set of latitude/longitude coordinates $L_c \subset L$ :
$$
L_c= \{ l \in L : l \ \mathrm{belongs \ to \  } c \}
$$
We define the subset of distances $D_c(x) \subset D(x)$ as:
\begin{equation}
D_c(x) = \lbrace \mathrm{dist}(l_{t_i}, l_{t_j}) \in D(x) : l_{t_i} \in L_c \lor l_{t_j} \in L_c \rbrace
\end{equation} 

The distribution of maximum distance for $c$ is then:
\begin{equation}
D(c) = \lbrace \mathrm{max}(D_c(x)) : D_c(x) \neq \emptyset, x \in A \} \rbrace
\label{eq:distance-metric-location}
\end{equation}
and the mean of this distribution is:
\begin{equation}
\mathrm{mean} \lbrace \mathrm{log_{10}}(\mathrm{max}(D_c(x))) : D_c(x) \neq \emptyset, x \in A \} \rbrace
\label{eq:distance-metric-location-average}
\end{equation}
which permits sorting geographic entities by their maximum distance distribution. 
Note the presence of
$\mathrm{log_{10}}$ for the same reason 
described in \S\ref{sec:methodology:impact:distance}.

Note that from Eq.~\ref{eq:distance-metric-location},
if an IP address moves from one country to another, we account it
for both countries.

%% file: data.tex
\section{Data}
\label{sec:data}

Before delving into the details of each finding, we present our datasets in 
Sec.~\ref{sec:maxmind-snapshots} and~\ref{sec:ip-types}.
Ethical considerations of our work are provided in
Appendix~\ref{appendix:ethics}.

\subsection{\maxmind snapshots}\label{sec:maxmind-snapshots}

We collect 214 \maxmind snapshots spanning the ten year period from
January 2010 to December 2019.  There are two primary challenges in
the raw data: (1) the snapshots we obtain are not uniformly
distributed in time; and (2) IP addresses appear within prefixes of
different networks and lengths over time.  To utilize this data within
the framework of our methodology and metrics, we pre-processed it.

\subsubsection{Sampling the snapshots for time uniformity}
\label{sec:time-unit} 

Sec.~\ref{sec:methodology} assumes a uniform distribution of snapshots
in time.
Our evaluation examines a ten year span from 2010-2019.  Within this 
ten year period, we have at
least one snapshot per month, but sometimes as many as one snapshot
per week.   
Therefore to ensure
uniformity, we simply down-sample so that the ten year period
includes one snapshot per month.
Our evaluation is conducted on this subset of
snapshots such that they are uniformly distributed in time.




\subsubsection{Prefixes of different lengths}

A \maxmind snapshot contains a mapping of prefixes to geolocation.
However, as with prefixes in a routing table, prefixes may split, be
aggregated, or even overlap in time.  While our analysis is at the
per-IP address granularity, rather than prefix, maintaining the
geolocation for all IP addresses over time is inefficient.  Our first
step then is to find a data structure to efficiently store and query
the snapshots.  Over all prefixes in all snapshots, we construct the
set of covering longest length prefixes and construct a Patricia
trie~\cite{sklower1991tree}.  We build one Patricia trie for each
geolocation granularity: country, city, and coordinates.  The Patricia
trie contains, per prefix, all its locations over the period of time.

To handle prefix variation over time, we insert into the Patricia trie
the longest prefixes.  For example, if for one snapshot, a prefix has
a length of 24, and for another snapshot it is split into two prefixes
with a length of 25, the two /25 prefixes are placed in the trie and
the first snapshot populates both prefixes with the location of the
/24.

\subsection{Different types of IP addresses}
\label{sec:ip-types}

Sec.~\ref{sec:survey} has shown that researchers use \maxmind to locate
three classes of IP addresses: end users, end hosts infrastructure and
routers.  We therefore collect and label three sets of IP addresses
corresponding to these three types. 

\begin{itemize}
\pt{End users}
M-Lab~\cite{mlab} performs and records measurements to end users
requesting performance tests (\ie a ``speedtest'').  From the M-Lab
public datasets we extract targets in the year 2019.  We randomly
sample these targets to obtain
6.7M IPv4 addresses in approximately 2M unique /24 prefixes.

\pt{End host infrastructure}
For end host infrastructure, we extract the daily top list made
available by~\cite{10.1145/3278532.3278574}.  We perform an
intersection of all 2019 lists in order to minimize the number of IP
addresses that could be reassigned for other purpose.  Because these
top lists are volatile, our filtering for high-confidence end host
infrastructure addresses produces 26,231 IP addresses in 16,942 /24
prefixes.

\pt{Routers}
We leverage both \textsc{CAIDA} ITDK dataset~\cite{caida-itdk} and 
Diamond-Miner~\cite{vermeulen2020diamond}
public Internet topology datasets to 
collect IP addresses belonging to router interfaces. Both datasets 
are the result of Internet-wide traceroute style probing.
We take the intersection of 2019-01, 2019-04 ITDK and 
2019-08 Diamond-Miner
datasets and obtain 
730k IP addresses in more than 177k different /24 prefixes.
By taking the intersection over time, the aim is again to ensure
the likelihood that the addresses indeed belong to routers.
\end{itemize}


%% file: evaluation.tex

\section{Evaluation}\label{sec:evaluation}
This section presents an evaluation of ten years of 
\maxmind data using the metrics defined in the methodology.
Our study analyzes if the dynamics depend on three different axes: time,
type of IP address and country. Primary results are presented here,
while a more exhaustive evaluation along all three of these
analysis dimensions, omitted due to space constraints, is available in 
an accompanying
technical report~\cite{vermeulen20}.

\subsection{Prevalence and persistence}\label{sec:evaluation:prev-pers}

As a first step to understanding the \maxmind data, we utilize our primary 
metrics of prevalence and persistence.  Recall that,
informally, prevalence is a measure of how frequently an IP address 
has a given geolocation across snapshots, while persistence is a measure
of how long the IP address has a geolocation before it changes.
Further recall that for each IP address, the \maxmind data includes a
country, city, and latitude/longitude.  Thus, these metrics can be computed 
relative to different location granularities.

\begin{figure*}[t]
\begin{subfigure}{.36\textwidth}
\centering
  \includegraphics[width=\linewidth]{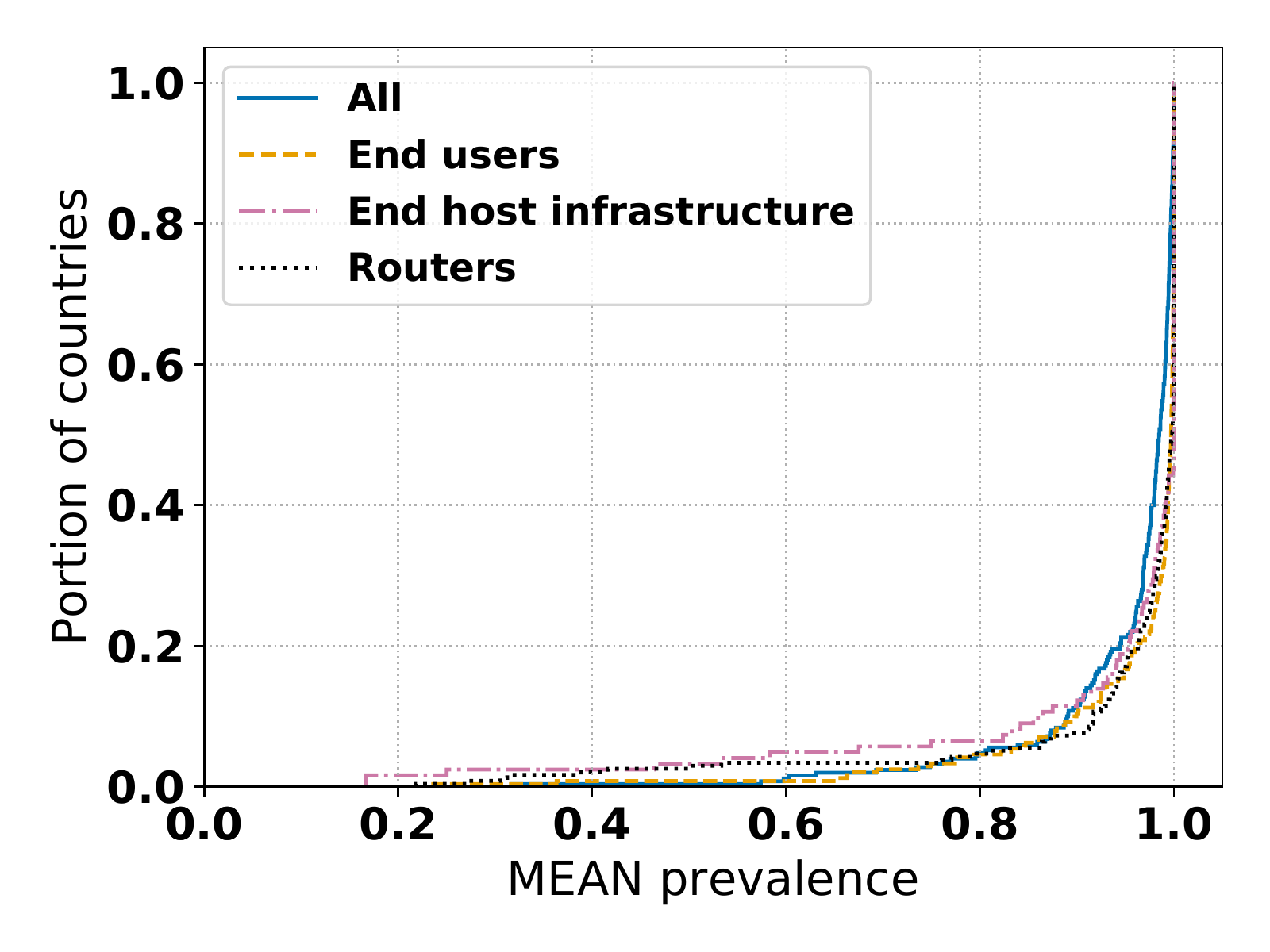}
  \caption{By country (Eq.~\ref{eq:prevalence-location-average})}
  \label{fig:prevalence-by-country}
\end{subfigure}%
\begin{subfigure}{.36\textwidth}
\centering
  \includegraphics[width=\linewidth]{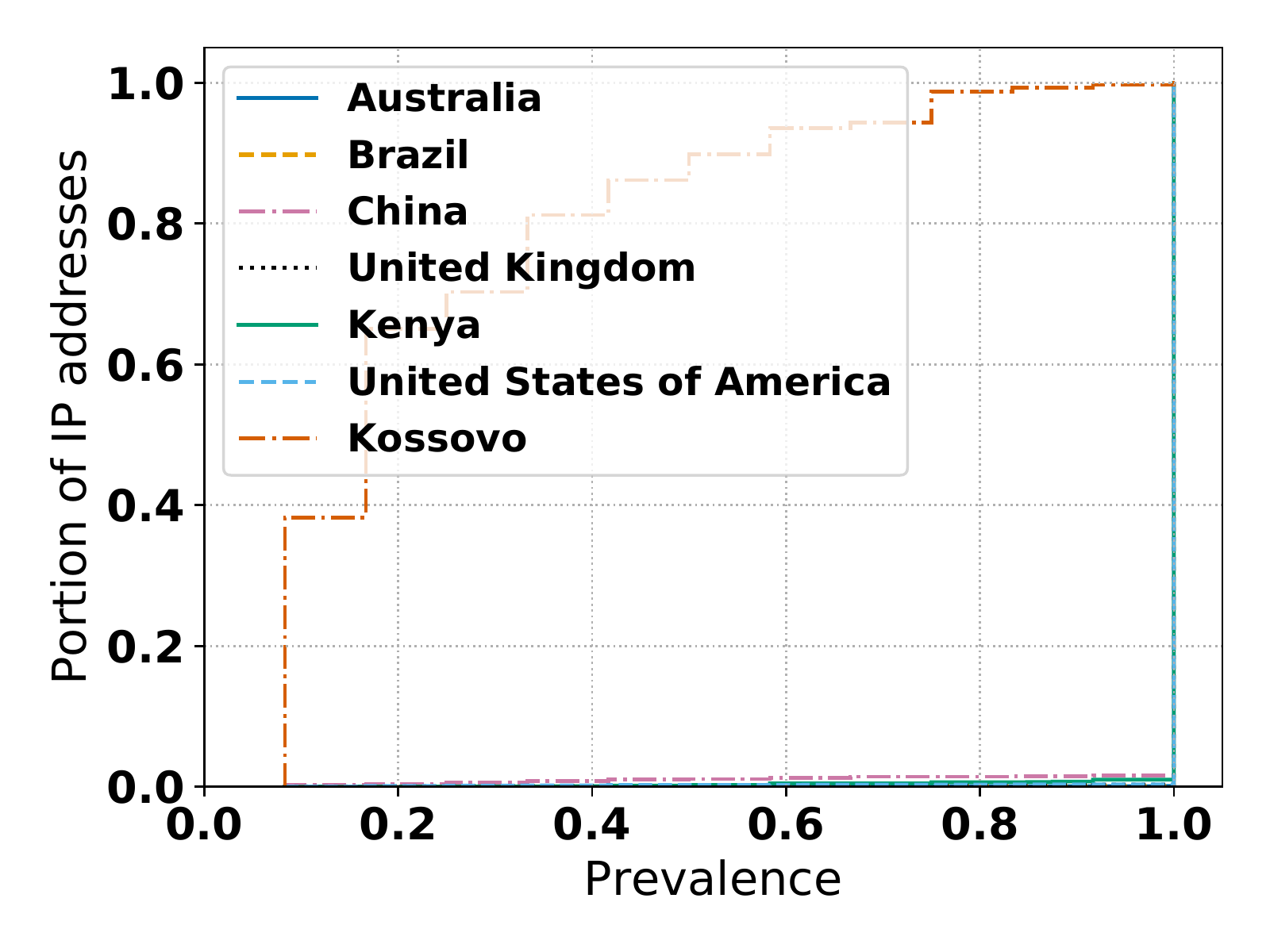}
  \caption{End users (Eq.~\ref{eq:prevalence-location})}
  \label{fig:prevalence-by-country-end-users}
\end{subfigure}%
\\
\begin{subfigure}{.36\textwidth}
\centering
  \includegraphics[width=\linewidth]{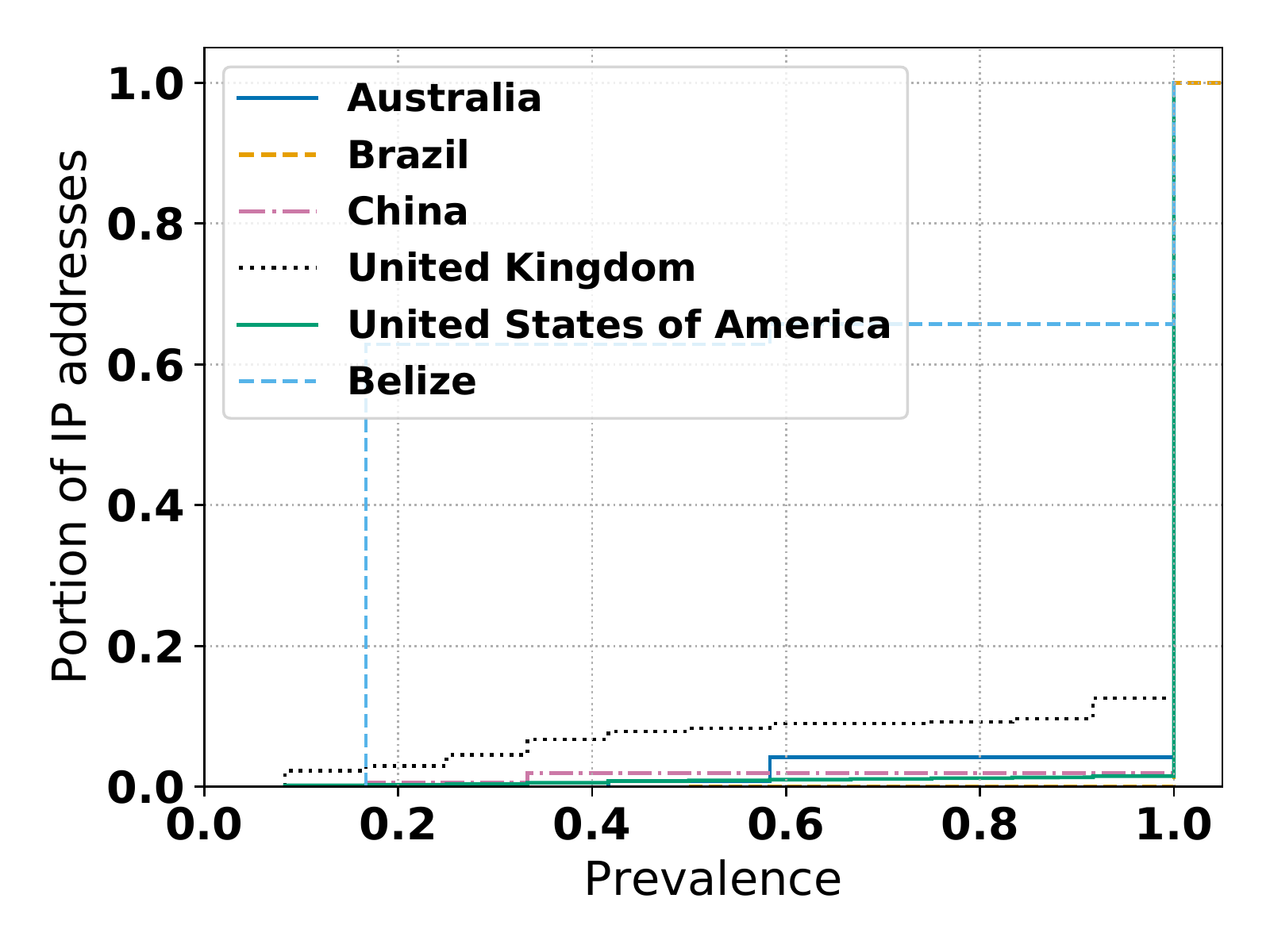}
  \caption{End host infra.\ (Eq.~\ref{eq:prevalence-location})}
  \label{fig:prevalence-by-country-infrastructure}
\end{subfigure}%
  \vspace{-2mm}
  \caption{Country prevalence measures how frequently an IP address has a
given country across snapshots.  While prevalence is generally 
high, it varies significantly for particular subsets of addresses
and countries.}
  \label{fig:country-dynamics}
\end{figure*}

\subsubsection{Country}
We find that for all years, except 2012, 
more than 98\% of 
the IP addresses have a country-level prevalence of 1, meaning that their
country did not change.
Note that because the prevalence is very close to 1, the persistence is also 
close to 1. 

However, the prevalence can vary significantly for particular subsets
of addresses and countries.  
Fig.~\ref{fig:prevalence-by-country} shows the cumulative fraction of
countries as a function of mean prevalence.
Across all classes of IP addresses, most countries have a
mean prevalence of more than 0.9, however 
some in the tail of the 
distribution have a low mean prevalence, with a minimum of 0.18. 
Kossovo, on Fig.~\ref{fig:prevalence-by-country-end-users} is the country
with the lowest prevalence. For end users, countries with low prevalence are 
mainly found in small Islands.
Fig.~\ref{fig:prevalence-by-country-end-users} 
and~\ref{fig:prevalence-by-country-infrastructure} also shows the prevalence
distribution 
for one big country per continent. Note that for end host
infrastructure, the prevalence for United Kingdom is lower than for
the other big countries:
8\% of IP addresses once located in UK had a prevalence for UK  of less than 
0.6, which is not negligible. 

\begin{figure*}[t]

\begin{subfigure}{.36\textwidth}
\centering
  \includegraphics[width=\linewidth]{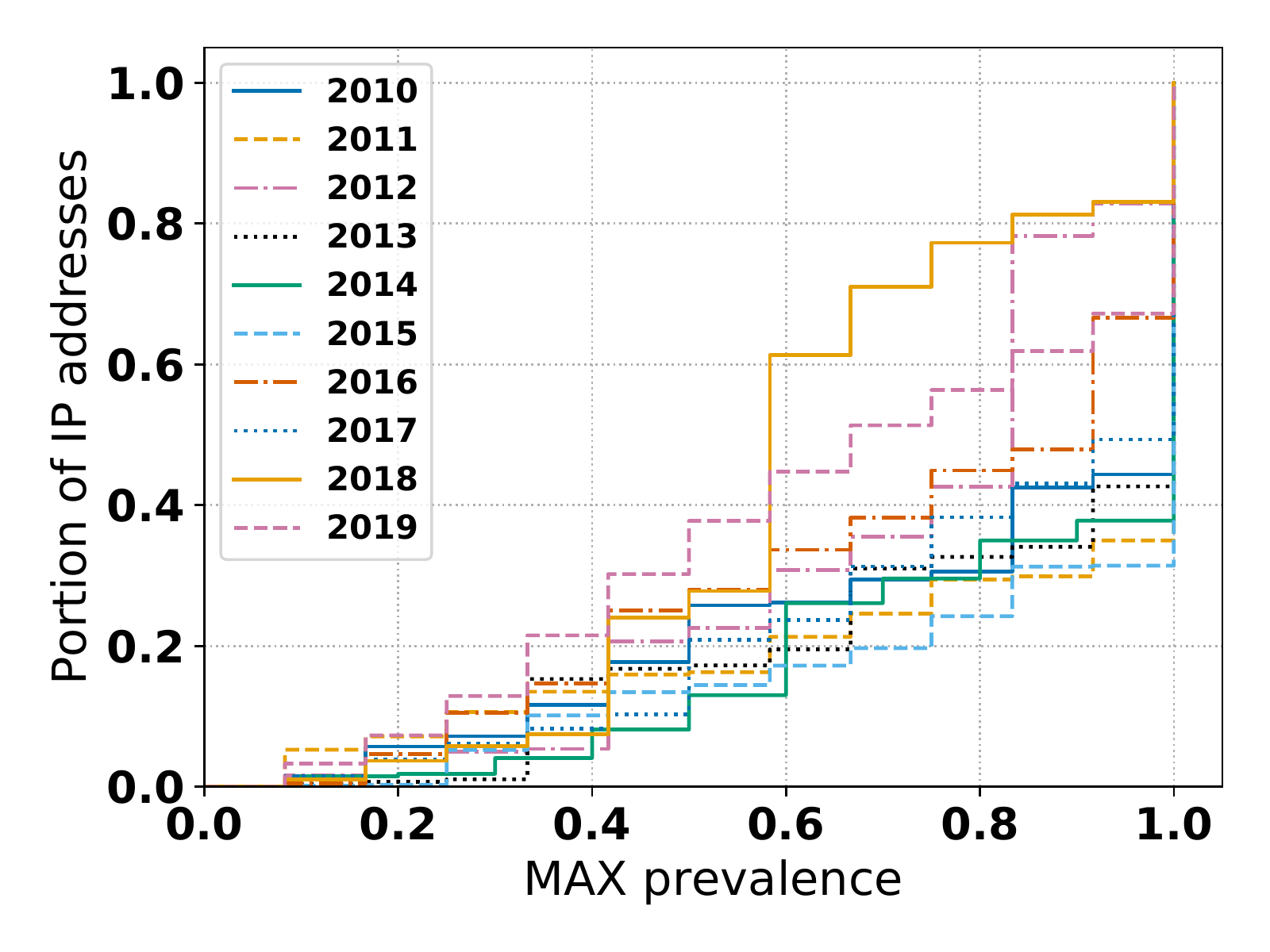}
  \caption{Over time (Eq.~\ref{eq:max-prevalence-ip})}
  \label{fig:prevalence-over-time-city}
\end{subfigure}%
\begin{subfigure}{.36\textwidth}
\centering
  \includegraphics[width=\linewidth]{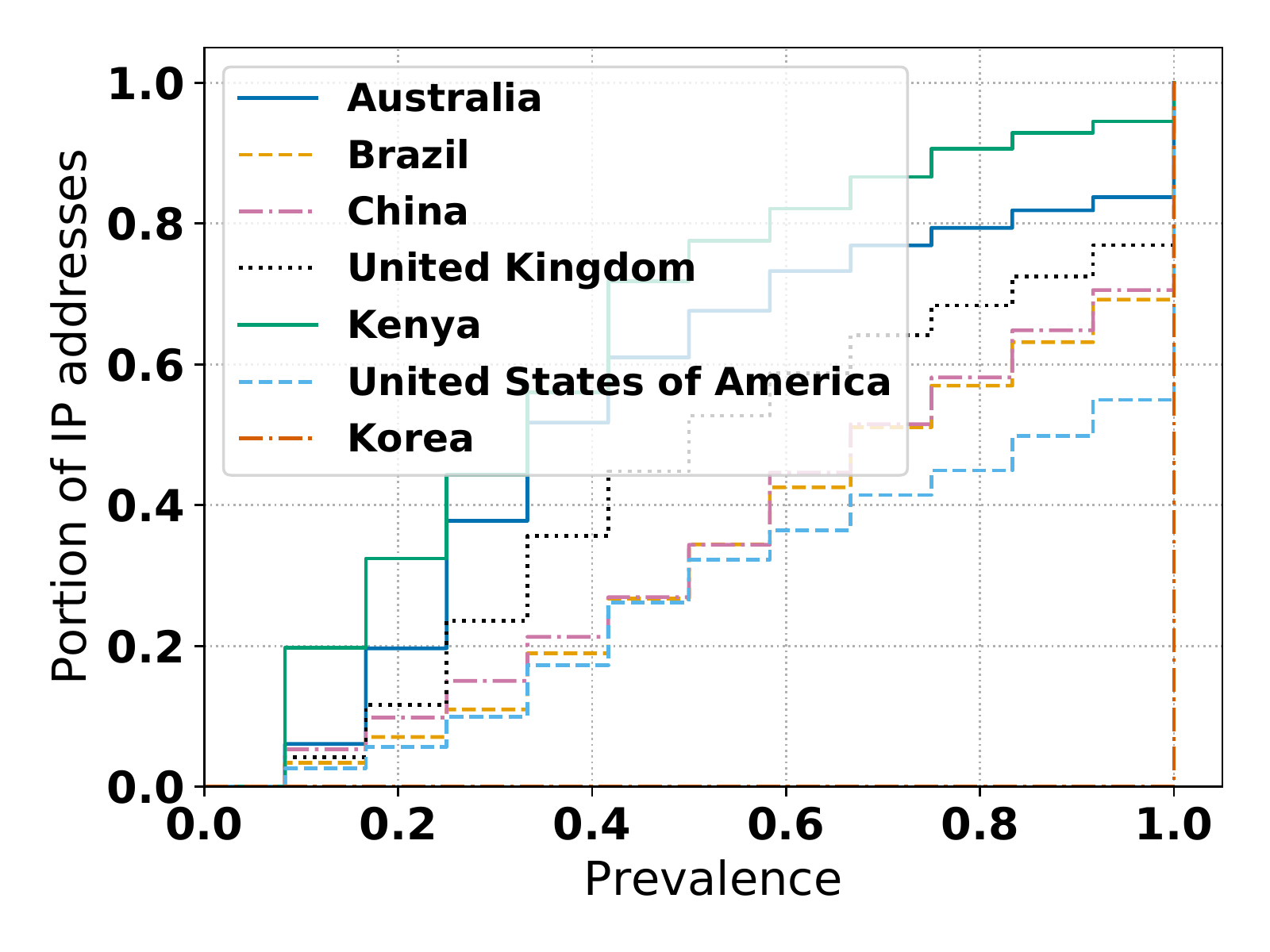}
  \caption{All (Eq.~\ref{eq:prevalence-location})}
  \label{fig:prevalence-by-country-all-city}
\end{subfigure}%
\\
\begin{subfigure}{.36\textwidth}
\centering
  \includegraphics[width=\linewidth]{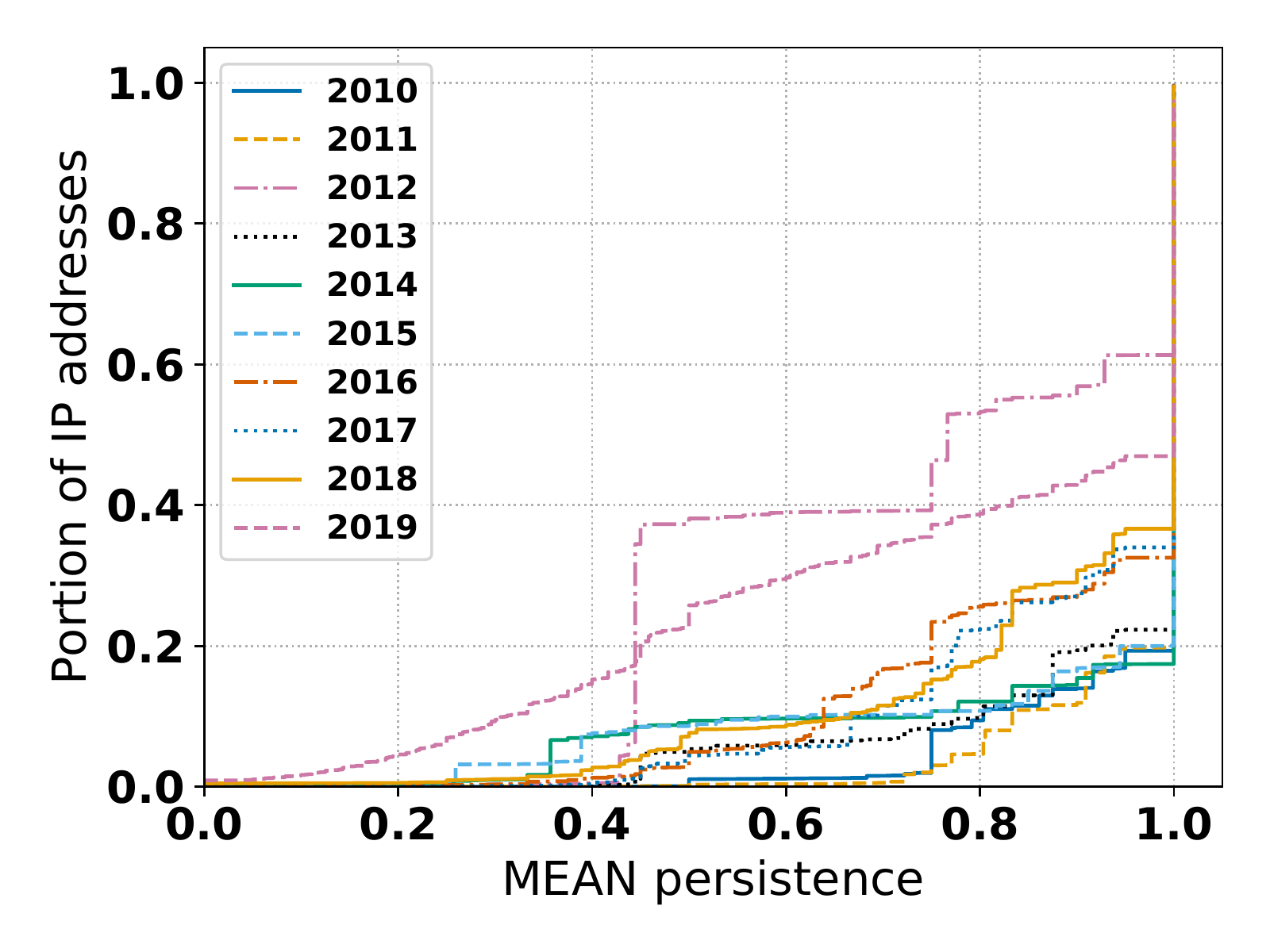}
  \caption{Over time (Eq.~\ref{eq:avg-persistence-ip})}
  \label{fig:persistence-over-time-city}
\end{subfigure}%
\begin{subfigure}{.36\textwidth}
\centering
  \includegraphics[width=\linewidth]{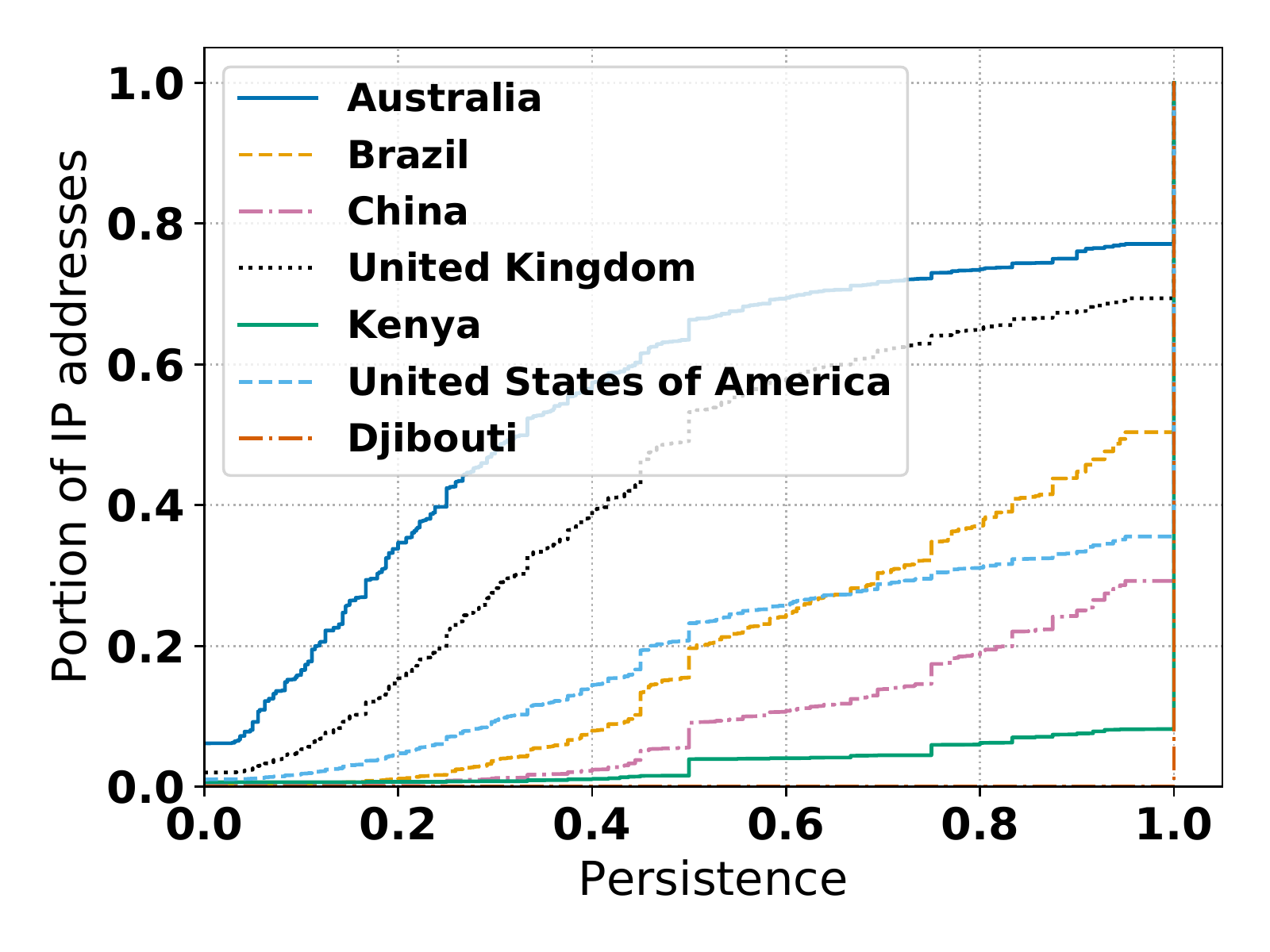}
  \caption{All (Eq.~\ref{eq:prevalence-location})}
  \label{fig:persistence-by-country-all-city}
\end{subfigure}%
\vspace{-2mm}
\caption{Decomposing city-level geolocation prevalence and
persistence, over time and
per country.  As compared to country locations, cities exhibit
both short and long term dynamics.}
  \label{fig:city-dynamics}
\end{figure*}

\paragraph{\textbf{Lesson}}
Statistically, \textbf{almost all IP 
addresses 
have a single country geolocation. However, when restricting
by orthogonal axes such as the class of IP address and country, 
this result is more nuanced}. 
\maxmind locations in such countries 
should be used with caution. There is a significant chance that the 
country may change for these IP addresses during the year. 

\subsubsection{City}
Fig.~\ref{fig:prevalence-over-time-city} 
shows the CDF of maximum prevalence at the city-granularity 
from 2010 to 2019 per
IP address. 
In contrast to the country-level, at the city-level
granularity, the CDF is not concentrated at a prevalence of 1.
In 2019, we observe that 38\% of the IP
 addresses have a maximum prevalence of less than 0.5, meaning that they do
 not have a dominant city.
Note that this result holds for all types of IP addresses. 
 
As we did for prevalence at country 
level,  Fig.~\ref{fig:prevalence-by-country-all-city} 
shows the results for the same five countries, plus the country with, 
this time, the
highest mean prevalence. The prevalence distribution for the five countries is 
not anymore all concentrated near one, and that at maximum, only 54\%
of IP addresses have a prevalence of one, corresponding to US.
Conversely, notice that for Korea, the prevalence distribution is located near 
one. We observed that other countries with high city mean prevalence were likely 
African countries. One hypothesis is that \maxmind have less information
for these countries and then IP addresses are less likely to change their location.


Fig.~\ref{fig:persistence-over-time-city} shows the CDF of the 
mean persistence per year from 2010 to 2019. 
We observe that the curve of 2019 is above all of the others, except 2012, 
so the mean persistence in 2019 was lower. 
30\% of the IP addresses have a mean persistence
under 0.5, while 50\% have a mean persistence of 1. 
This wide spectrum of mean persistence values show that both short 
and long term city location changes exist. 
Note that this result holds for all types of IP addresses, although we noticed 
that end host infrastructure had a higher persistence in general than the other
types of IP addresses.  

Again, we dig into the data to study if the city-level persistence 
depends on the country.
Fig.~\ref{fig:persistence-by-country-all-city} shows mean persistence
for the five countries, plus the one having the
highest mean persistence. 
Notice the example of Kenya that has a higher mean persistence than the other
countries. We observed that again, African countries tend to have a higher mean
persistence than countries on other continent.

\paragraph{\textbf{Lesson}}
The prevalence results show that \textbf{the majority of IP addresses 
covered by \maxmind are not mapped to a single city during the 2019 year.}
The wide range of values for persistence 
show that there are \textbf{both short and long term city changes}.

\subsection{Distance}\label{sec:evaluation:distance}
\begin{figure*}[t]

\begin{subfigure}{.36\textwidth}
\centering
  \includegraphics[width=\linewidth]{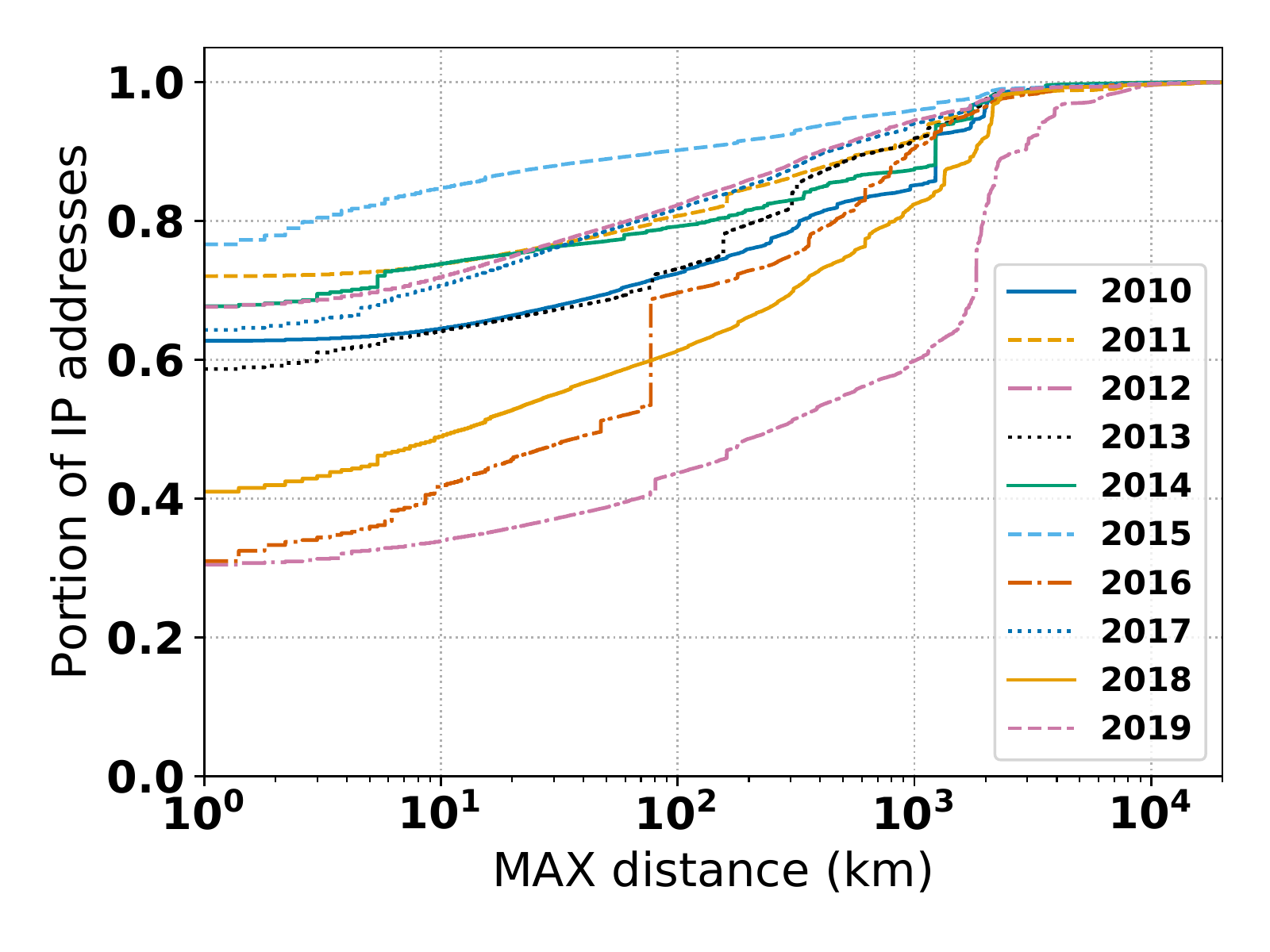}
  \caption{Over time (Eq.~\ref{eq:max-distance-metric})}
  \label{fig:distance-over-time}
\end{subfigure}%
\begin{subfigure}{.36\textwidth}
\centering
  \includegraphics[width=\linewidth]{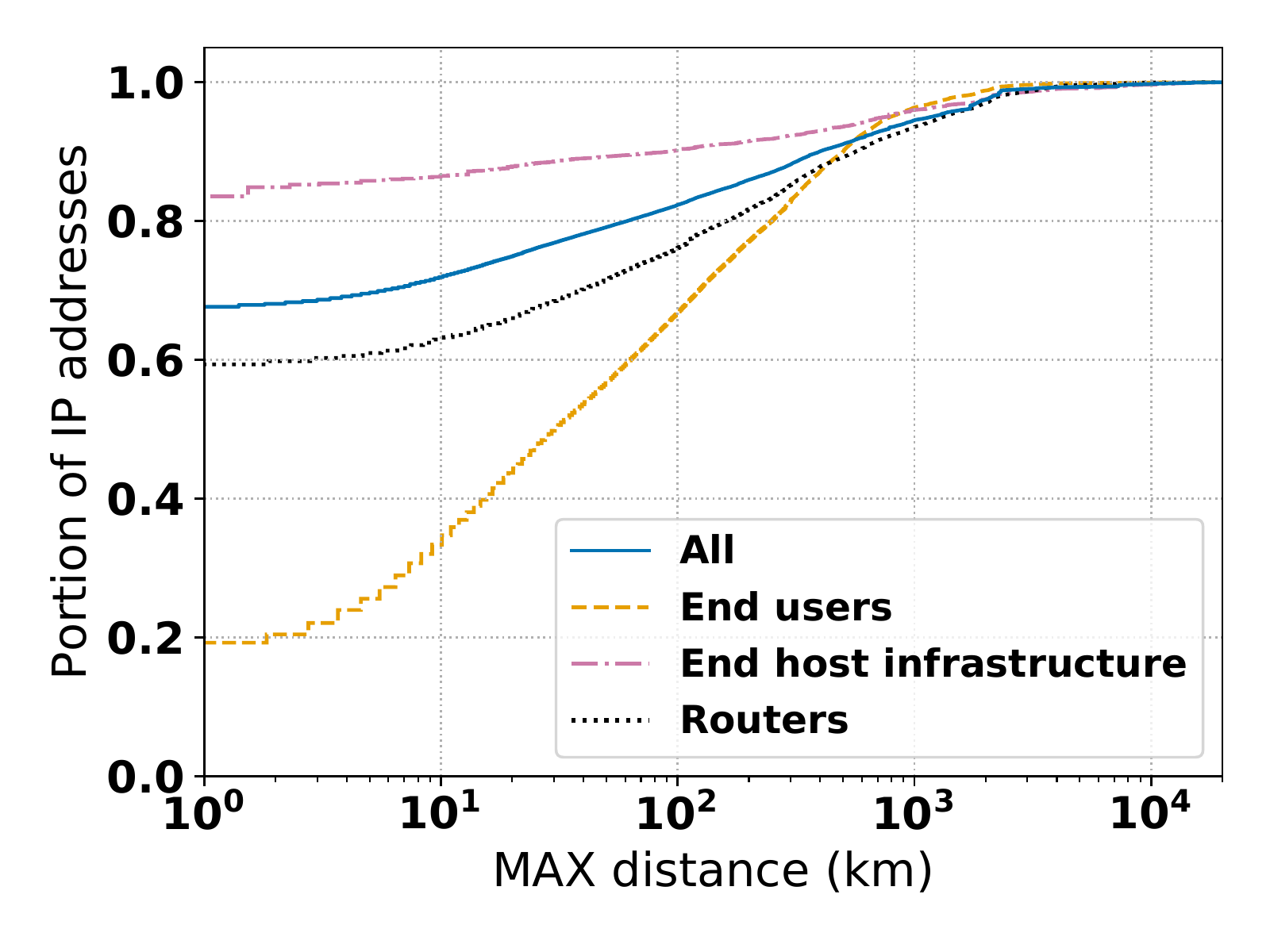}
  \caption{By IP class (Eq.~\ref{eq:max-distance-metric})}
  \label{fig:distance-by-ip-type}
\end{subfigure}%
\\
\begin{subfigure}{.36\textwidth}
\centering
  \includegraphics[width=\linewidth]{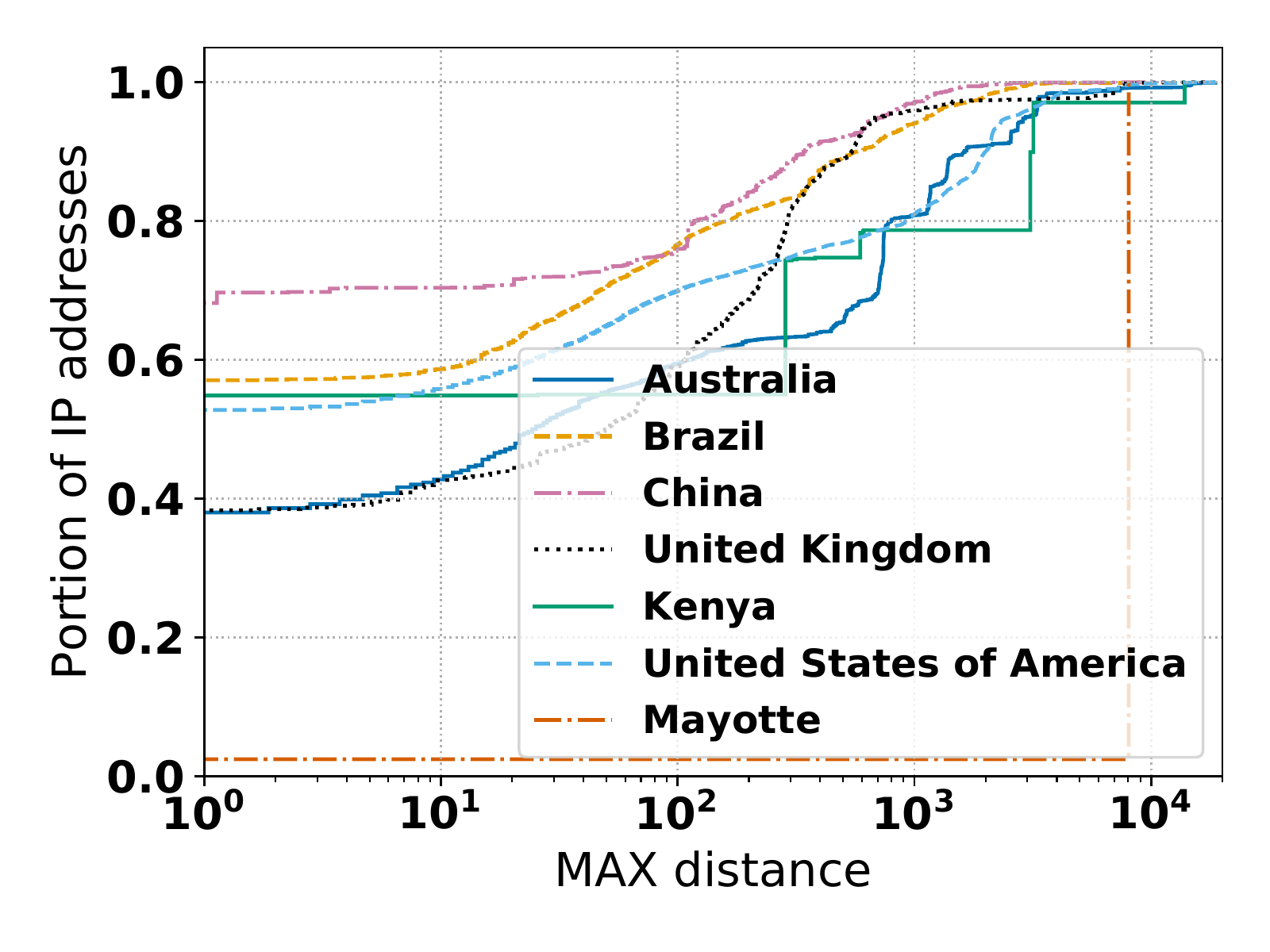}
  \caption{Routers (Eq.~\ref{eq:distance-metric-location})}
  \label{fig:distance-by-country-routers}
\end{subfigure}
 \vspace{-2mm}
 \caption{Quantifying distance (km) dynamics over time, IP class,
and selected countries.  While most distance changes are 
$\le$40km, there are classes of addresses and countries with significantly
higher movement.}
  \label{fig:distance-coordinates}
\end{figure*}

Results of the previous section on prevalence and persistence of the 
\maxmind city database show that per-IP city geolocations are 
highly dynamic. In this subsection, we quantify the distance of these
changes.
In contrast to Section~\ref{sec:evaluation:impact} that analyzes the distribution of
distance differences between two snapshots separated by different
hypothetical intervals of time, here we analyze the complete dynamics of  
the latitude / longitude coordinates over time as present in \maxmind.
For instance, whereas the previous analysis considers the distance
change over all addresses in the two snapshots, here we consider the
maximum distance change for each IP address across all snapshots.


\subsubsection{The maximum distance distribution is 
highly variable between years}
Fig.~\ref{fig:distance-over-time} 
shows the distribution of the maximum distance between 
two 
locations of an IP address during each year from 2010 to 2019. 
We do not observe any trend in distance change over time and 
each year exhibits a different distribution.
For instance, in 2018, 42\% of the IP addresses had a maximum distance change of more than 
40km, whereas only 22\% had this much movement in 2019.
The year 2012 is an outlier with 
40\% of the IP addresses having a 
maximum distance change of more than 1000km.

\subsubsection{The maximum distance depends on the type of IP address and
the country}
Fig.~\ref{fig:distance-by-ip-type} shows the maximum distance change
distribution in 2019 as a function of IP address class.
We observe that the distribution 
depends on the IP address type for distances under 1000km.
End user IP addresses
experience the most significant changes; 47\% have a maximum 
distance of more than 40km, as compared to 11\% for end 
host infrastructure and 28\% for routers.

We investigate the maximum distance of our five countries.
Fig.~\ref{fig:distance-by-country-routers} show the distance change
distributions for routers. 
Here again, we observe that the results are largely dependent on the country:
28\% of IP addresses in China have a maximum 
distance of more than 40 km, whereas it is 36\% for IP addresses in US, and
50\% for the United Kingdom.


\paragraph{\textbf{Lesson}}
Distance change results bring additional insights to country and city 
dynamics. 
Again, if statistically, 
\textbf{a majority of IP addresses have a maximum distance change less than
40km}, 
digging into the data
reveals a \textbf{high disparity between IP address type and country
with respect to geolocation change distance.}

\subsection{Visualizing Internet-wide geo movement}
\begin{figure}[t]
\begin{subfigure}{.33\textwidth}
\centering
  \includegraphics[width=0.95\linewidth]{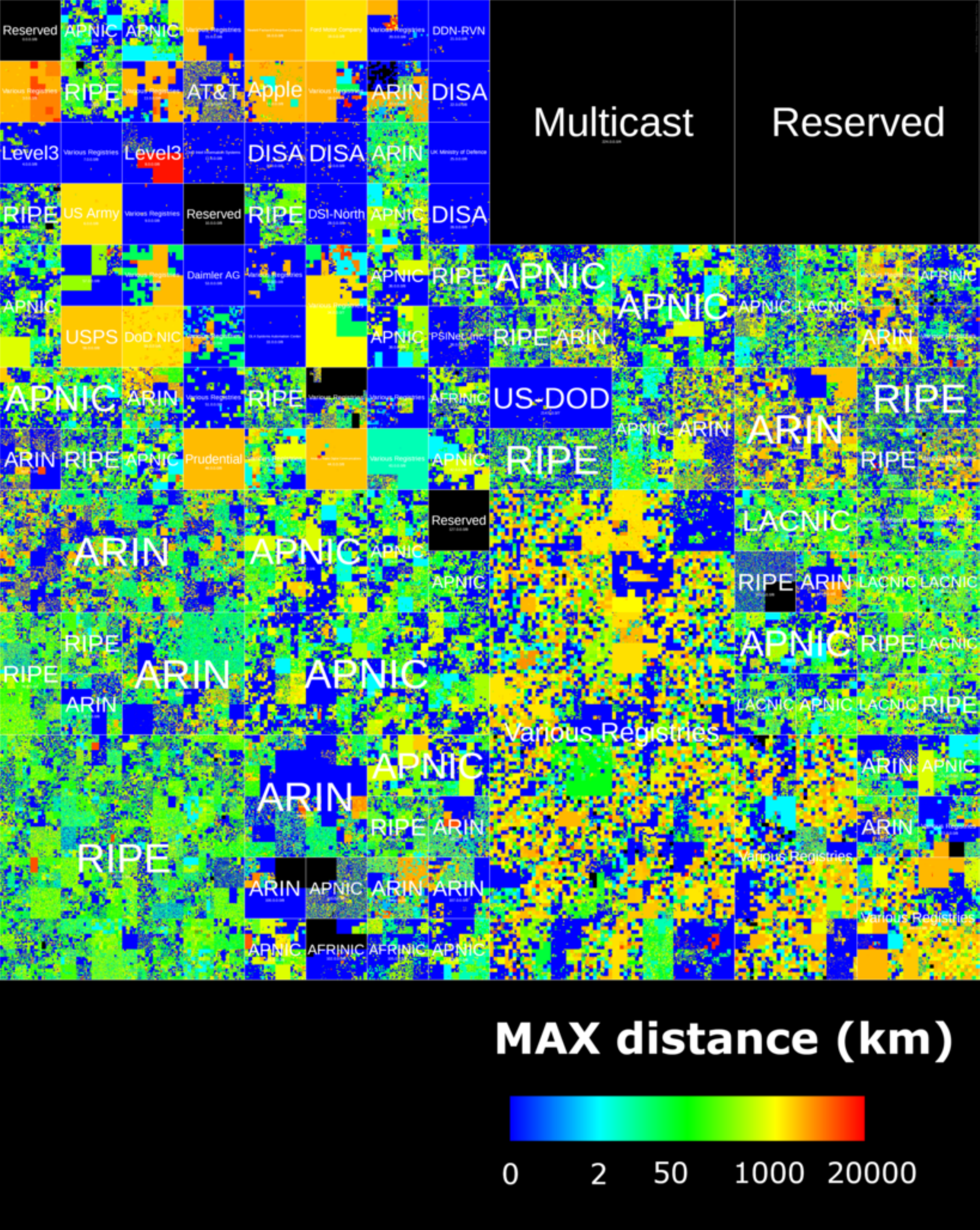}
  \end{subfigure}%
  \begin{subfigure}{.33\textwidth}
\centering
  \includegraphics[width=0.95\linewidth]{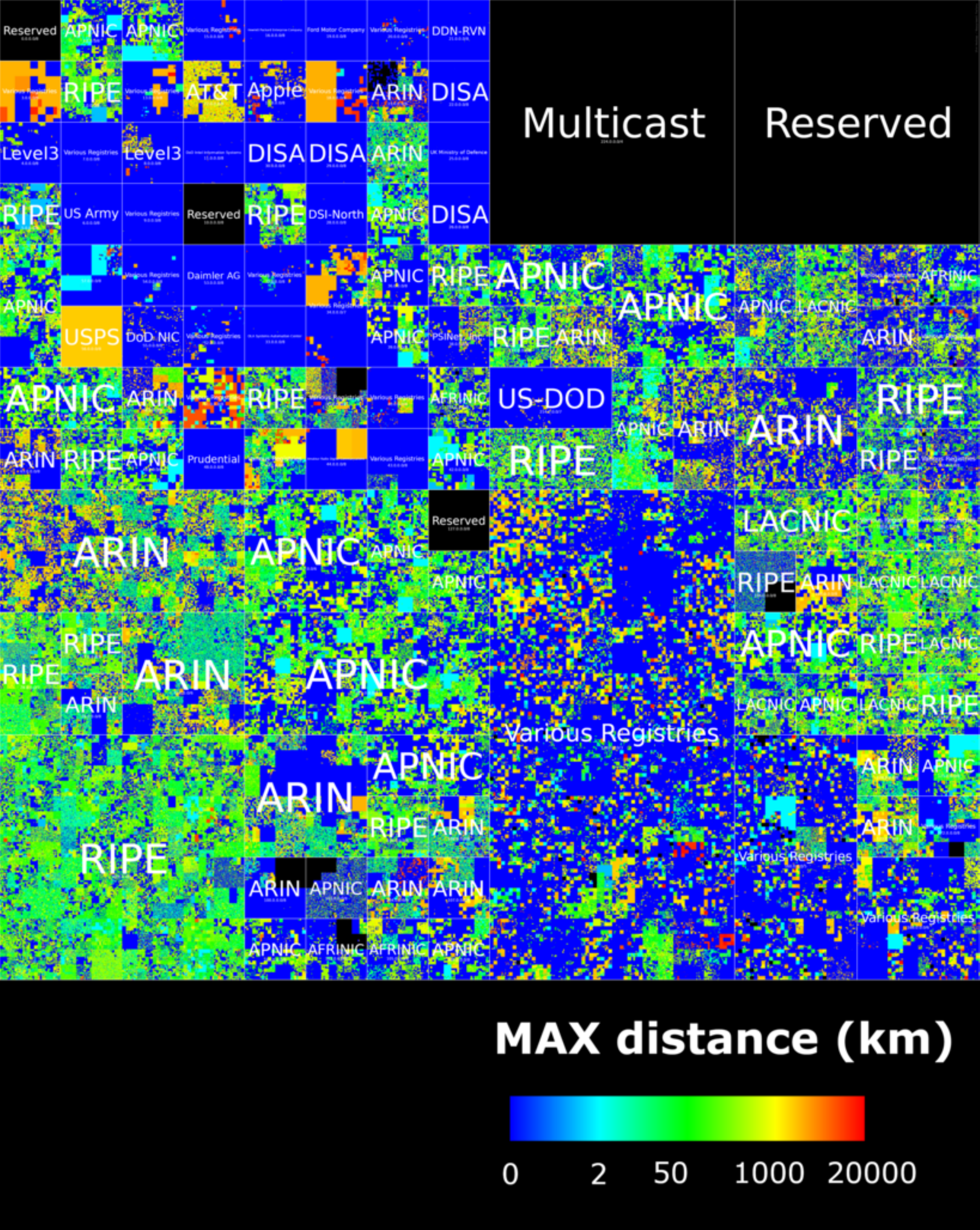}
  \end{subfigure}%
  \begin{subfigure}{.33\textwidth}
\centering
  \includegraphics[width=0.95\linewidth]{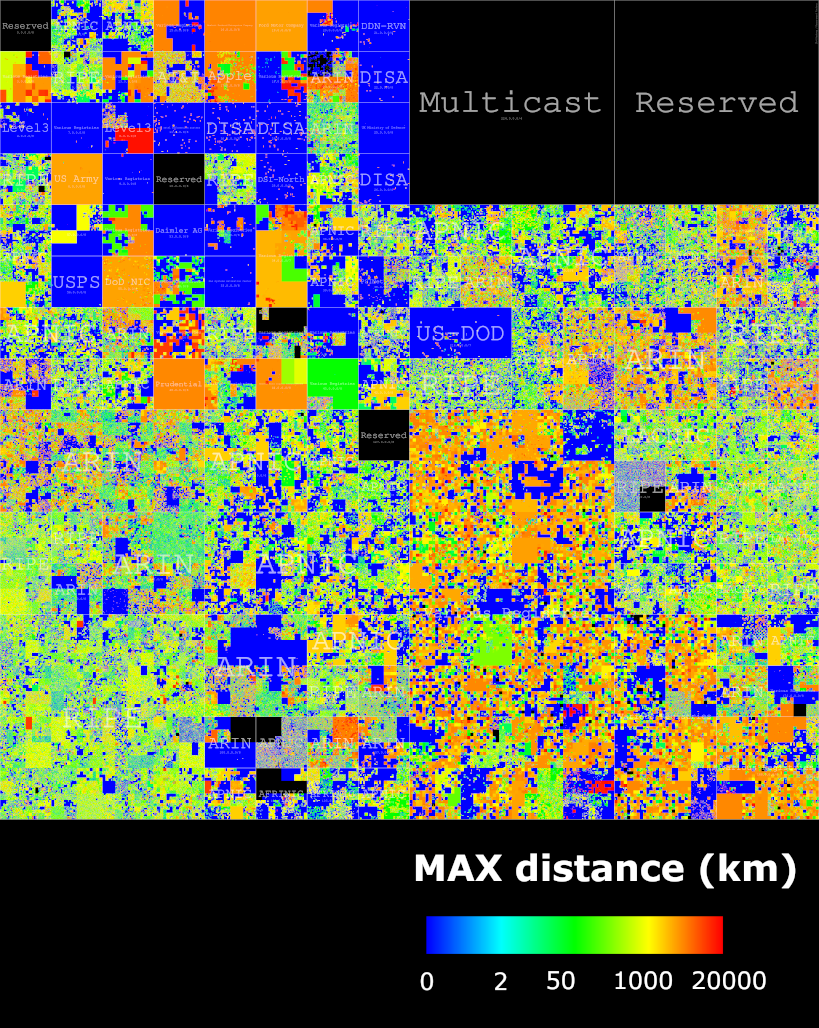}
\end{subfigure}%
  \caption{Hilbert heatmap of maximum distance 
  on a logarithmic scale (Eq.~\ref{eq:max-distance-metric}) in 
  2018 (left) 2019 (center) and absolute difference (right). There is a high degree of global
  geolocation dynamics and year-to-year variation.}
    \label{fig:hilbert}  
\end{figure}

Fig.~\ref{fig:hilbert} shows an exhaustive representation of the 
maximum distance change
for each /24 of the entire IPv4 space for 2018 and 2019, as well as 
the absolute difference between the two years. 
Each pixel represents a /24, and the color represents the maximum distance 
between two locations; black pixels indicate that the IP address is not
present in the database. 
If the /24 contains more specific entries in the \maxmind database, 
we take
the maximum of the maximum distance of the IP addresses within the /24. 

We see that the visualization of 2019 differs from 2018:
Many of the various registries in the bottom center right and top left part of 
the plots have a maximum distance of more than 1000km in 2018, whereas 
they did not move in 2019.
There is also a red square in the prefixes belonging to Level3, that had a 
maximum distance of 20,000km in 2018 but did not move in 2019.

Surprisingly, there are also some IP addresses that were covered in 2018
 (\ie in this case, having lat/long coordinates) which are not covered in 2019. 
This is the case of some blocks of IP addresses in the bottom center left of the 
graph belonging to APNIC and AFRINIC. 

All these differences between the two years are highlighted by the  map 
on the right: we clearly see the center 
and the bottom left mainly colored in orange and red as well as some big 
prefixes on the top right. 
It reveals a significant dynamic changes not only along the prefixes but also 
through time. 

Overall, by looking at the Hilbert representations of each year 
over the 10 years dataset, it is difficult to perceive a trend that could lead 
us to say that prefixes are experiencing bigger or smaller distance
change over years.

\begin{figure}[t]
\includegraphics[width=.36\linewidth]{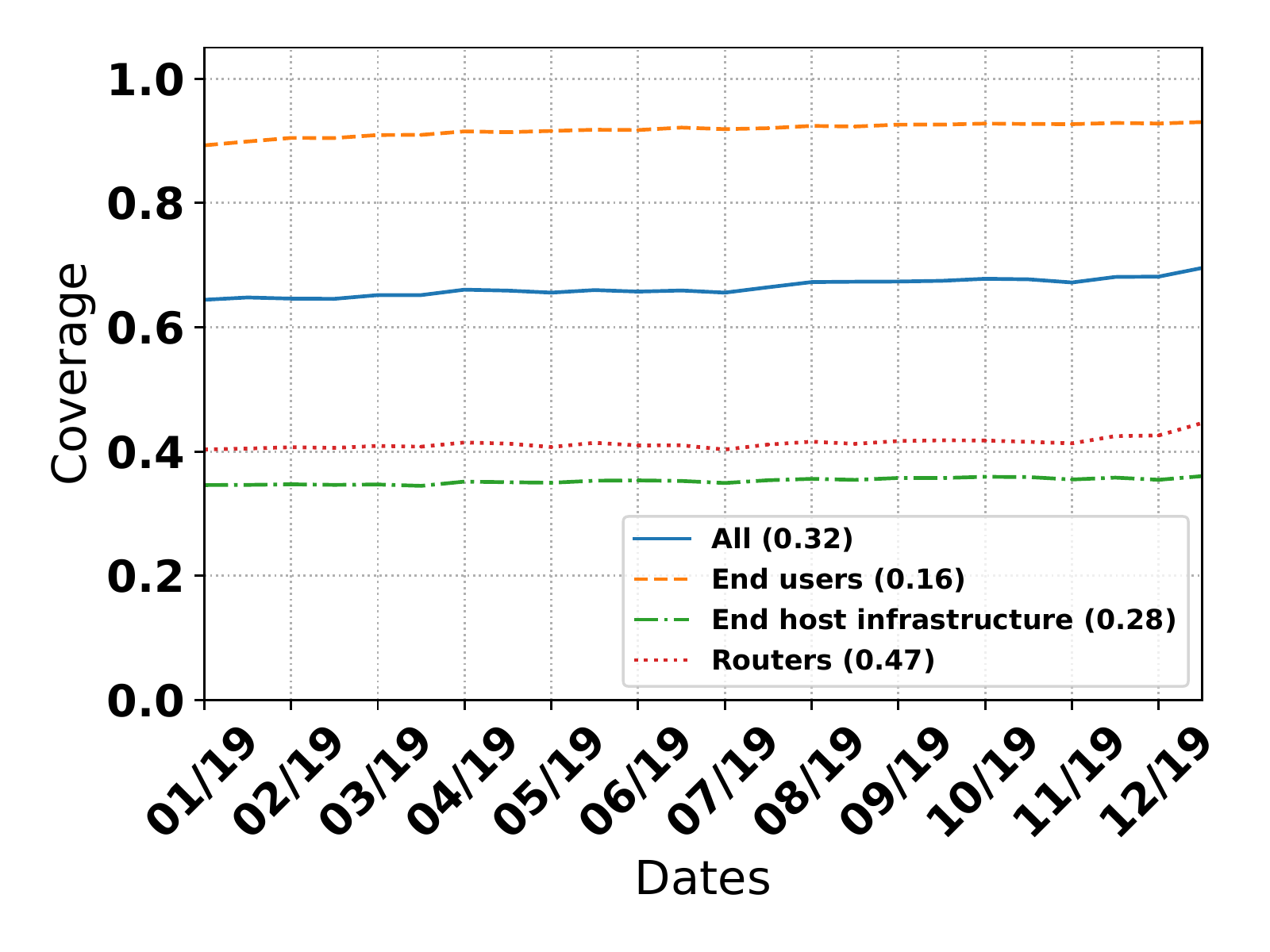}
 \caption{Coverage by address type}
  \label{fig:coverage-full-with-jaccard}
\end{figure}

Finally, we consider a different notion of coverage -- the coverage
relative to routed IP address space.  To assess routed
address space coverage, we utilize the global BGP tables available
from Route Views~\cite{routeviews} coinciding with the date of each
\maxmind snapshot.
Fig.~\ref{fig:coverage-full-with-jaccard} shows two things: first, 
the overall coverage is relatively constant over time for all classes of IP addresses.
However, the coverage value strongly 
depends on the type of address. There is about 90\% coverage 
for end users, 37\% for end host infrastructure and 41\% for routers. 
The Jaccard distance is interesting: going from 0.16 
for end users, to 0.47 for routers. 
This means that even if the coverage is constant over time, the set of 
IP addresses covered varies significantly.

\paragraph{\textbf{Lesson}}
These visualizations confirm not only the \textbf{high degree of global geolocation
dynamics}, but also the presence of \textbf{year-to-year variation in
geolocation movement}.
An IP address can experience a maximum distance change of 0km in 2018 
and more than 1000km in 2019, and vice versa.
Also, depending of the address type, the BGP coverage will change significantly.

\subsection{Impact}\label{sec:evaluation:impact}

While the preceding analysis demonstrates how our metrics can shed 
light on the underlying dynamics of a geolocation database, we
conclude this section with an analysis of the potential \emph{impact}
of selecting a particular snapshot of \maxmind versus a different
snapshot, for instance as a researcher seeking to geolocation a 
population of IP addresses under study.

To bound our
results, we compare pairs of snapshots from 2019 within three time windows:
when the snapshots differ by
less than 3 months, between 3 to 6 months,
and between 6 to 12 months.
We evaluate the impact across the four IP classes:
all IP addresses (``All''), 
end users, end host infrastructure, and routers.

\begin{figure*}[h]
\centering

\begin{subfigure}{.33\textwidth}
\centering
  \includegraphics[width=\linewidth]{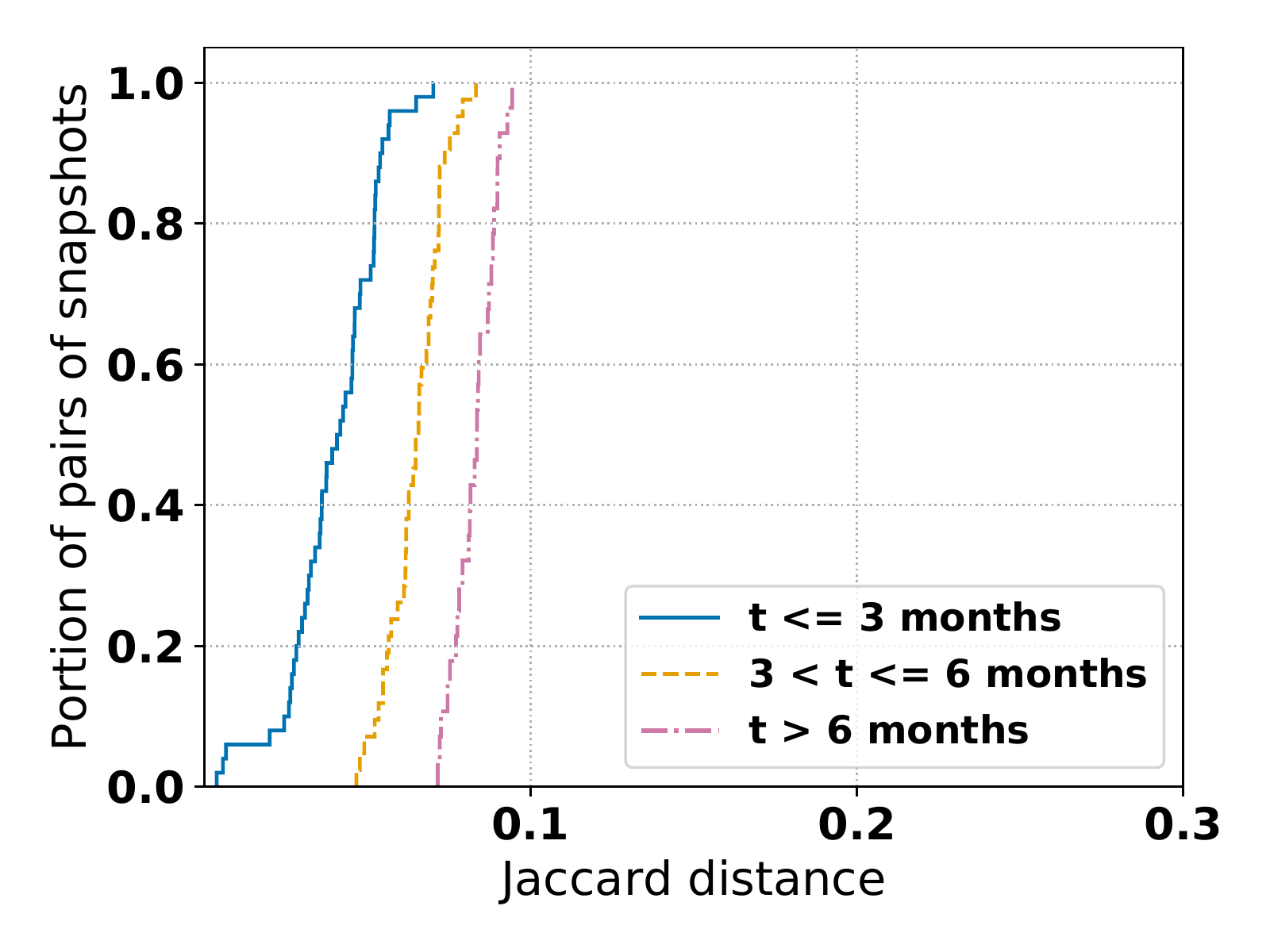}
  \caption{End users}
  \label{fig:coverage-variation-end-users}
  \end{subfigure}%
\begin{subfigure}{.33\textwidth}
\centering
  \includegraphics[width=\linewidth]{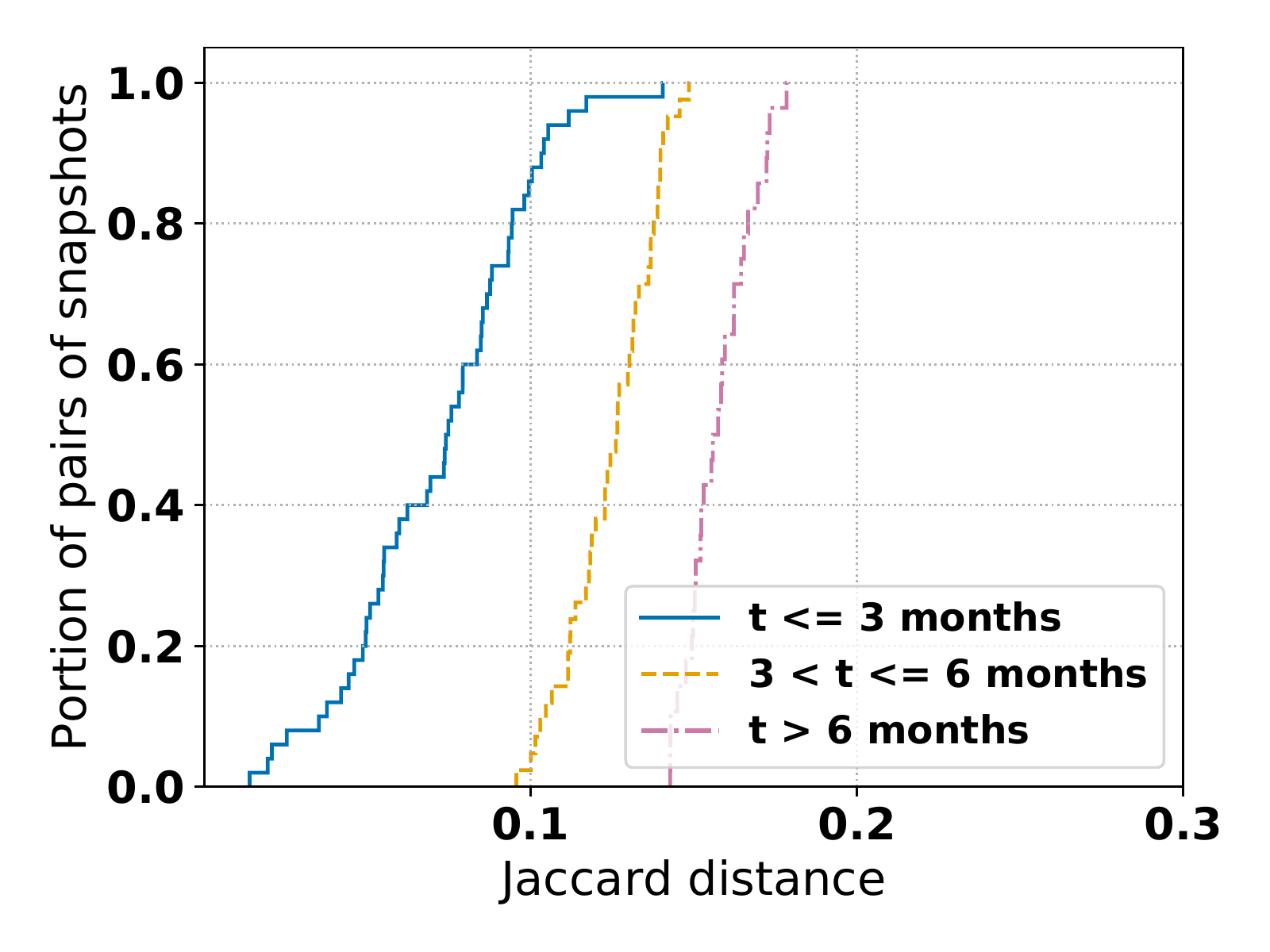}
  \caption{End host infrastructure}
  \label{fig:coverage-variation-infrastructure}
  \end{subfigure}%
\begin{subfigure}{.33\textwidth}
\centering
  \includegraphics[width=\linewidth]{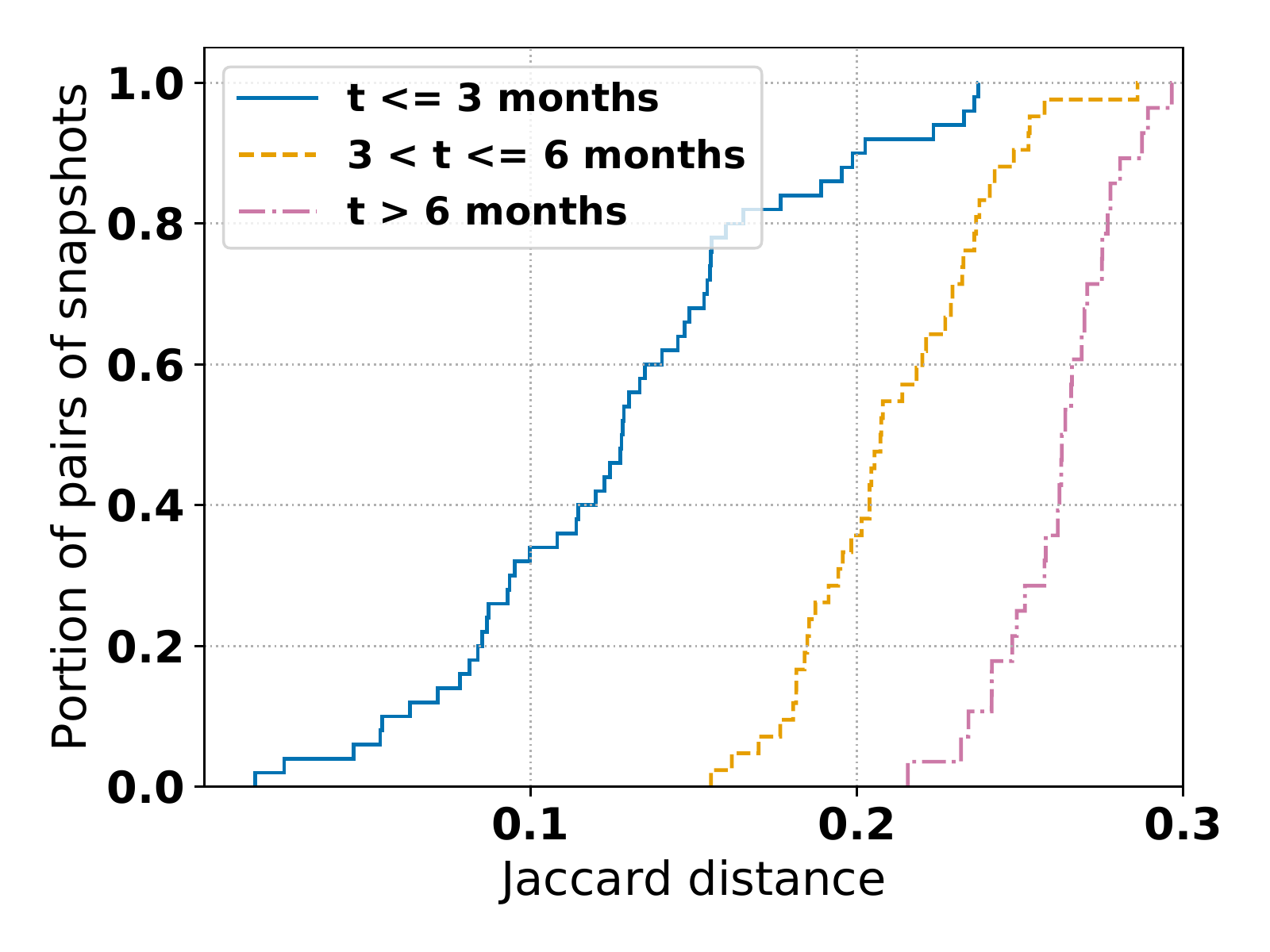}
  \caption{Routers}
  \label{fig:coverage-variation-routers}
\end{subfigure}
\caption{Comparing pairs of database snapshots by city coverage 
         difference (Eq.~\ref{eq:delta-coverage}).  Across all 
         classes of IP addresses, there are significant coverage
         differences, even on among snapshots closely separated in
         time.}
\label{fig:coverage-variability}
\end{figure*}

\begin{figure*}[h]
\begin{subfigure}{.33\textwidth}
\centering
  \includegraphics[width=\linewidth]{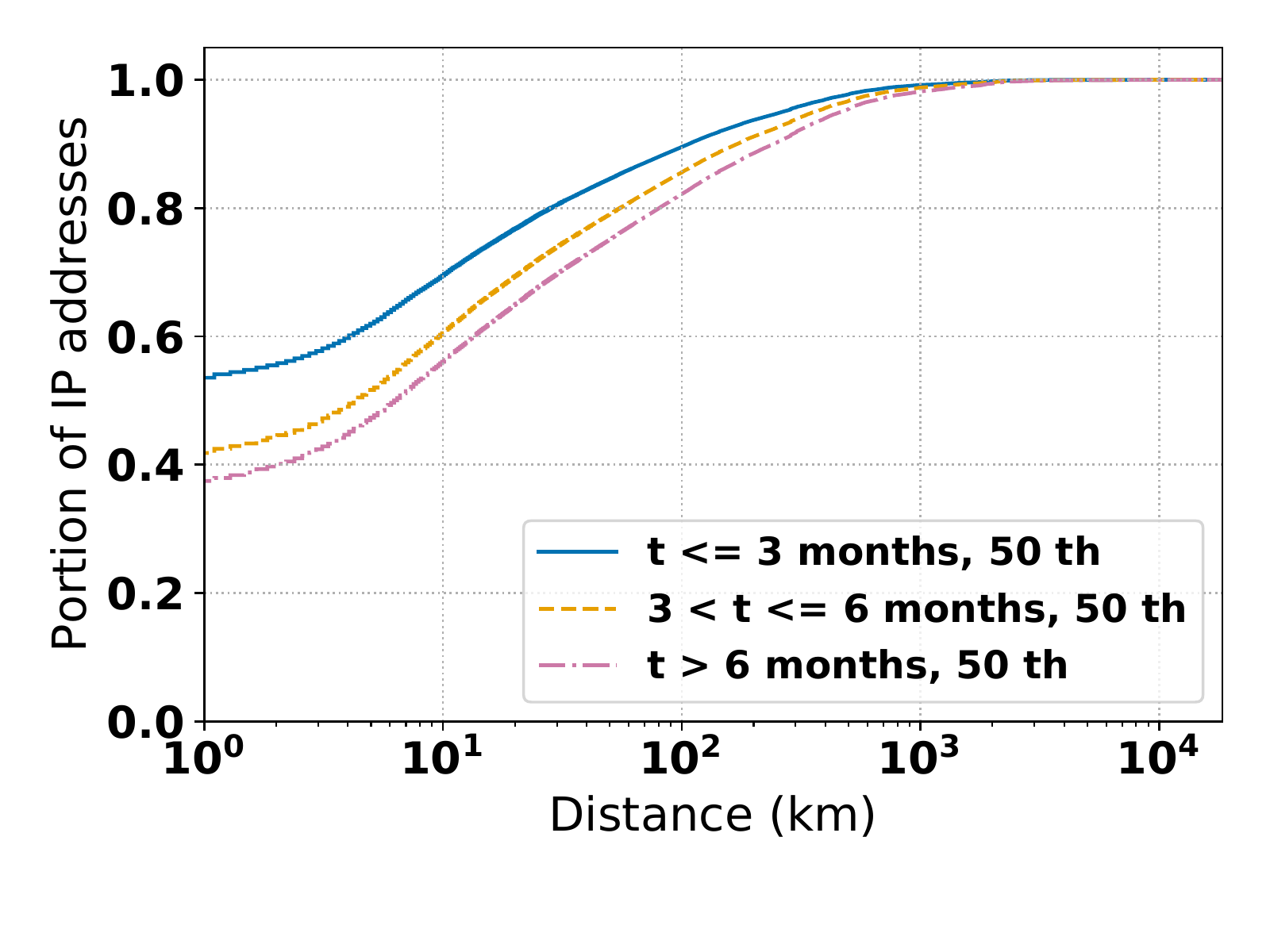}
  \caption{End users}
  \label{fig:distance-variation-end-users}
\end{subfigure}%
\begin{subfigure}{.33\textwidth}
\centering
  \includegraphics[width=\linewidth]{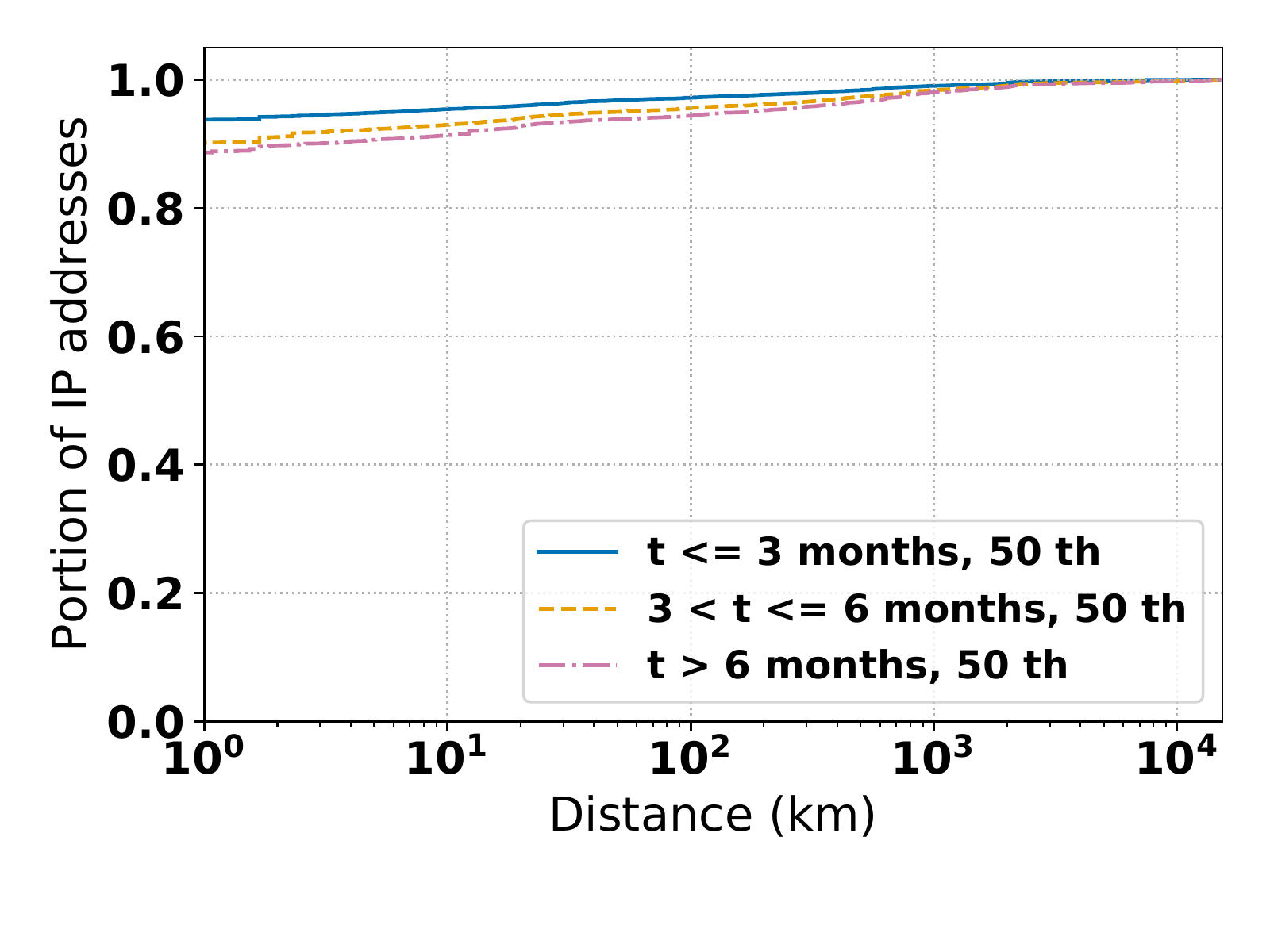}
  \caption{End host infrastructure}
  \label{fig:distance-variation-infrastructure}
\end{subfigure}%
\begin{subfigure}{.33\textwidth}
\centering
  \includegraphics[width=\linewidth]{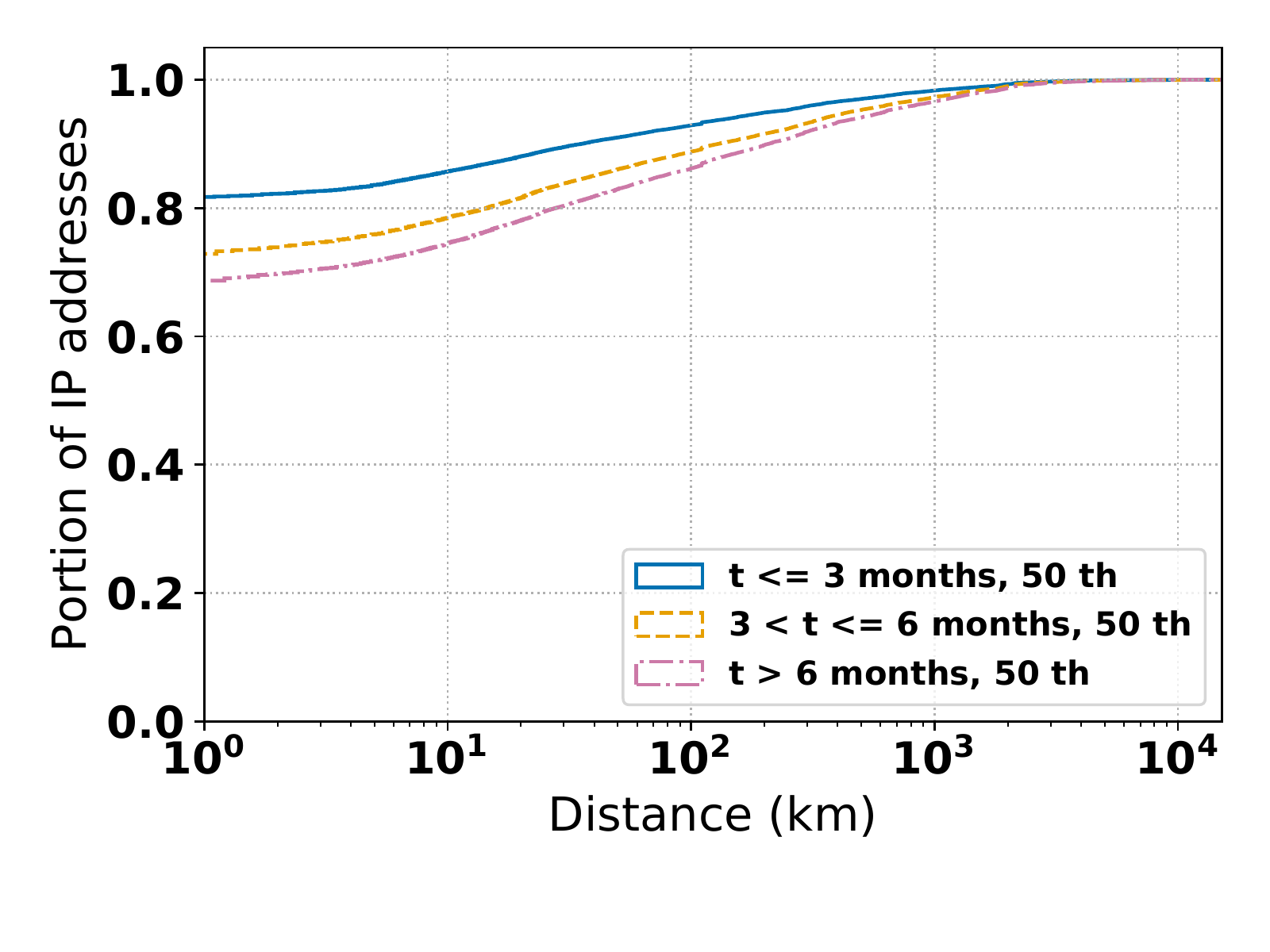}
  \caption{Routers}
  \label{fig:distance-variation-routers}
\end{subfigure}
\caption{Comparing pairs of database snapshots by IP address distance 
         difference (Eq.~\ref{eq:delta-distance-distribution}).  Up to
         22\% of addresses move more than 40km among snapshots in a 
         three month window.}
\label{fig:movement-variability}
\end{figure*}

\subsubsection{Coverage (Fig.~\ref{fig:coverage-variability})}
\label{sec:impact-coverage-evaluation}
For coverage we show results at the city level as we find 
no 
country level coverage differences between snapshots; almost all IP
addresses, across all classes of addresses,
have a country geolocation present in the database.
%

Not unexpectedly, 
for all types of IP addresses, we observe that the coverage 
difference increases with time
between the two 
snapshots. 
The overall 
coverage being globally constant (see Fig.~\ref{fig:coverage-full-with-jaccard}), 
this cannot be imputed to an increase of the
total coverage. 

We see that even for two 
snapshots created within less than three months of each other,
there 
is a significant coverage difference, 
up to 6\%, 11\% and 20\% for end users, 
end host infrastructure
and routers respectively.  As seen in Fig.~\ref{fig:coverage-variation-routers},
there is a 50\% probability of more than 12\% coverage difference
between two snapshots created less than three months apart.
Between two snapshots of more than six months and less than a year, the 
difference can be even worse, up to 9\%, 17\% and 30\%.

\subsubsection{Distance (Fig.~\ref{fig:movement-variability})}\label{sec:impact-distance-evaluation}

We first sort the pairs of snapshots by the metric defined in 
Eq.~\ref{eq:distance-metric-average}, the mean of the logarithmic 
distances (MLD).
Recall, the higher the MLD, the more the snapshots differ.
We compute distance across pairs of snapshots within the same time
ranges as for
coverage: less than three months, between three and
six months, and between six and twelve months.
From the MLD distribution,
we then show
the pair of snapshots corresponding to the median.  
For example, in Fig.~\ref{fig:distance-variation-end-users}, one should read:
On the end users dataset, 15\% of the IP addresses moved 
by at least
40km. This corresponds to the median result for a pair 
of snapshots that are less than three months apart.

Fig.~\ref{fig:movement-variability} shows two trends. First, as one might expect, 
for all types of IP addresses, the more time between two snapshots, 
the more IP addresses move.
Then, the percentage of IP addresses moving depend on 
the type of IP addresses. 
We observe that end users tend to move more than routers and end host 
infrastructure.
In details, for a pair of snapshots that are more than six months apart, 
we have 28\%, 8\% and 18\% of IP addresses that move 
more than 
40km 
for respectively end users, end host infrastructure, and routers.

If we consider that 40km corresponds to most
metropolitan areas~\cite{gharaibeh_2017}, this implies that a non-trivial 
portion of IP addresses 
experience a location change out of the metro area -- a significant change.
However, distances greater than 1000km are rare, accounting for less than 
5\% of IP addresses across all addresses classes.


\paragraph{\textbf{Lesson}}
There are non negligible differences in both coverage and distance of
movement even for
database snapshots created closely in time (< 3 months).
Therefore:
\textbf{we recommend, insofar as possible, aligning geolocation
database snapshots with the 
measurements that produced them}, for instance by programatically
using an 
API to lookup IP addresses on-demand as they are gathered.  Further, one should 
look at several snapshots closely spaced in time over the measurement
period
and more deeply \textbf{investigate IP addresses
that experienced significant changes}.

%% file: usecases.tex
\section{Use Case}
\label{sec:usecases}
Previous sections have shown two things.
\maxmind is a widely used database (Sec.~\ref{sec:survey}),
and selecting a particular snapshot in a time period can have a significant impact
on the results (Sec.~\ref{sec:evaluation}). In this section, we
concretely demonstrate the potential impact on research that depends on
\maxmind 
by reproducing the results from 
Gharaibeh \etal's IMC 2017 work~\cite{gharaibeh_2017} 
with different
\maxmind snapshots.
Gharaibeh \etal study the accuracy of different databases 
for router geolocation, including \maxmind.
Using the author's publicly available ground truth, 
we reproduce their
accuracy results (see Sec.\ 5.2, Fig.\ 2 of \cite{gharaibeh_2017}).

Surprisingly, the \maxmind snapshot that produces the largest impact
on the results was created within only two months of the snapshot used
by the authors.
This snapshot shifts the median of the distance distribution to ground truth
from more than 100 km
to 40 km, which is close to the results of the paid version.
Given this variability, the claim that the free version 
of \maxmind is worse than the paid version
seems to depend on the specific instance studied.

\subsection{Dataset}
The Gharaibeh dataset consists of 16,586 router interface IP addresses with 
corresponding ground truth locations 
inferred either with RTT-based measurements or DNS-based techniques. 
The authors do not mention which specific snapshot 
of \maxmind
they used, however: ``The databases are accessed again on early July 
2016, to geolocate the ground truth.'' 
We inquired with the authors for the exact snapshot date, 
but unfortunately they could not be more specific.
We therefore select the closest snapshot as our reference,
from July 8, 2016.  Sec.~\ref{sec:use-case:results} confirms that 
the results of this snapshot are very close to those presented in the
original paper.

The measurement period for the ground-truth collection and validation, however,
spanned a larger time period.  As stated in the paper:
``Overall, between May 2016 and September 2017, 
8,197 (69.1\%) [...] have different hostnames, and 6.9\% no longer have 
rDNS records.''  
We therefore restrict our comparison between snapshots belonging
to this period of time, on which the authors consider that 
the ground truth is valid.

\subsection{Results}\label{sec:use-case:results}
\begin{figure}
\centering
\captionsetup[subfigure]{position=t}
\begin{subfigure}{.35\textwidth}
\centering
  \includegraphics[width=\linewidth]{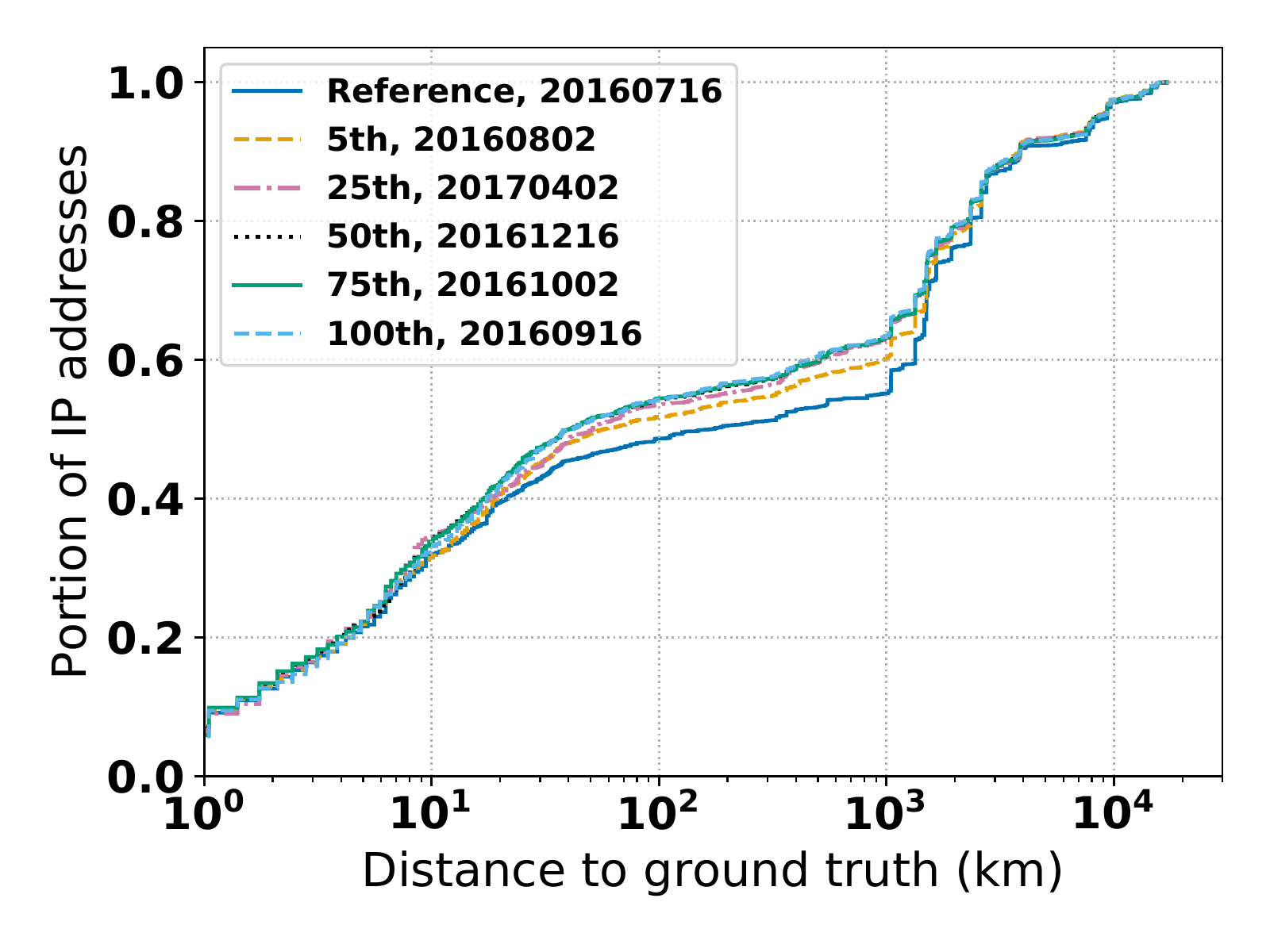}
  \caption{Distance to ground truth CDF on different snapshots}
  \label{fig:distance-ground-truth-huffaker}
\end{subfigure}%
\begin{subfigure}{.35\textwidth}
\centering
  \includegraphics[width=\linewidth]{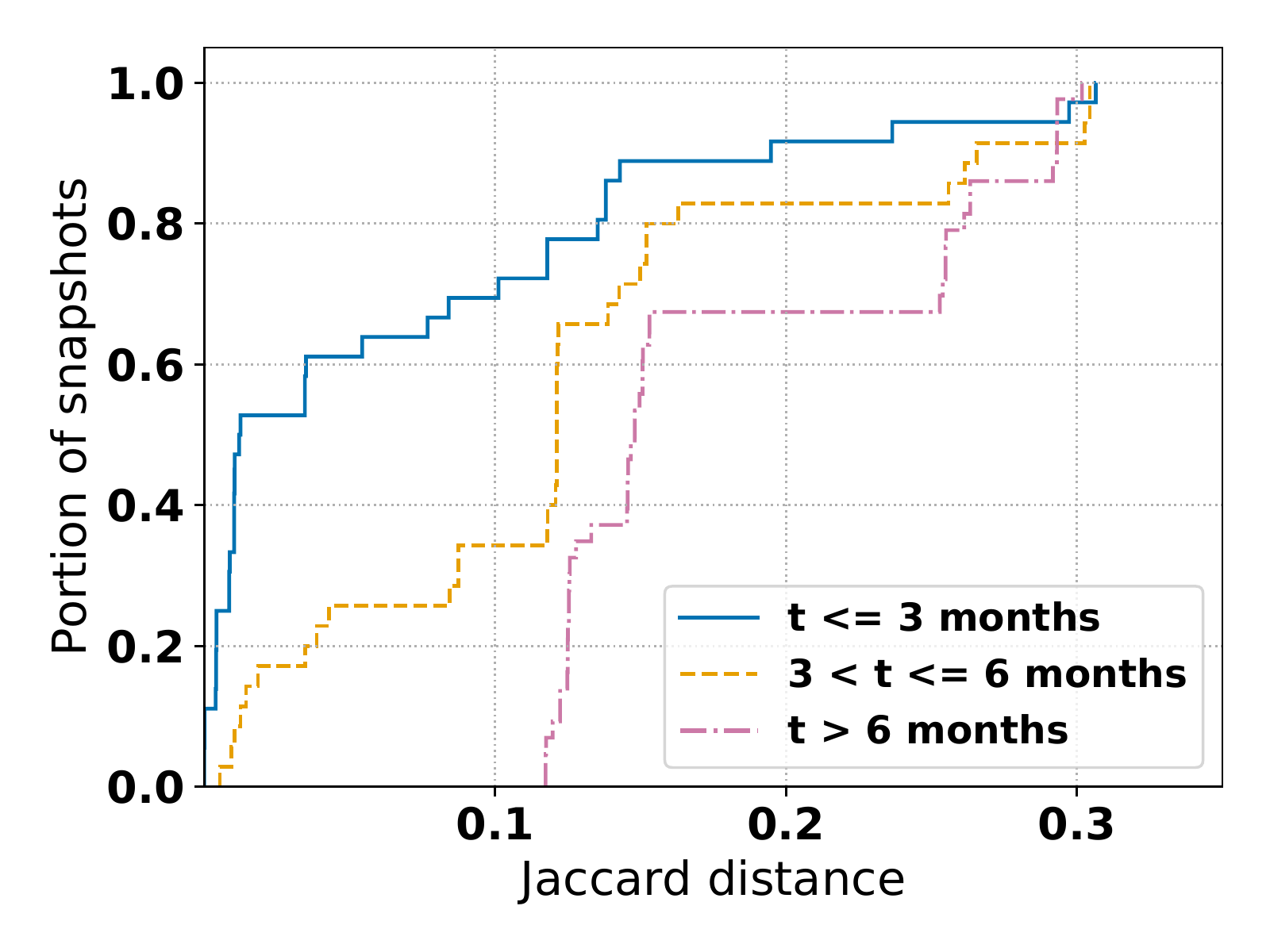}
  \caption{Coverage difference on city with the reference snapshot}
  \label{fig:coverage-variation-huffaker}
\end{subfigure}
\vspace{-3mm}
\caption{Reproducing result of~\cite{gharaibeh_2017} with different snapshots}
\end{figure}

\subsubsection{Distance to ground truth}
We compute the distance to ground truth distribution
of all the snapshots from May 2016 to September 2017.
We then compare each of these distributions to the distribution of the reference
snapshot, using the Kolmogorov-Smirnov (KS) test~\cite{lilliefors1967kolmogorov}.
The KS test quantifies how 
dissimilar are two distributions, with higher values indicating less
similarity.
Fig.~\ref{fig:distance-ground-truth-huffaker} shows the distance to 
ground truth 
distribution of the snapshots 
corresponding to the 5th, 25th, 50th, 75th, and 
100th percentiles of the KS distribution. We
also show the
reference snapshot. 

First, we compare Fig.2 of Gharaibeh \etal.  with
our reference snapshot.
We infer that on Fig.2 of Gharaibeh \etal., 
$\sim$8\%, 47\%, 50\%, 55\%, and 96\%,   
are located at less than respectively 1, 40, 100, 1,000, and 10,000km
from the ground truth, whereas it is 8\%, 46\%, 49\%, 56\% and 97\% in 
our reference snapshot.
Overall, the shape of the distribution looks the same, so 
this give us confidence that our July 8th snapshot is a good reference.

Then, we observe that there are significant differences between 
the other snapshots and the reference. 
The median shifts from 167km in the reference snapshot to respectively 
57, 51, 41, 40 and 40km for the 5th, 25th, 50th, 75th and 100th percentile.

When we consider the snapshots dates with these percentiles, it is surprising to observe
that the 100th percentile was created only two months after the reference 
snapshot, whereas the 5th percentile is a snapshot taken one month later and 
the 25th percentile corresponds to a snapshot taken nine months later. 
This implies that \maxmind did not improve over time for these
addresses,
but also that there are significant differences in the results within 
a relatively short time.

Finally, we look at the comparison between the free and paid
version of \maxmind.
On Fig.2 of Gharaibeh \etal
we infer that the paid version has a median between 30 and 
40km, so that the difference between this distribution and the different
snapshots of Fig.~\ref{fig:distance-ground-truth-huffaker}
is less pronounced than the difference between the free and paid version
of their graph. 
Therefore the conclusion that the paid version performs better than the 
free one should be taken with caution.

\subsubsection{Coverage}
Finally, we are interested to compute the variability in coverage as defined in 
\S\ref{sec:methodology:coverage}. 
In Gharaibeh \etal, the authors only compute the distance to ground 
truth if the IP address is covered by \maxmind at the city level. 

Fig.~\ref{fig:coverage-variation-huffaker} shows the distribution of the 
coverage difference (Eq.~\ref{eq:delta-coverage}) 
as we did in \S.~\ref{sec:impact-coverage-evaluation}, but only comparing 
snapshots with the reference snapshot.
We observe that even with snapshots taken 3 months apart from the reference 
snapshot, 77\% of the snapshots have more than 23\% of coverage difference.
It is even worse for snapshots between 3 and 6 months and snapshots with more
than 6 months of difference, with a coverage difference of about 30\%.
This means that if the authors had used a different snapshot, the
set of IP addresses over which they would have computed their accuracy
measures would have significantly changed.

%% file: related-work.tex
\section{Related work}

Mapping IP addresses to the physical world is an important topic that
has seen two decades of research.  Early efforts used landmarks, hosts
with known position, to assign locations to unknown targets at coarse
granularity~\cite{10.1145/383059.383073}.  Landmark-based geolocation
was subsequently enhanced to use latency
constraints~\cite{gueye2006constraint}, network
topology~\cite{katz_2006}, and population
densities~\cite{eriksson_2010} to improve accuracy.  Because the
accuracy of latency-based techniques is often proportional to the
distance between the target and its nearest landmark, Wang \etal
developed techniques to find and utilize additional
landmarks~\cite{wang_2011}.

IP geolocation has since matured, with several competing commercial
offerings including~\cite{edgescape,maxmind,hexasoft}.  While the
exact methodology of these commercial services is proprietary, they
likely use a combination of databases (\eg whois and DNS), topology,
latency, and privileged data feeds from providers~\cite{geofeeds}.

Even so, the inference-based nature of IP geolocation imparts errors
and inaccuracies even in commercial
databases~\cite{huffaker2011,shavitt_2011}.  demonstrated by several
prior analyses.  For instance, Poese \etal found 50-90\% of
ground-truth locations to be geolocated with greater than 50km of
error~\cite{poese2011ip}; most recently Komosn{\`y} \etal studied
eight commercial geolocation databases and found mean errors ranging
from 50-657km~\cite{komosny2017location}.  Geolocation of network
infrastructure, including routers, is known to be particularly
problematic~\cite{huffaker2014drop,gharaibeh_2017}.
However, as shown in Sec.~\ref{sec:survey}, \maxmind is still widely used, 
for the simple reason that there exists no other alternative than geolocation 
database to get an 
Internet scale IP geolocation mapping.

Our work looks at IP geolocation through a novel lens by analyzing the
longitudinal characteristics of a popular geolocation database.  By
showing the stability of locations at different granularities and
timescales, we offer a first look at the error bounds for particular
classes of applications that utilize geolocations, as well as offer
practical lessons for consumers of IP geolocation data.

%% file: conclusion.tex
\section{Conclusion}
\label{sec:conclusion}

Physical mapping of Internet hosts and resources is critical in this
day and age.  Techniques to perform IP geolocation have matured into
commercial offerings.  While the accuracy of these geolocation
databases has been extensively studied, little attention has been paid
to understand the way they have evolved over time.  Our work
demonstrates that a commonly used geolocation database, \maxmind,
exhibits significant changes in coverage, persistence, and prevalence,
especially when considering particular subsets of addresses.  

These changes can occur even on short timescales, including on the
order of a typical measurement study duration.  In this way we
highlight the importance of geolocation lookups that are
contemporaneous with the time an IP address is measured, observed, or
gathered.  Via a case study, we demonstrate the potential for a large
discrepancy in results depending on the particular date of a
geolocation snapshot.  Similar large variances in auxiliary data
sources at short time scales have been demonstrated in the past, \eg
for DNS and Internet top lists~\cite{10.1145/3278532.3278574}.  Thus,
a take-away of our work is to encourage alignment of geolocation
lookups with measurements, publishing the exact date of a geolocation
snapshot or lookup methodology, and rigorously investigating addresses
that change geolocation significantly over the course of a measurement
study.  In the spirit of similar measurement best
practices~\cite{10.1145/1028788.1028824}, we hope to encourage more
sound and reproducible measurement research. Because \maxmind does 
not provide access to historical data, we provide historical snapshots on demand 
to the community.

In future work, we plan to more deeply investigate the root causes of
the geolocation movement we observe, characterize IPv6 geolocation,
and work toward integrating our findings into more robust geolocation
services.  Finally, we believe the metrics we developed generalize and
could be applied to not only other geolocation databases, but also to
understand the longitudinal behavior of other dynamic Internet
resources such as reverse DNS records.

%% file: appendix.tex
../code/resources/
\appendix
\section{Ethical Considerations}
\label{appendix:ethics}

Our work does not involve human subjects, questionnaires, or
personally identifiable information, and, hence, does not meet the
standards for IRB review.  The \maxmind data we analyze is covered by
the Creative Commons Attribution-ShareAlike 4.0 International (CC
BY-SA 4.0) license which permits adaption of the database: ``remix,
transform, and build upon the material for any purpose, even
commercially.''  Beginning in January 1, 2020, \maxmind adopted a more
restrictive policy in order to comply with GDPR
requirements~\cite{maxmind_eula}; our research does not analyze any
data after 2019.

\section{Exhaustive Evaluation}
\label{appendix:exhaust}
In Sec.~\ref{sec:evaluation}, we only show the primary results that support 
our main findings. However, we have performed a thorough evaluation 
of each metric defined in Sec.~\ref{sec:methodology} 
over three axes: time, type of IP address and location. We put here 
the whole set of results. 
\subsubsection{Prevalence on \maxmind country database}
\begin{figure*}[h]
\begin{subfigure}{.24\textwidth}
\centering
  \includegraphics[width=\linewidth]{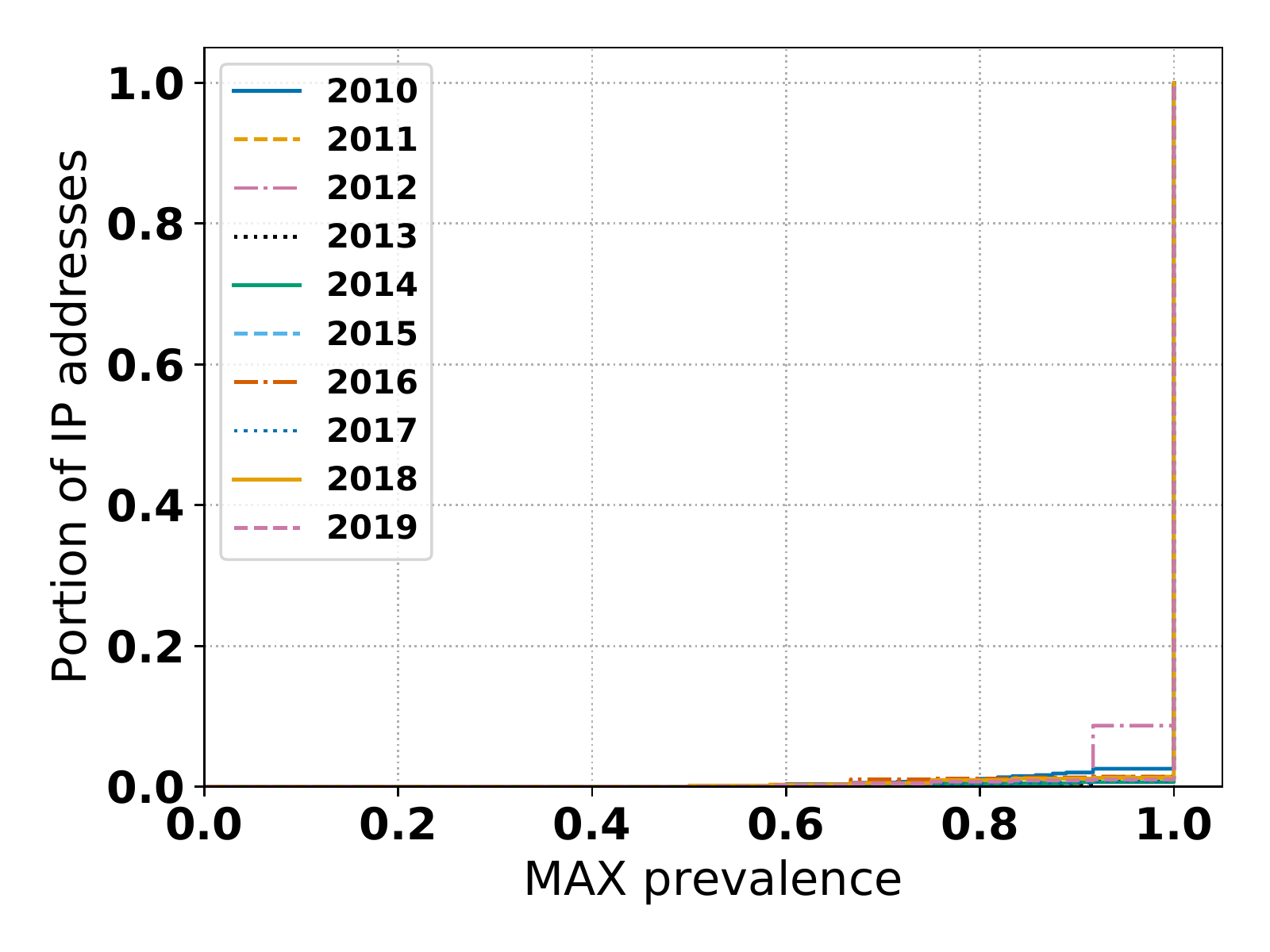}
  \caption{Over time (Eq.~\ref{eq:max-prevalence-ip})}
    \label{fig:appendix:prevalence-over-time}
\end{subfigure}%
\begin{subfigure}{.24\textwidth}
\centering
  \includegraphics[width=\linewidth]{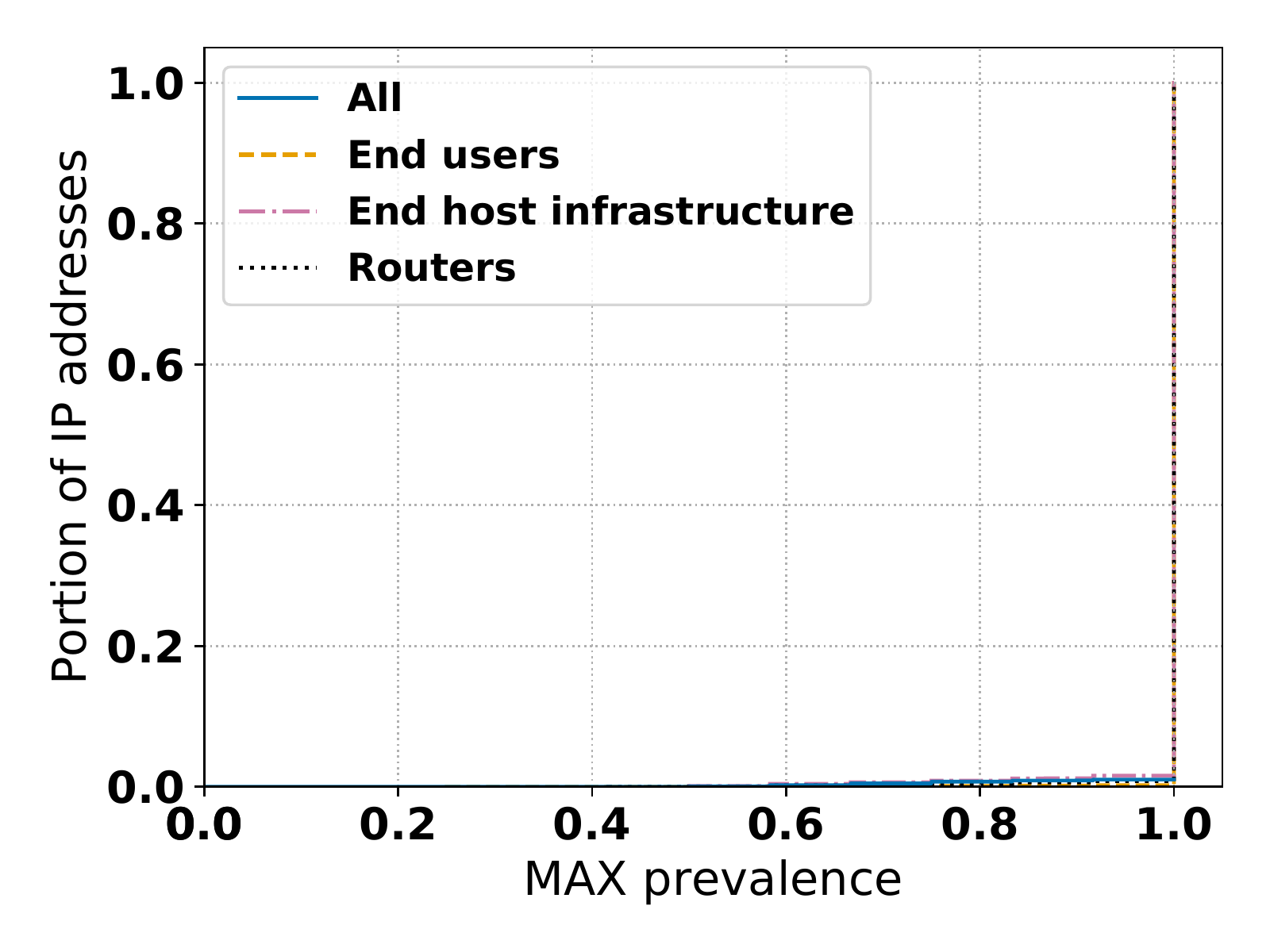}
  \caption{By IP class (Eq.~\ref{eq:max-prevalence-ip})}
  \label{fig:appendix:prevalence-by-ip-type}
\end{subfigure}%
\begin{subfigure}{.24\textwidth}
\centering
  \includegraphics[width=\linewidth]{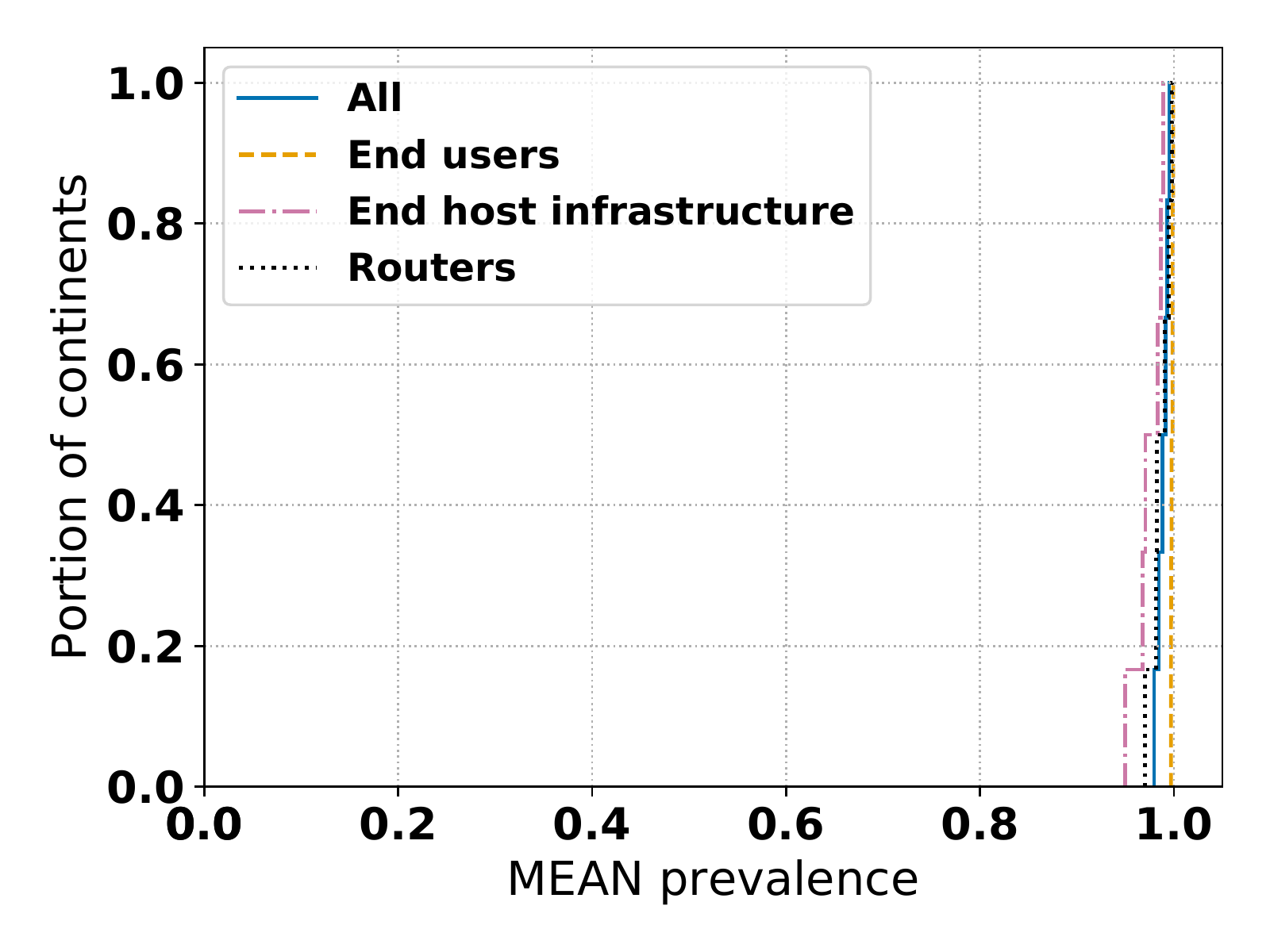}
  \caption{By continent (Eq.~\ref{eq:prevalence-location-average})}
  \label{fig:appendix:prevalence-by-continent}
\end{subfigure}
\begin{subfigure}{.24\textwidth}
\centering
  \includegraphics[width=\linewidth]{figures/metrics/prevalence/country/prevalence_per_aggregate_country_country_database_aggregate_country_max_year.pdf}
  \caption{By country (Eq.~\ref{eq:prevalence-location-average})}
  \label{fig:appendix:prevalence-by-country}
\end{subfigure}


\begin{subfigure}{.24\textwidth}
\centering
  \includegraphics[width=\linewidth]{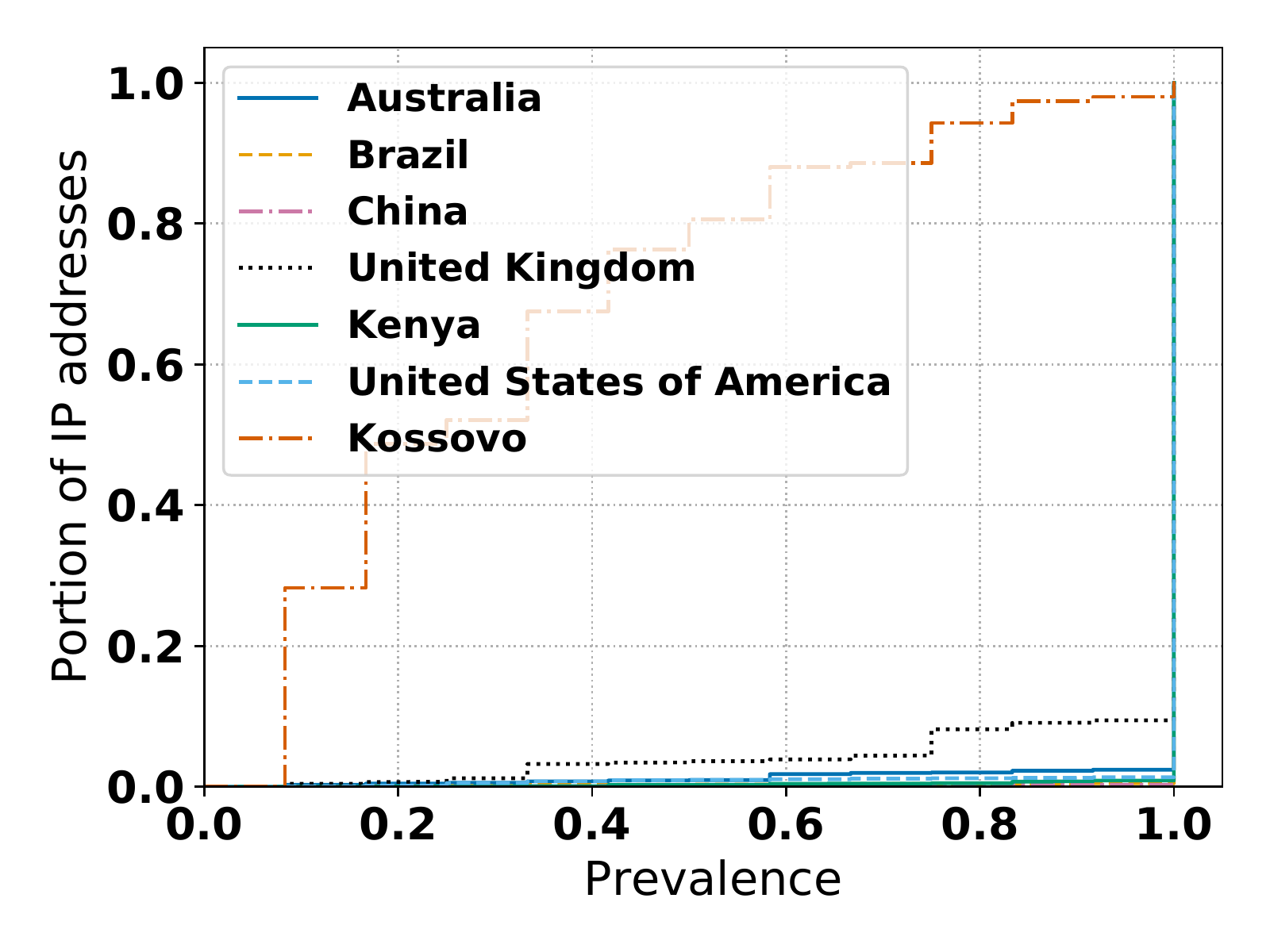}
  \caption{All (Eq.~\ref{eq:prevalence-location})}
  \label{fig:appendix:prevalence-by-country-all}
\end{subfigure}%
\begin{subfigure}{.24\textwidth}
\centering
  \includegraphics[width=\linewidth]{figures/metrics/prevalence/country/prevalence_snapshots_country_mlab_per_country_country_max_year.pdf}
  \caption{End users (Eq.~\ref{eq:prevalence-location})}
  \label{fig:appendix:prevalence-by-country-end-users}
\end{subfigure}%
\begin{subfigure}{.24\textwidth}
\centering
  \includegraphics[width=\linewidth]{figures/metrics/prevalence/country/prevalence_snapshots_country_end_hosts_infrastructure_sample_per_country_country_max_year.pdf}
  \caption{End host infra. (Eq.~\ref{eq:prevalence-location})}
  \label{fig:appendix:prevalence-by-country-infrastructure}
\end{subfigure}%
\begin{subfigure}{.24\textwidth}
\centering
  \includegraphics[width=\linewidth]{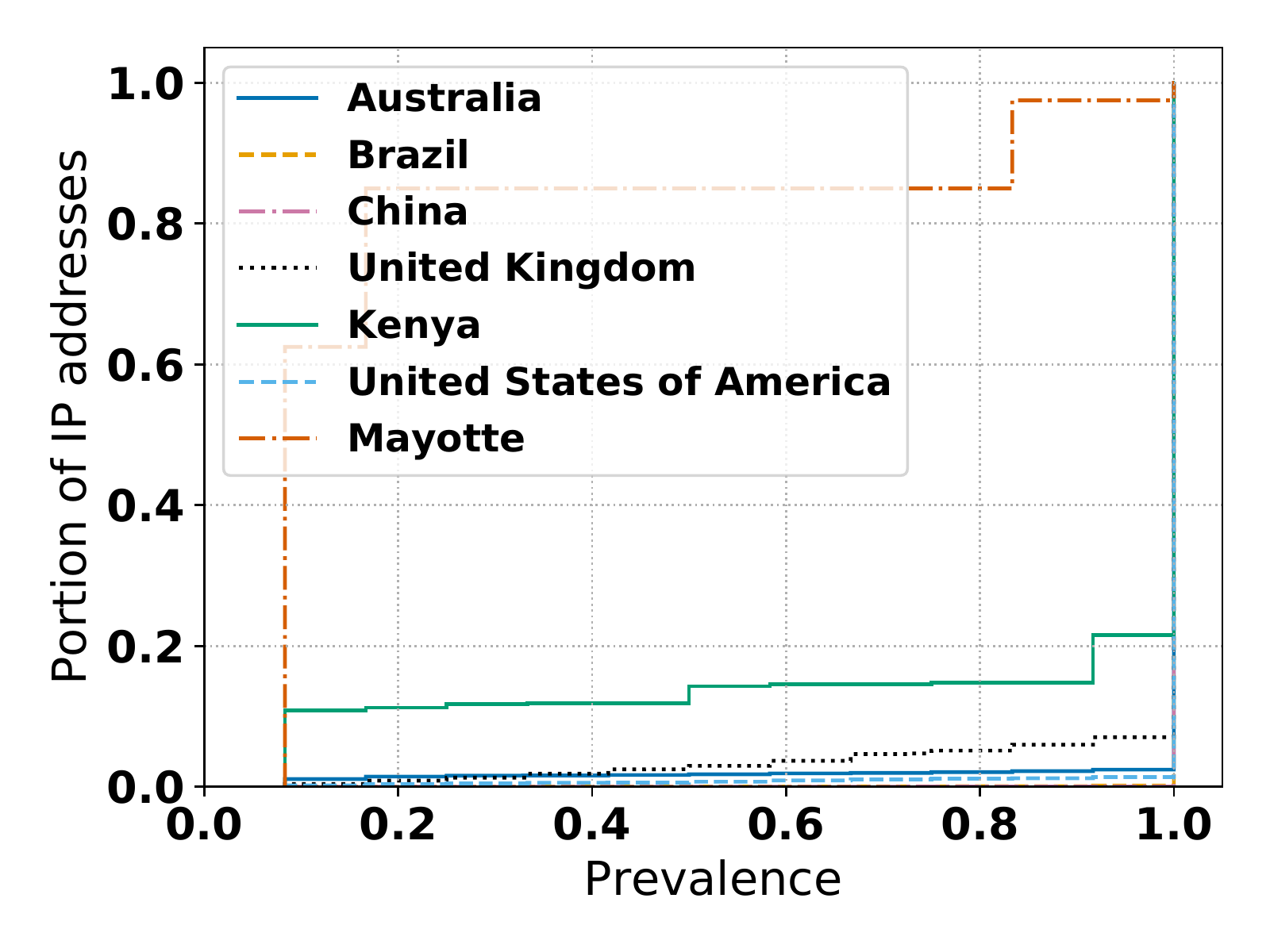}
  \caption{Routers (Eq.~\ref{eq:prevalence-location})}
  \label{fig:appendix:prevalence-by-country-router}
\end{subfigure}
  \caption{Prevalence results on \maxmind country geolocation}
  \label{fig:appendix:prevalence}
\end{figure*}

\begin{figure*}[h]

\begin{subfigure}{.24\textwidth}
\centering
  \includegraphics[width=\linewidth]{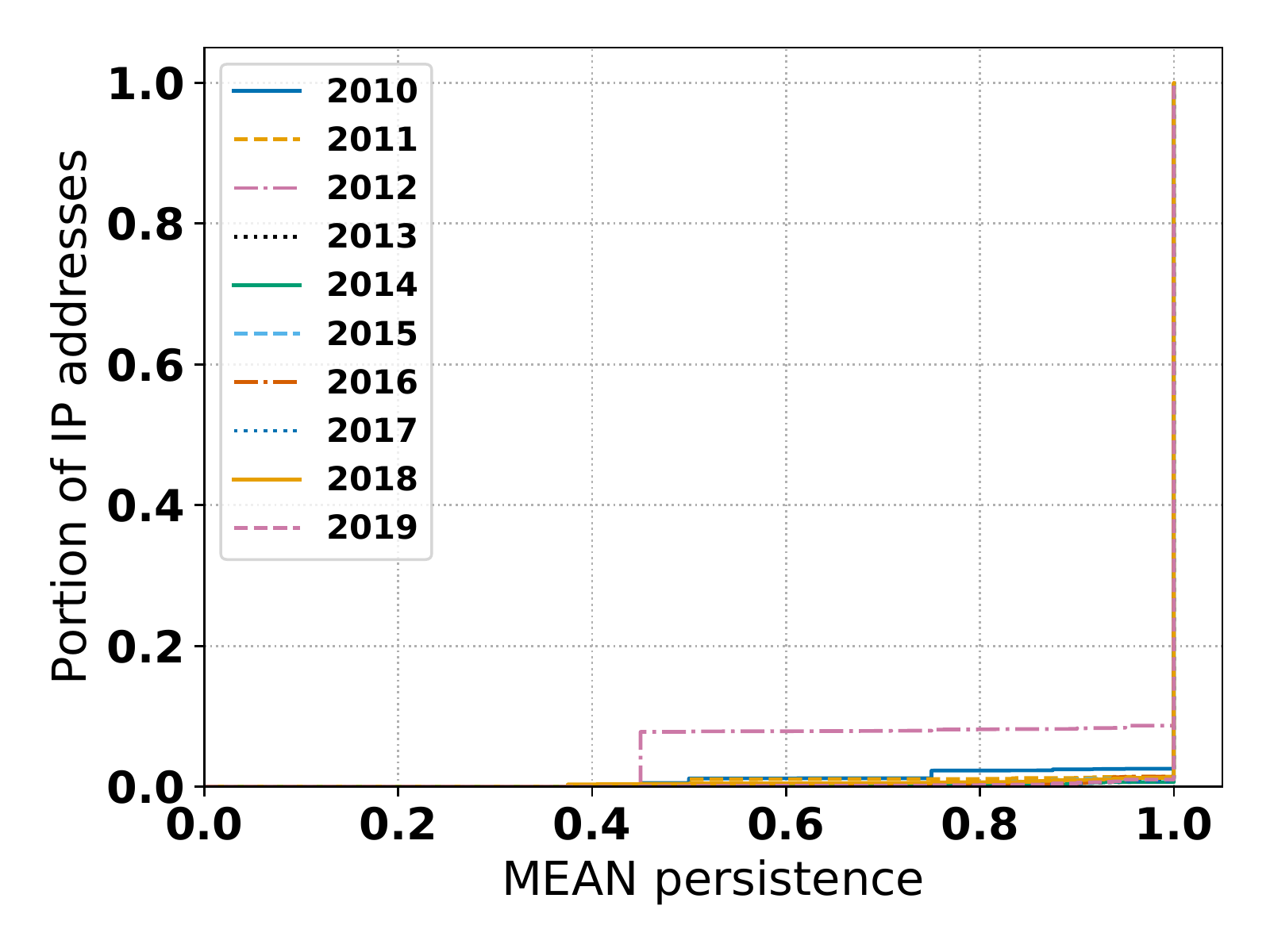}
  \caption{Over time (Eq.~\ref{eq:avg-persistence-ip})}
  \label{fig:appendix:persistence-over-time}
\end{subfigure}%
\begin{subfigure}{.24\textwidth}
\centering
  \includegraphics[width=\linewidth]{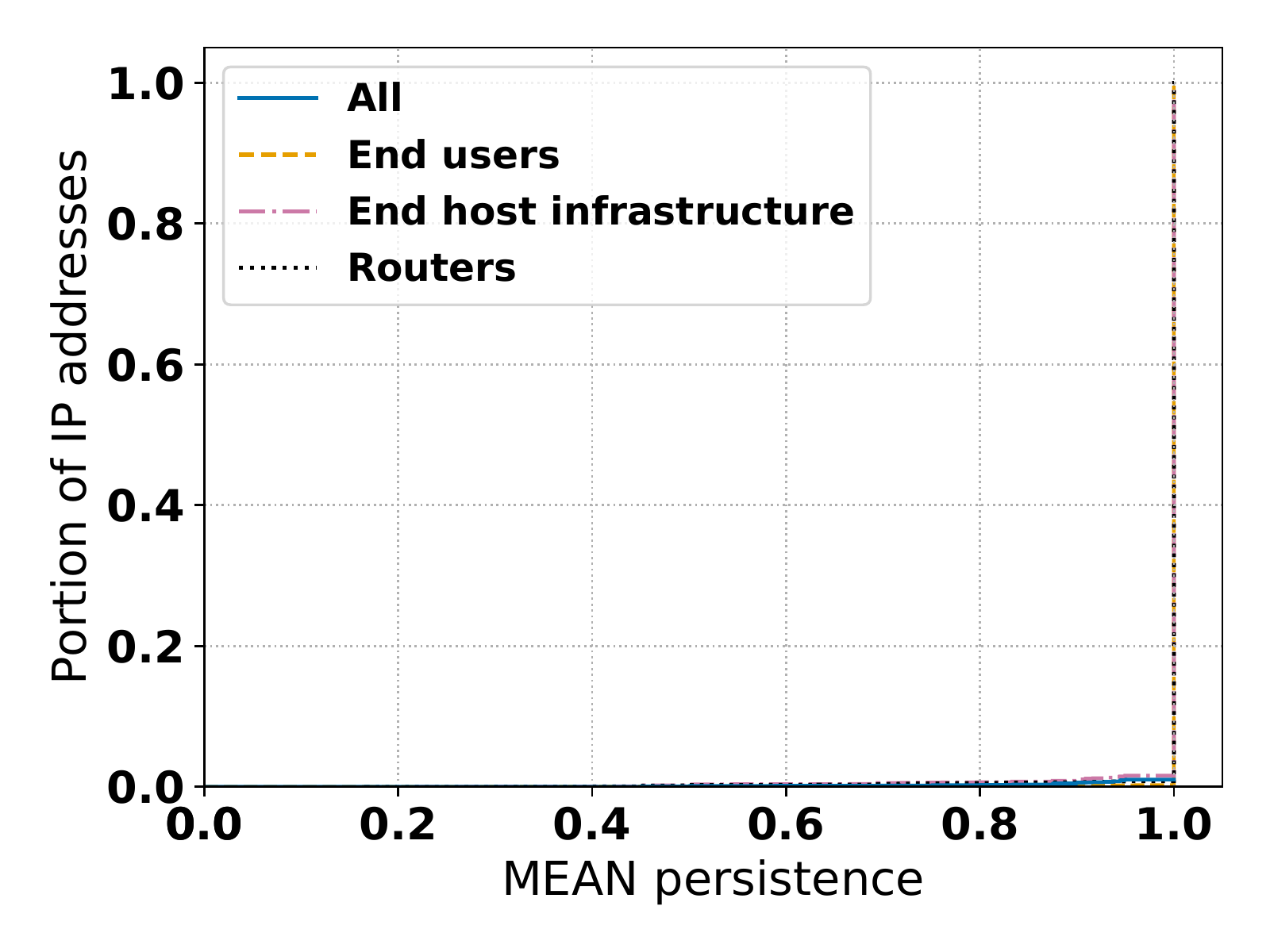}
  \caption{By IP class (Eq.~\ref{eq:avg-persistence-ip})}
  \label{fig:appendix:persistence-by-ip-type-country}
\end{subfigure}%
\begin{subfigure}{.24\textwidth}
\centering
  \includegraphics[width=\linewidth]{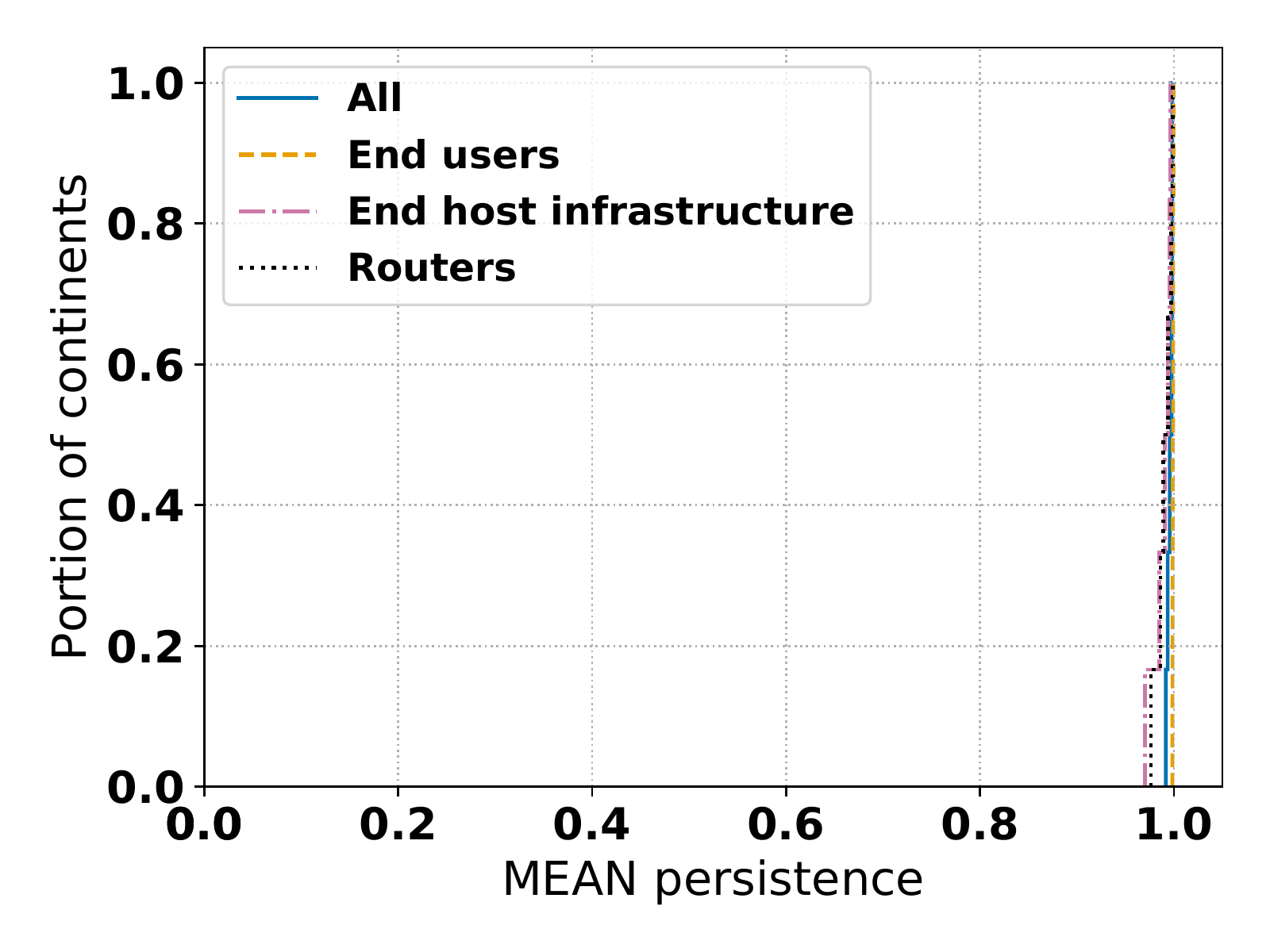}
  \caption{By continent (Eq.~\ref{eq:persistence-location-average})}
  \label{fig:appendix:persistence-by-continent-country}
\end{subfigure}
\begin{subfigure}{.24\textwidth}
\centering
  \includegraphics[width=\linewidth]{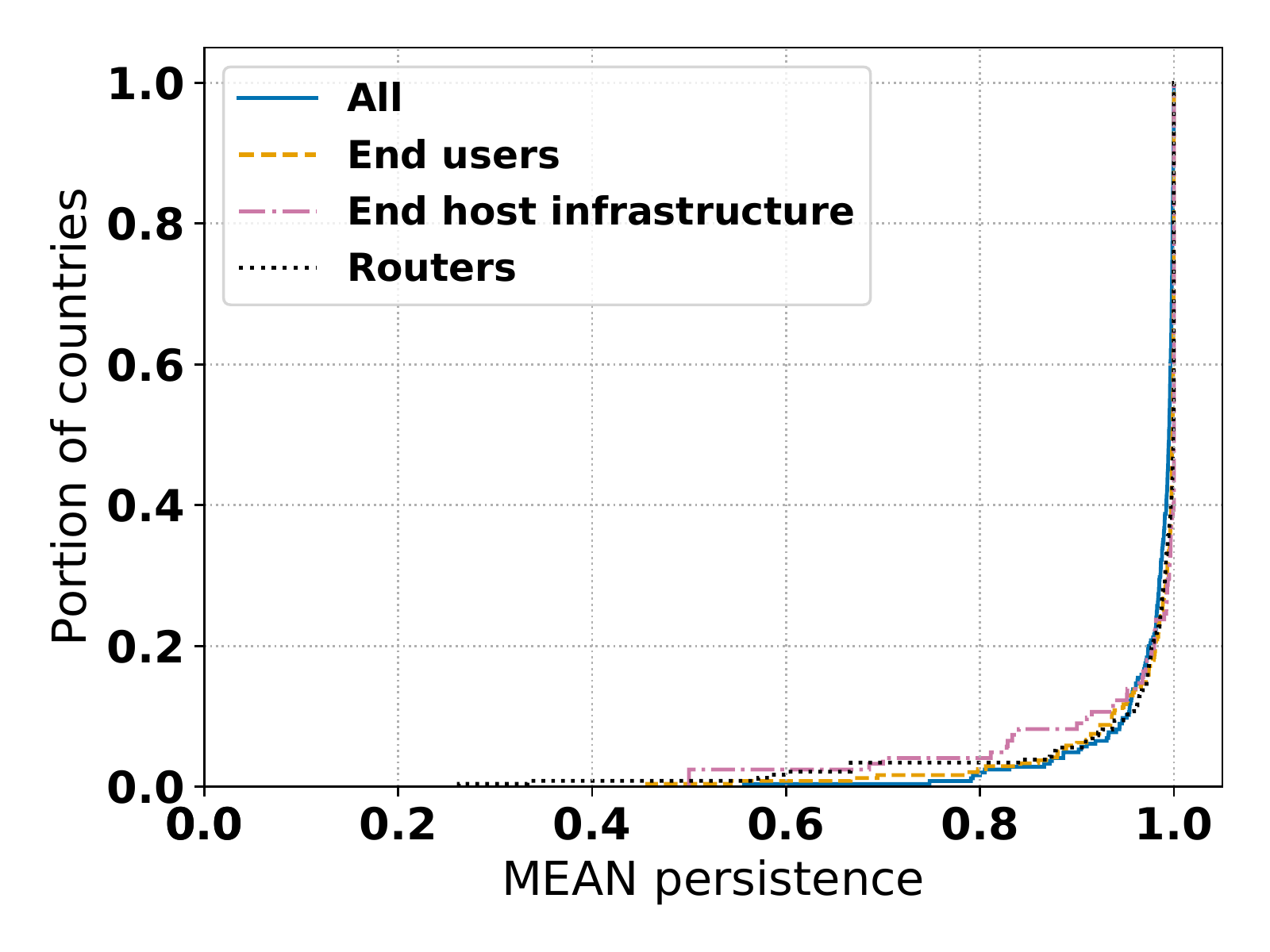}
  \caption{By country (Eq.~\ref{eq:persistence-location-average})}
  \label{fig:appendix:persistence-by-country-country}
\end{subfigure}

%
%

\begin{subfigure}{.24\textwidth}
\centering
  \includegraphics[width=\linewidth]{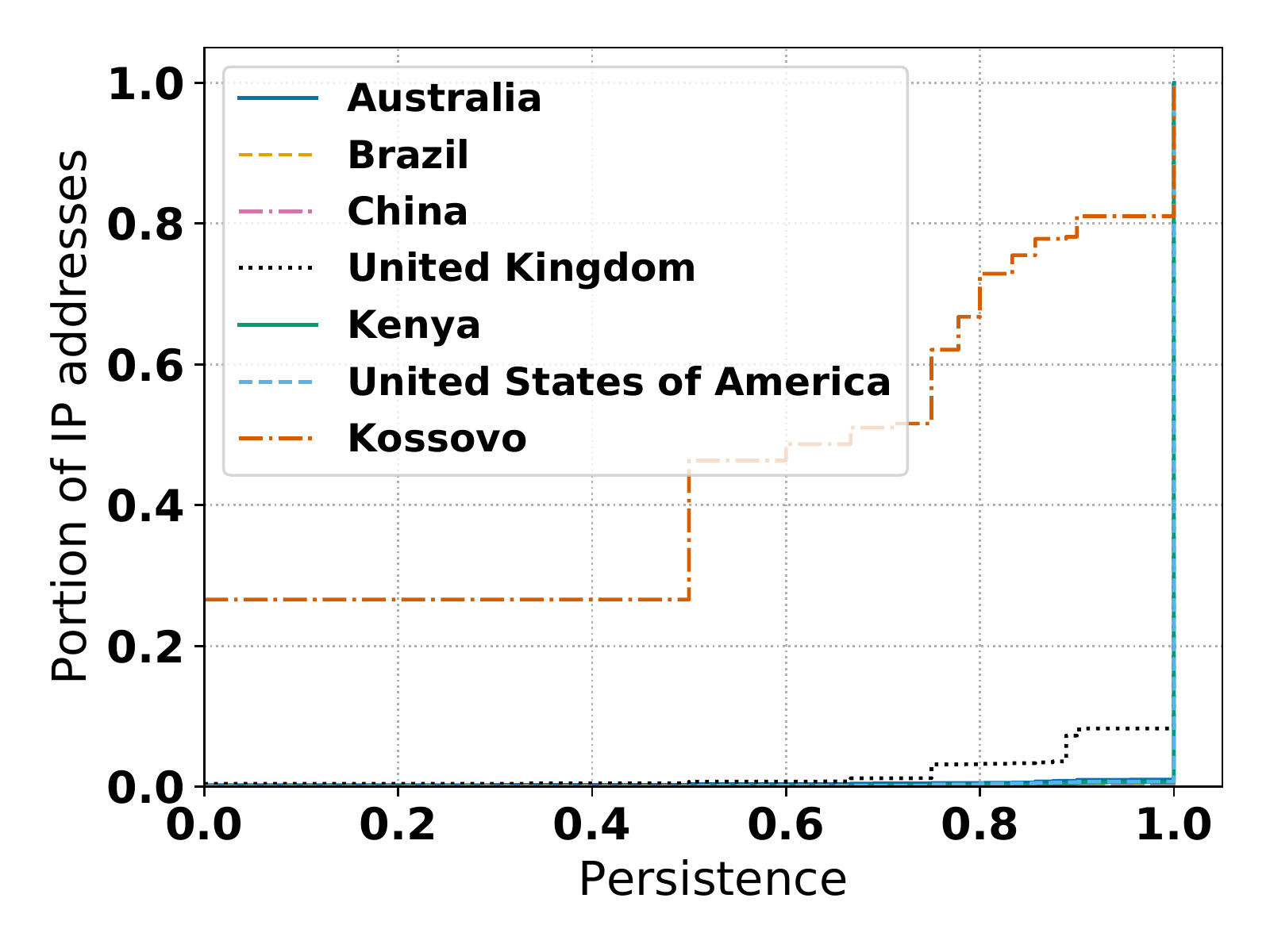}
  \caption{All (Eq.~\ref{eq:persistence-location})}
  \label{fig:appendix:persistence-by-country-all}
\end{subfigure}%
\begin{subfigure}{.24\textwidth}
\centering
  \includegraphics[width=\linewidth]{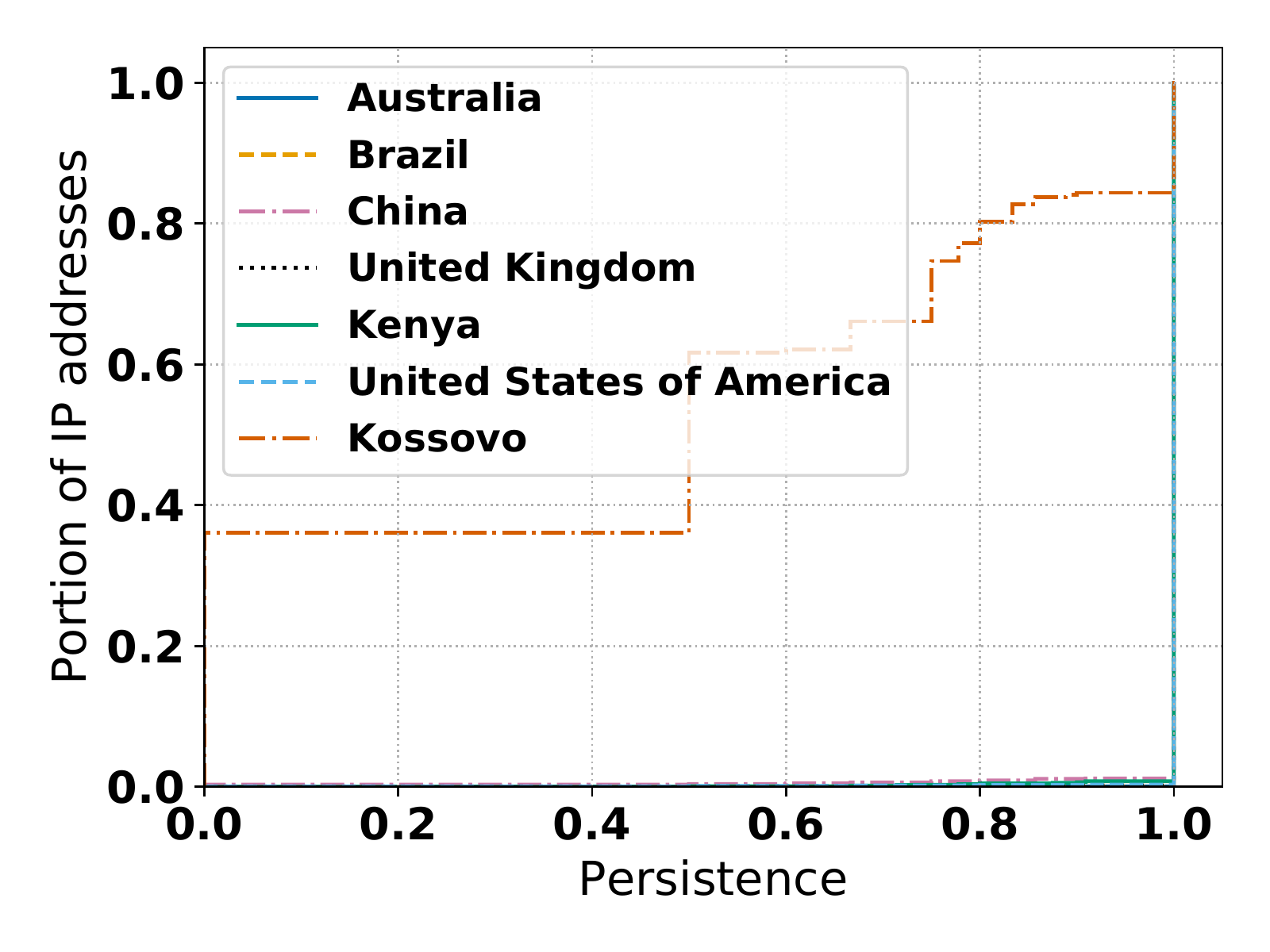}
\caption{End users (Eq.~\ref{eq:persistence-location})}  
  \label{fig:appendix:persistence-by-country-end-users}
\end{subfigure}%
\begin{subfigure}{.24\textwidth}
\centering
  \includegraphics[width=\linewidth]{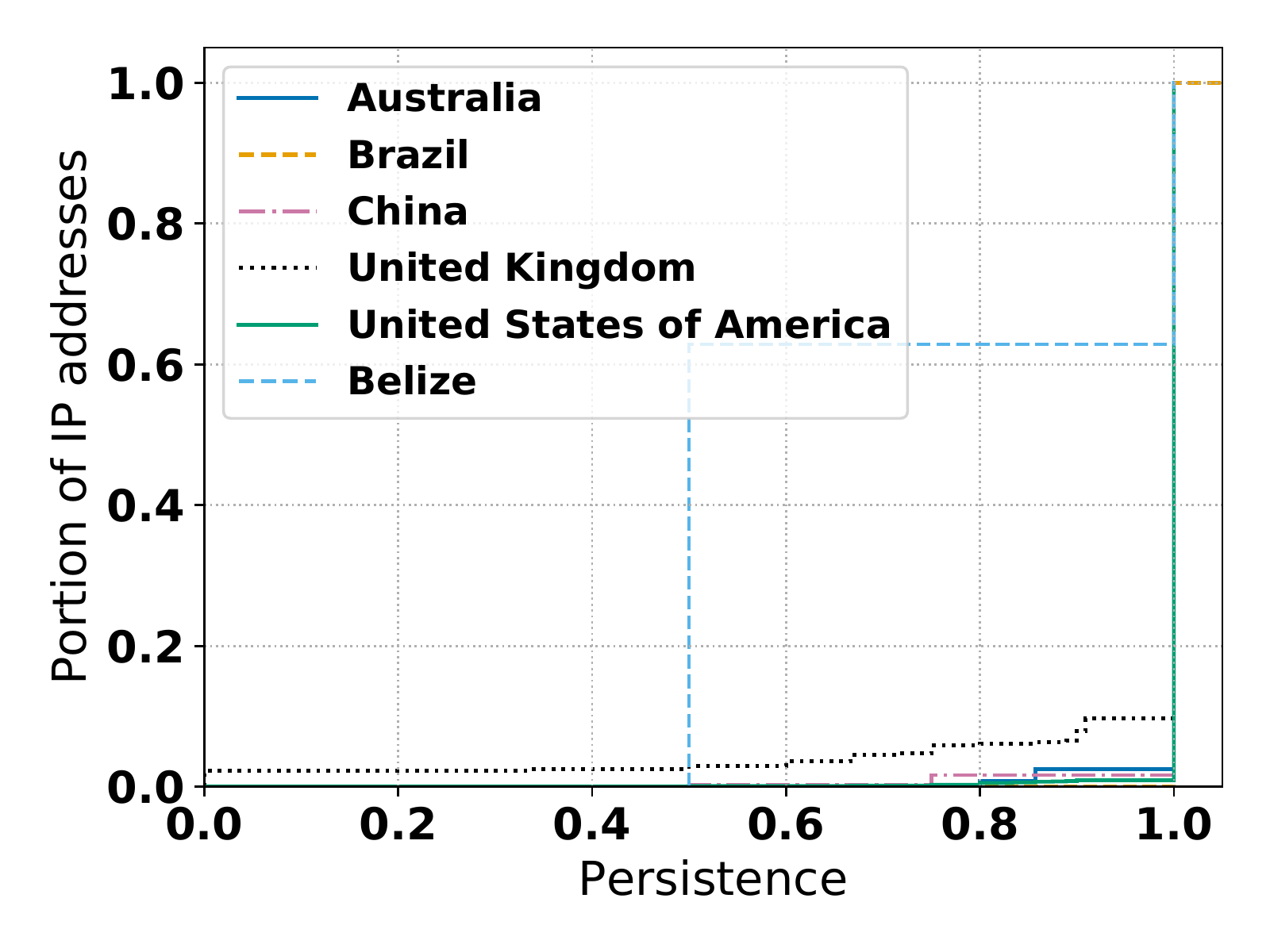}
\caption{End host infra. (Eq.~\ref{eq:persistence-location})}
  \label{fig:appendix:persistence-by-country-end-host-infrastructure}
\end{subfigure}%
\begin{subfigure}{.24\textwidth}
\centering
  \includegraphics[width=\linewidth]{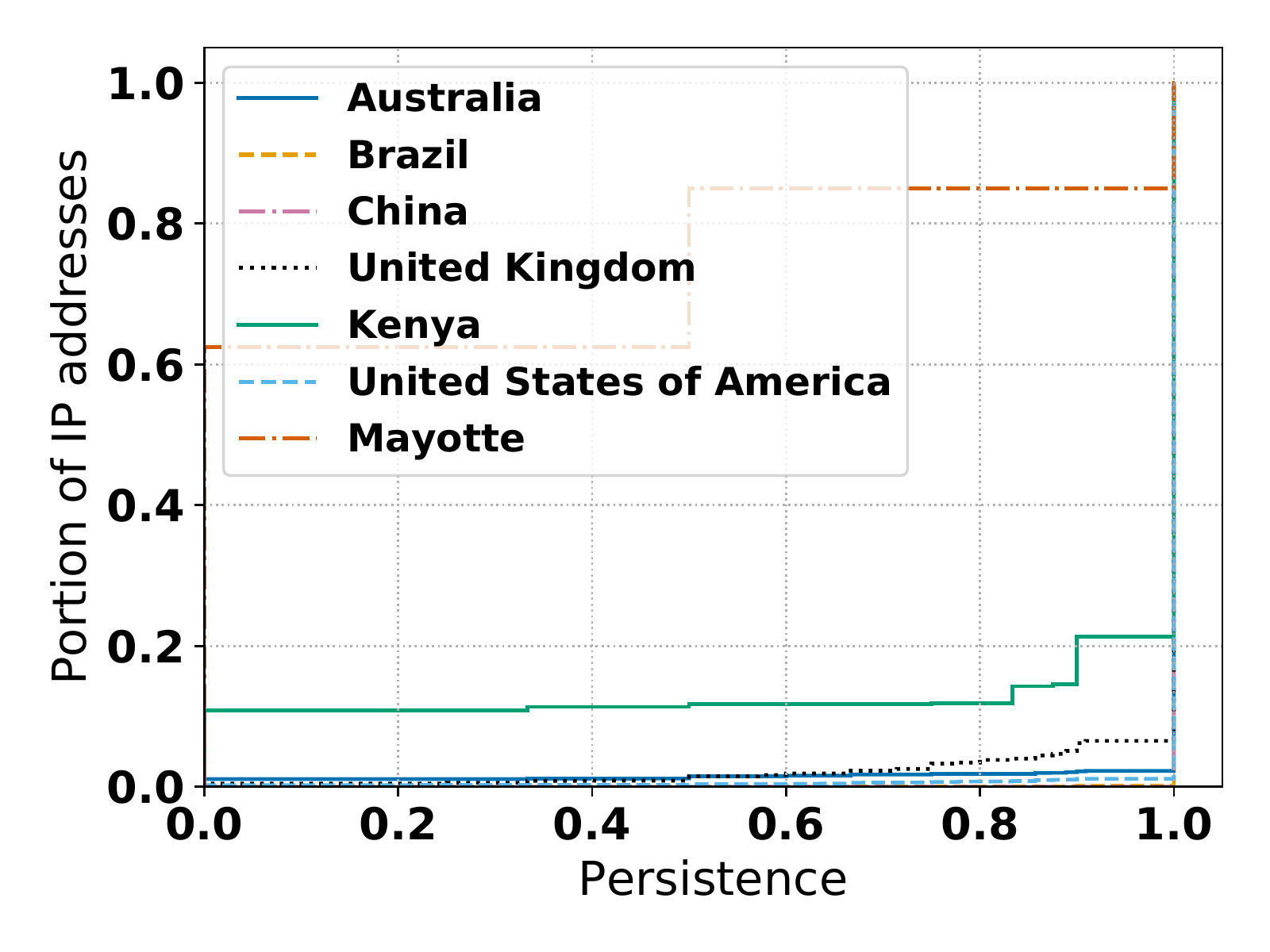}
  \caption{Routers (Eq.~\ref{eq:persistence-location})}  
  \label{fig:appendix:persistence-by-country-routers}
\end{subfigure}
 \caption{Persistence results on \maxmind country geolocation}
  \label{fig:appendix:persistence-country}
\end{figure*}

Fig.~\ref{fig:appendix:prevalence} shows our results on the prevalence for the 
\maxmind country database.
As our metrics are built to do so, we vary three orthogonal axis
to evaluate the prevalence: the time (2010-2019 and 2019)
the prevalence
on different types of IP addresses, and 
the prevalence per geographic location. 
The two latter focus on the 2019 year. 

On each title of each paragraph, we put in parenthesis the corresponding 
figure and
metric defined in Sec.~\ref{sec:methodology:dynamics} that it 
analyzes.
\paragraph{Evolution of the prevalence (Fig.~\ref{fig:appendix:prevalence-over-time}, 
Eq.~\ref{eq:max-prevalence-ip})}
We see that for all years, except for 2012, the
maximum is concentrated close to 1. 
This means that IP addresses located by \maxmind have always had 
a single country.

\paragraph{Per type of IP addresses (Fig.~\ref{fig:appendix:prevalence-by-ip-type}, Eq.~\ref{eq:max-prevalence-ip})}
There is no noticeable difference between all 
types of IP addresses, the maximum prevalence being close to 1.

\paragraph{Per geographical locations (Fig.~\ref{fig:appendix:prevalence-by-continent}
,~\ref{fig:appendix:prevalence-by-country},
 Eq.~\ref{eq:prevalence-location-average})}
We analyze if the prevalence of country 
depends on the geographic locations at two levels:
the continent and the country.

Fig.~\ref{fig:appendix:prevalence-by-continent} and~\ref{fig:appendix:prevalence-by-country} 
shows the CDF of the average value of the prevalence for each
location for each of our dataset (``All'' being the full \maxmind
snapshot of the year 2019.).
We observe that at continent level, all the continents have a average prevalence
close to 1.

At country level, this distribution states that for all 4 types of IP addresses, 
the tail of the distribution has a low average prevalence, 
so that there are potentially important differences depending on the country.
We deeper investigate these differences in the next paragraphs.

\paragraph{Detailed investigation on continent (Second row of Fig.~\ref{fig:appendix:prevalence}, 
Eq.~\ref{eq:prevalence-location})}
At continent level, interestingly we can see that on the ``All'' dataset, the distribution 
for Europe is slightly above the others; 
4\% of IP addresses in Europe have a prevalence
of less than 0.6.
This means that 4\% of IP located in European countries do not have a 
single country.

Interestingly, the prevalence per continent also 
depend on the type of IP addresses: 
On end users, the prevalence is almost always 1. 
On end host infrastructure and routers, the curve of Africa is slightly above 
the other continents;
9\% of IP addresses located in Africa do not have a single country.

\paragraph{Detailed investigation on countries (Third row of Fig.~\ref{fig:appendix:prevalence},
 Eq.~\ref{eq:prevalence-location})}
These graphs show 
the 10 countries having the lowest average 
prevalence (Eq.~\ref{eq:prevalence-location-average}). 

First, we see that for all of these 10 countries, for each type of IP address, 
the prevalence is no 
longer concentrated near 1. 

On the ``All'', end users, and routers dataset, we can observe that the
countries having a low mean of maximum prevalence are rather small, typically
islands. Notice however the presence of Netherlands for the routers.

The results on end hosts infrastructure are more surprising. The 10 
countries 
are mainly located in Europe, with Norway, Luxembourg, and United Kingdom 
for Western Europe and 
Bulgaria, Lithuania, Moldova, and Czech Republic for Eastern 
Europe. Notice also the 
presence of big countries on other continents, such as Argentina.

We evaluate the persistence with the same
methodology as for the prevalence, with the three same axes.

The two first rows of graphs of Fig.~\ref{fig:appendix:persistence-country} have nearly 
the same shape at those on prevalence, meaning that:
The average persistence at country level is high. 
This is mainly due to the high maximum prevalence shown in the previous
paragraph; all IP addresses have a single country.

\paragraph{Detailed investigation on countries, third row 
of Fig.~\ref{fig:appendix:persistence-country}, Eq.~\ref{eq:persistence-location}}
The third row, focusing on countries having a lower average persistence, 
reveals the duration of these country changes.
First, the countries having the lower average persistence are also 
the ones that had the least prevalence. 
Noticeably, we find the same surprising European countries for end host 
infrastructure, Italy added.
Then, we observe that the persistence of these countries depend on the 
type of IP 
address. In general we observe that for end users, the curves of 
the countries are higher than the ones for end host infrastructure and routers, 
meaning that they have a lower persistence. 
In details, only Belize for end host infrastructure and Mayotte and Nothern
Mariana Islands for routers have more than 20\% of IP addresses 
having a persistence
of less than 0.5, whereas there are at least 5 countries for end users.

Our interpretation is that when there is a country change 
on countries having a low average persistence, 
\textbf{the changes are more likely to last for end host infrastructure
and routers, whereas they are more likely to be short 
for end users}.

\begin{figure*}[h]

\begin{subfigure}{.24\textwidth}
\centering
  \includegraphics[width=\linewidth]{figures/metrics/prevalence/country/prevalence_snapshots_city_over_time_all_max_year.pdf}
  \caption{Over time (Eq.~\ref{eq:max-prevalence-ip})}
  \label{fig:appendix:prevalence-over-time-city}
\end{subfigure}%
\begin{subfigure}{.24\textwidth}
\centering
  \includegraphics[width=\linewidth]{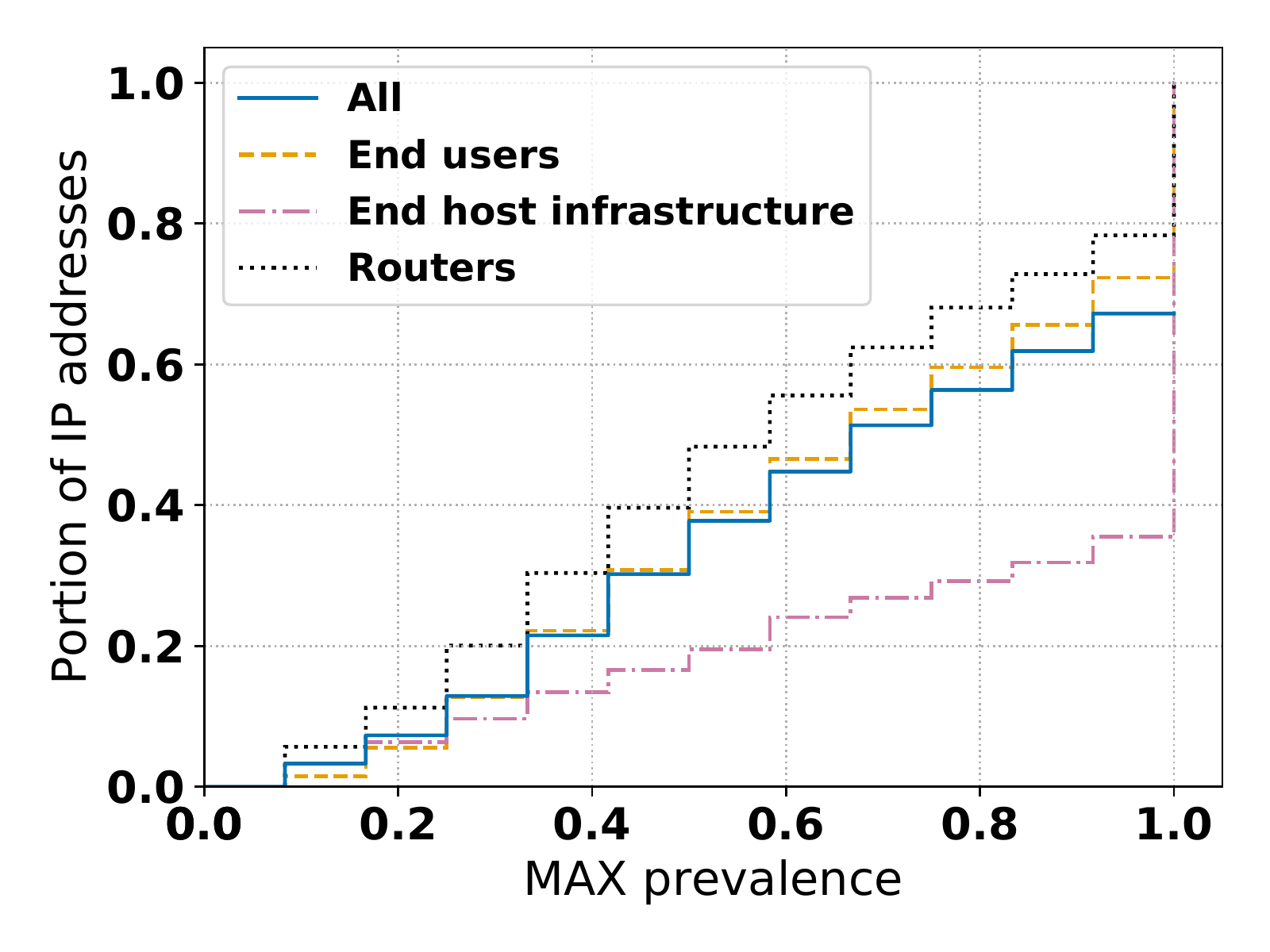}
  \caption{By IP class (Eq.~\ref{eq:max-prevalence-ip})}
  \label{fig:appendix:prevalence-by-ip-type-city}
\end{subfigure}%
\begin{subfigure}{.24\textwidth}
\centering
  \includegraphics[width=\linewidth]{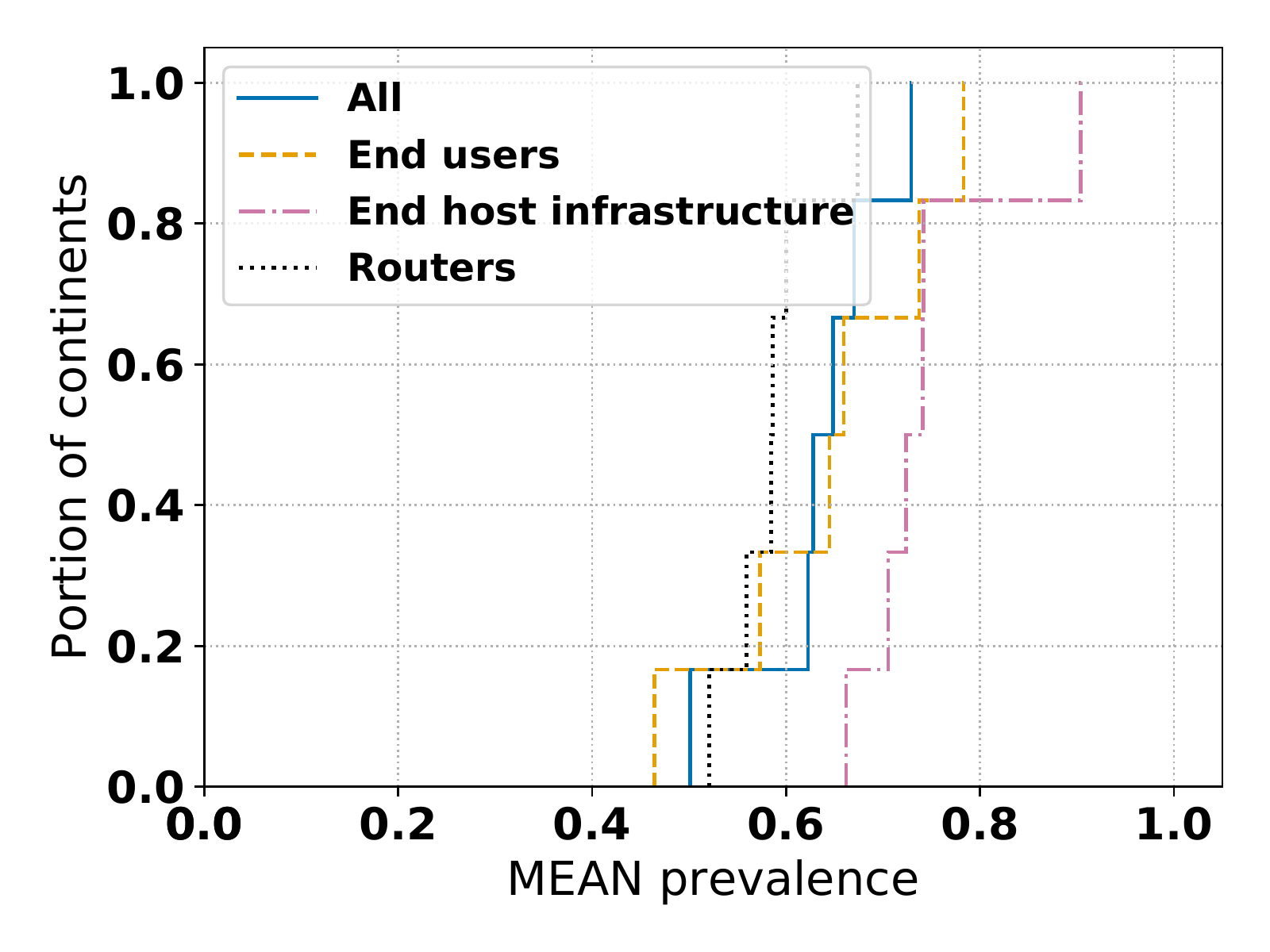}
  \caption{By continent (Eq.~\ref{eq:prevalence-location-average})}
  \label{fig:appendix:prevalence-by-continent-city}
\end{subfigure}
\begin{subfigure}{.24\textwidth}
\centering
  \includegraphics[width=\linewidth]{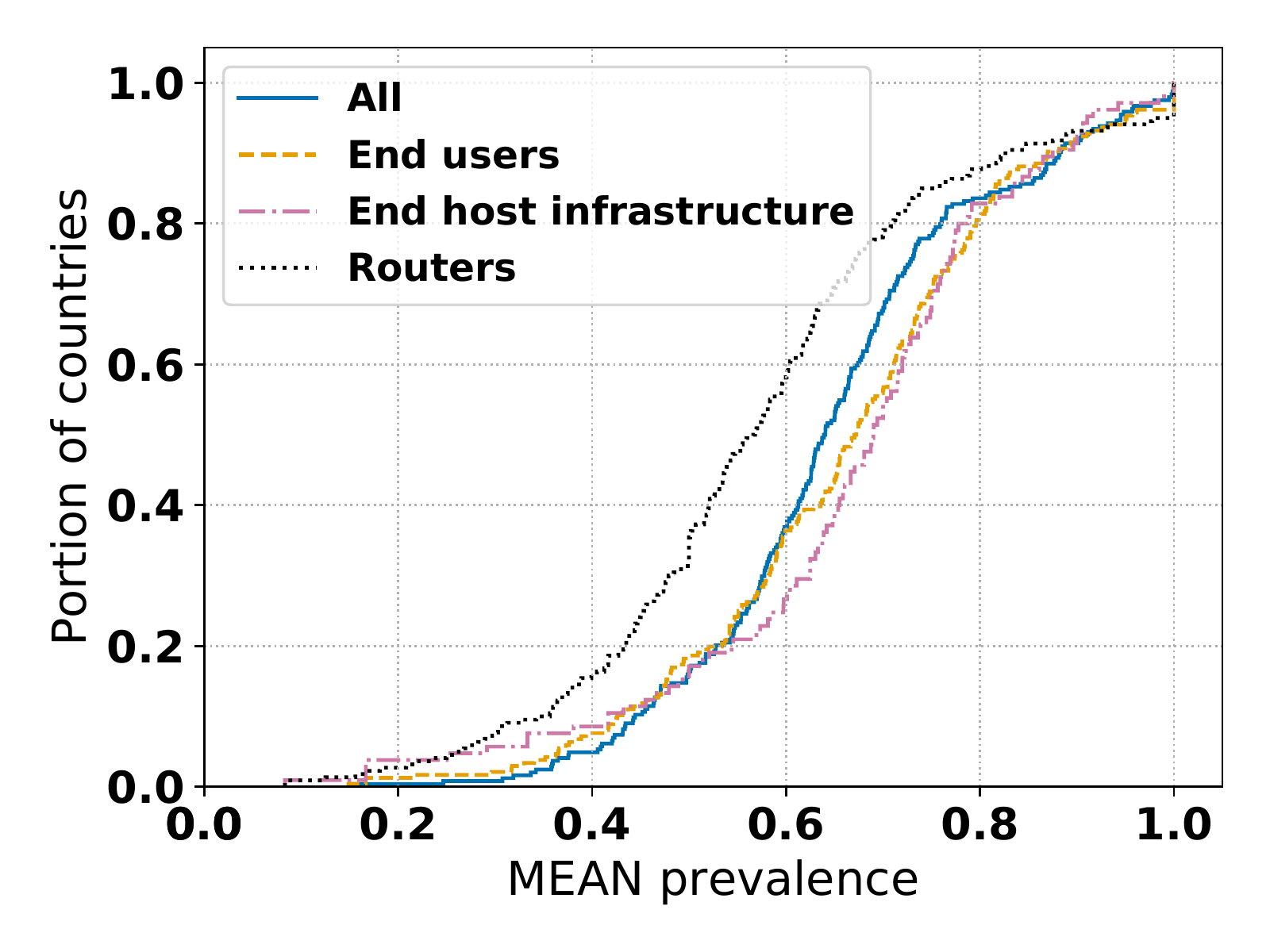}
  \caption{By country (Eq.~\ref{eq:prevalence-location-average})}
  \label{fig:appendix:prevalence-by-country-city}
\end{subfigure}



\begin{subfigure}{.24\textwidth}
\centering
  \includegraphics[width=\linewidth]{figures/metrics/prevalence/country/prevalence_snapshots_city_per_country_country_max_year.pdf}
  \caption{All (Eq.~\ref{eq:prevalence-location})}
  \label{fig:appendix:prevalence-by-country-city-all}
\end{subfigure}%
\begin{subfigure}{.24\textwidth}
\centering
  \includegraphics[width=\linewidth]{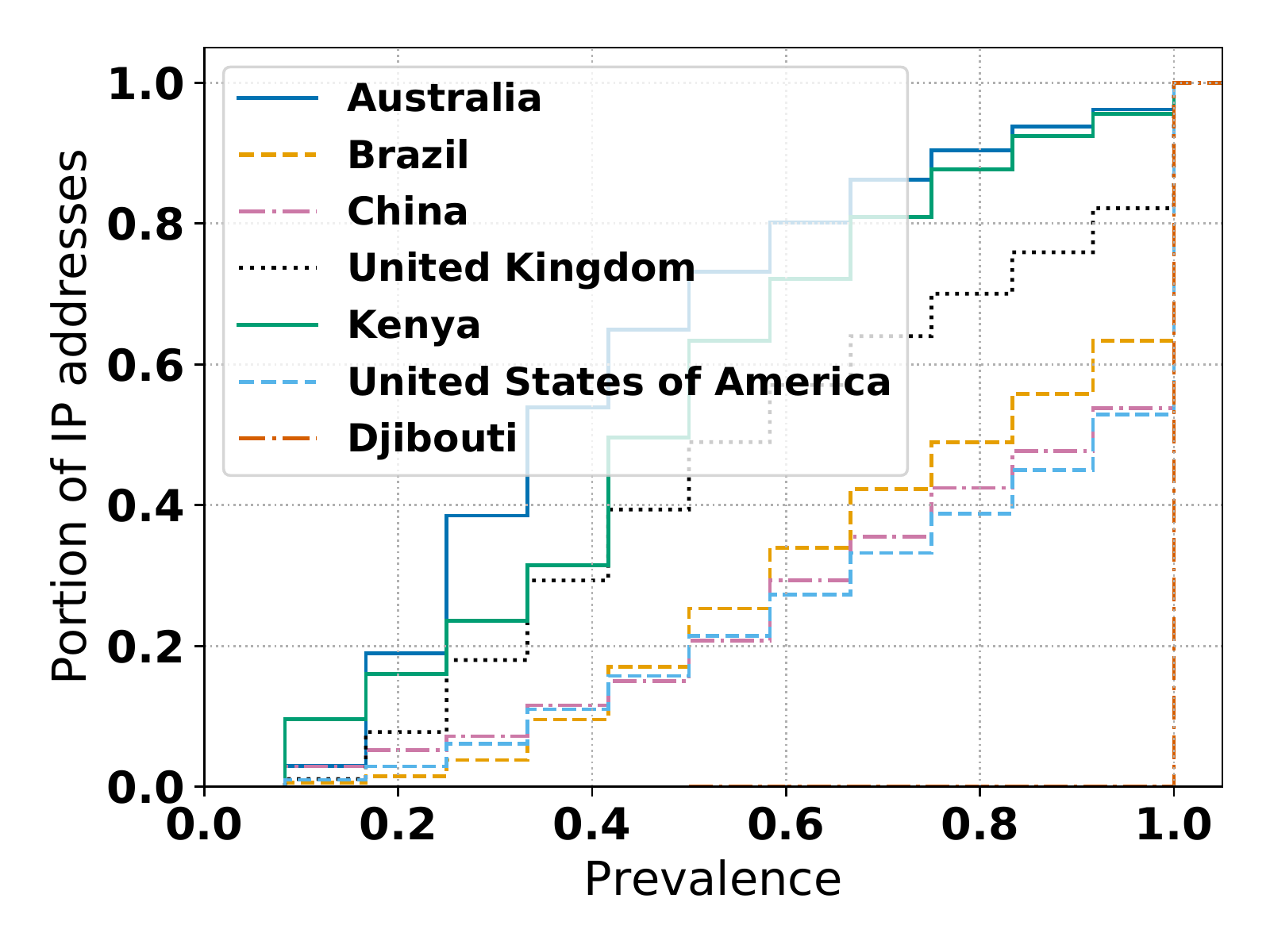}
  \caption{End users (Eq.~\ref{eq:prevalence-location})}
  \label{fig:appendix:prevalence-by-country-city-end-users}
\end{subfigure}%
\begin{subfigure}{.24\textwidth}
\centering
  \includegraphics[width=\linewidth]{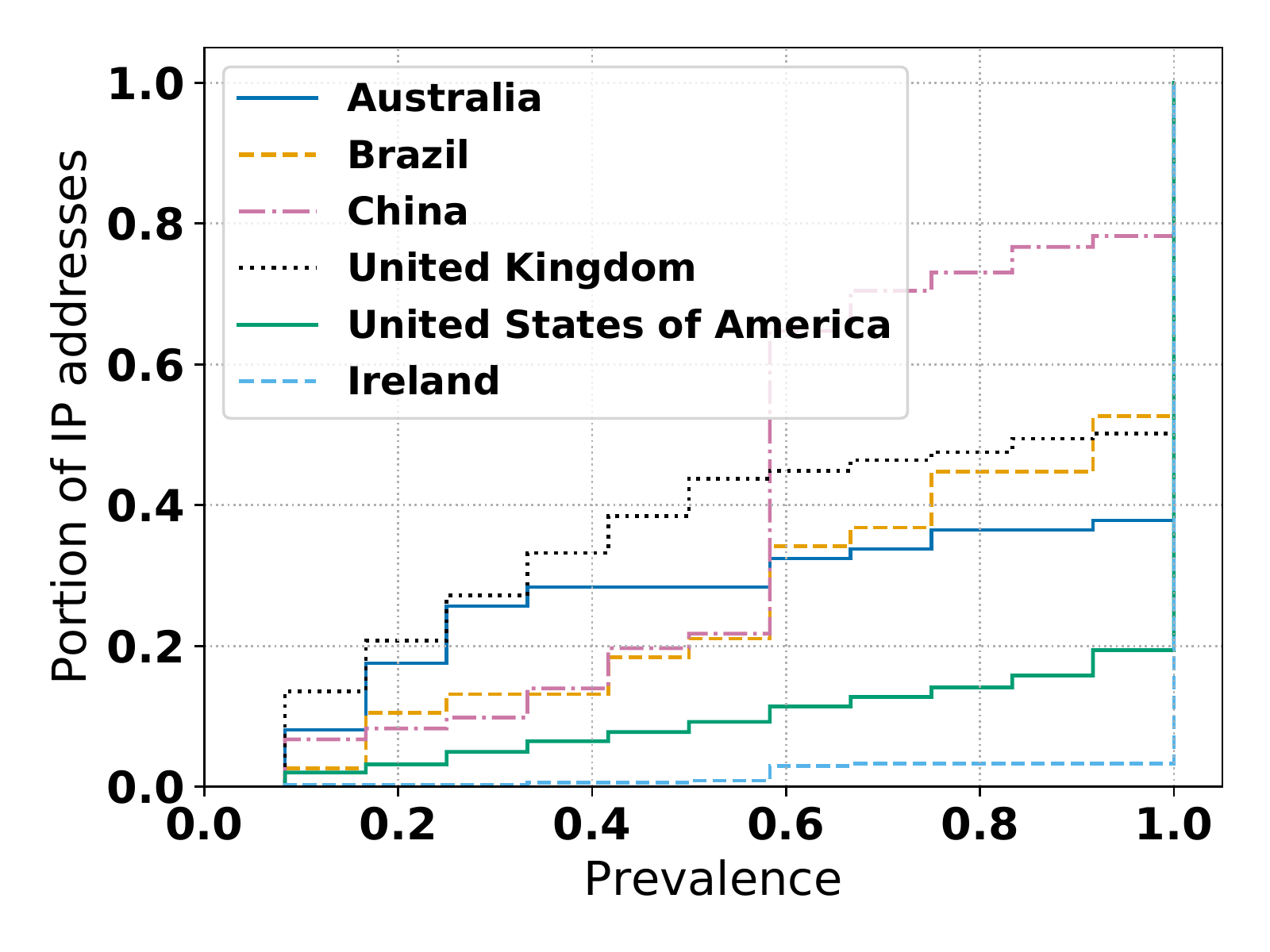}
  \caption{End host infra. (Eq.~\ref{eq:prevalence-location})}
  \label{fig:appendix:prevalence-by-country-city-infrastructure}
\end{subfigure}%
\begin{subfigure}{.24\textwidth}
\centering
  \includegraphics[width=\linewidth]{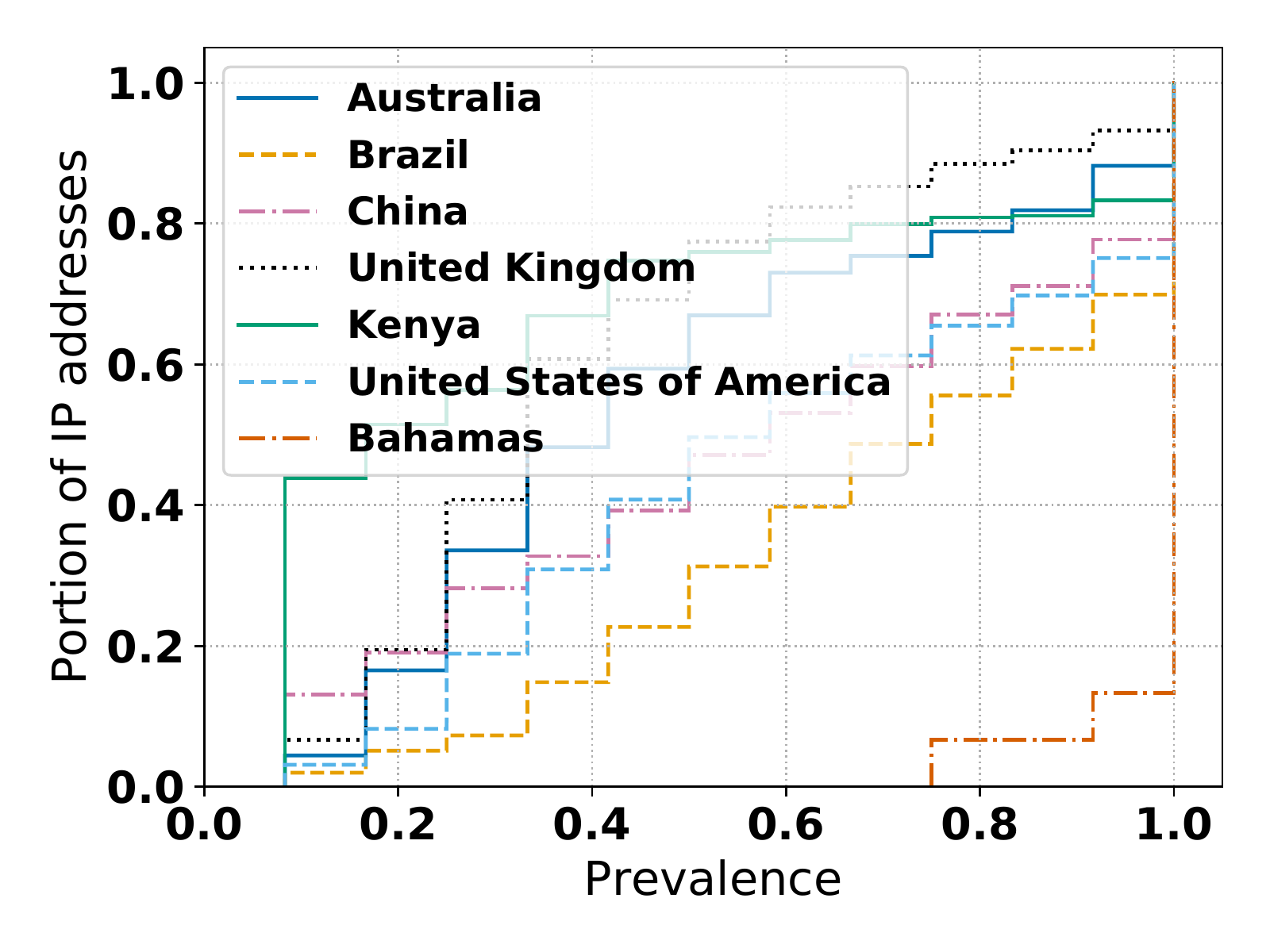}
  \caption{Routers (Eq.~\ref{eq:prevalence-location})}
  \label{fig:appendix:prevalence-by-continent-city-routers}
\end{subfigure}

  \caption{Prevalence results on \maxmind city geolocation}
  \label{fig:appendix:prevalence-city}
\end{figure*}

\subsection{Prevalence on \maxmind city database}

Fig.~\ref{fig:appendix:prevalence-city} shows our results on the prevalence for the 
\maxmind city database.
We evaluate the prevalence on cities with the same
methodology as we did for the \maxmind country database on our three axes.

\paragraph{Evolution of the prevalence, 
Fig.~\ref{fig:appendix:prevalence-over-time-city}, Eq.~\ref{eq:max-prevalence-ip}}
First of all,  
we observe that the shape of the distributions are totally different 
compared to the \maxmind country database. 
They are not concentrated near 1. 
This means that most IP addresses do not have 
a single dominant city.

Regarding the maximum prevalence over time, we do not identify any 
trend, except that 
we can notice that the two last years 2018 and 2019 have
their curves of their distributions above the
others, meaning that IP addresses 
tend to not have a dominant city even more for these two years.

\paragraph{Per type of IP addresses, Fig.~\ref{fig:appendix:prevalence-by-ip-type-city},
 Eq.~\ref{eq:max-prevalence-ip}}
We notice a clear difference between end host infrastructure
and the others; they have a higher maximum prevalence, but still, 
20\% of the IP addresses have a maximum prevalence of 0.5, meaning that
they do not have a single city, or even a highly dominant one.

\paragraph{Per geographical locations, Fig.~\ref{fig:appendix:prevalence-by-continent-city} 
and~\ref{fig:appendix:prevalence-by-country-city}, 
Eq.~\ref{eq:prevalence-location-average}}

We observe that the average prevalence depends on the continent:
If we take extreme values for all curves, it goes from 0.45 to 0.9.

At country level, the range of average prevalence goes from 0.08 to almost 1, 
with a median between 0.54 and 0.70 depending on the type of IP addresses.
This means that the prevalence on cities is highly dependent on the country.

Like we did for \maxmind country database, we further investigate these
differences.
\paragraph{Detailed investigation on continents, second row of Fig.~\ref{fig:appendix:prevalence-city},
 Eq.~\ref{eq:prevalence-location}}
 
We see that for all continent and for all types of IP addresses, 
the range of possibles maximum prevalence 
goes from almost 0.08 to 1. 
For ``all'', end users, and end host infrastructure, North America has the 
highest prevalence in general, while for routers,  South America has 
the highest prevalence.

\paragraph{Detailed investigation on countries, third 
row of Fig.~\ref{fig:appendix:prevalence-city},  Eq.~\ref{eq:prevalence-location}}
This time, for each type of IP address, we show the 
the 10 countries having, this time, the highest average 
prevalence (Eq.~\ref{eq:prevalence-location-average}).
There are considerable differences depending on the type of IP address.
On the ``All'' dataset, the 10 countries are either countries in Africa or small 
Islands, with the interesting exception of Korea, which has the highest average
prevalence.
On end users, we mostly find middle eastern and island countries.
On end host infrastructure, the 10 countries include countries 
from all continents, and are big countries, such as US, Canada, 
Finally, on routers, we find in the ten countries middle east countries, such as
Iran, and Irak, and also small countries located in islands.

\begin{figure*}[h]

\begin{subfigure}{.24\textwidth}
\centering
  \includegraphics[width=\linewidth]{figures/metrics/persistence/country/persistence_snapshots_city_over_time_all_avg_year.pdf}
  \caption{Over time (Eq.~\ref{eq:avg-persistence-ip})}
  \label{fig:appendix:persistence-over-time-city}
\end{subfigure}%
\begin{subfigure}{.24\textwidth}
\centering
  \includegraphics[width=\linewidth]{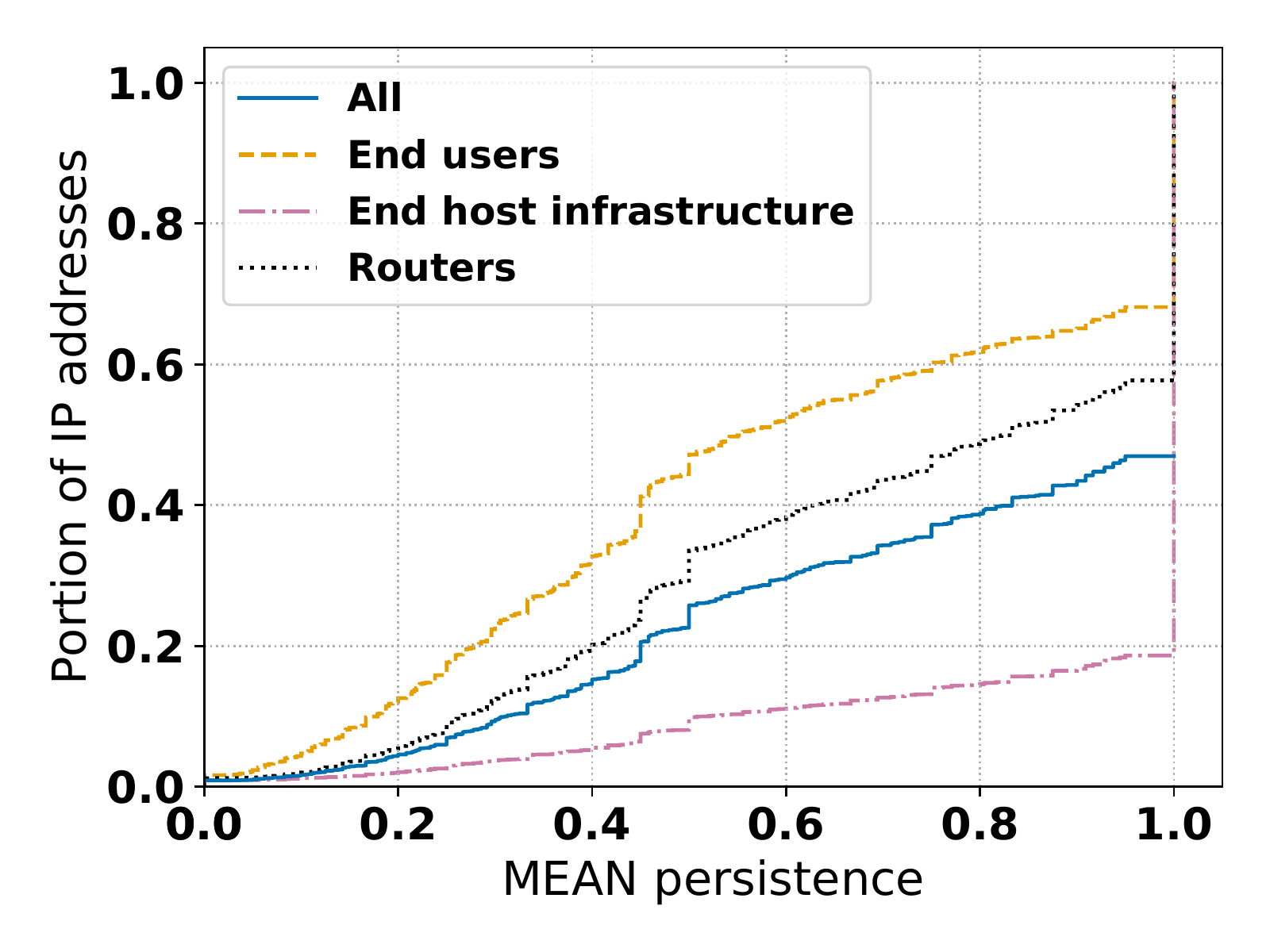}
  \caption{By IP class (Eq.~\ref{eq:avg-persistence-ip})}
  \label{fig:appendix:persistence-by-ip-type-city}
\end{subfigure}%
\begin{subfigure}{.24\textwidth}
\centering
  \includegraphics[width=\linewidth]{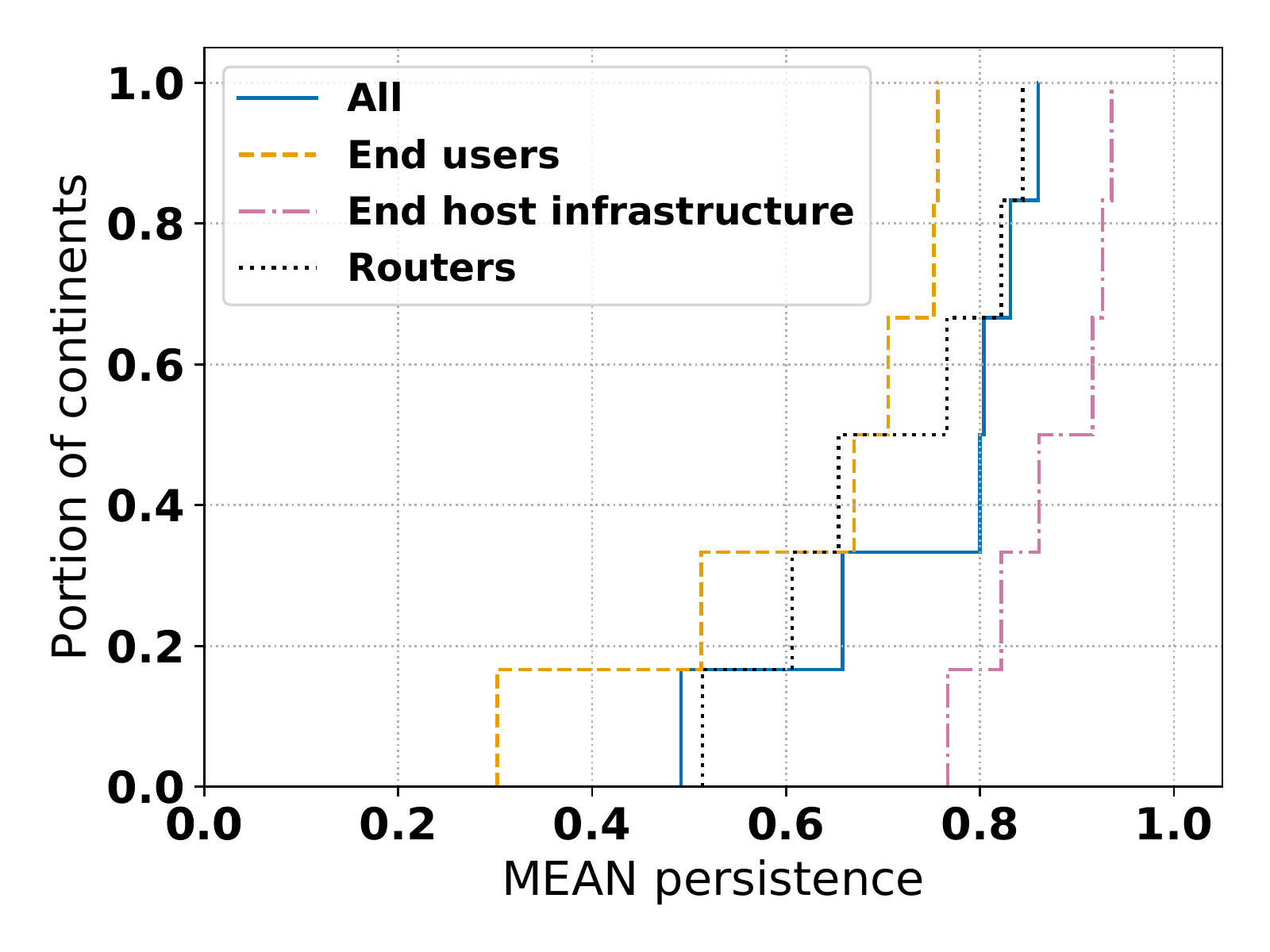}
  \caption{By continent (Eq.~\ref{eq:persistence-location-average})}
  \label{fig:appendix:persistence-by-continent-city}
\end{subfigure}
\begin{subfigure}{.24\textwidth}
\centering
  \includegraphics[width=\linewidth]{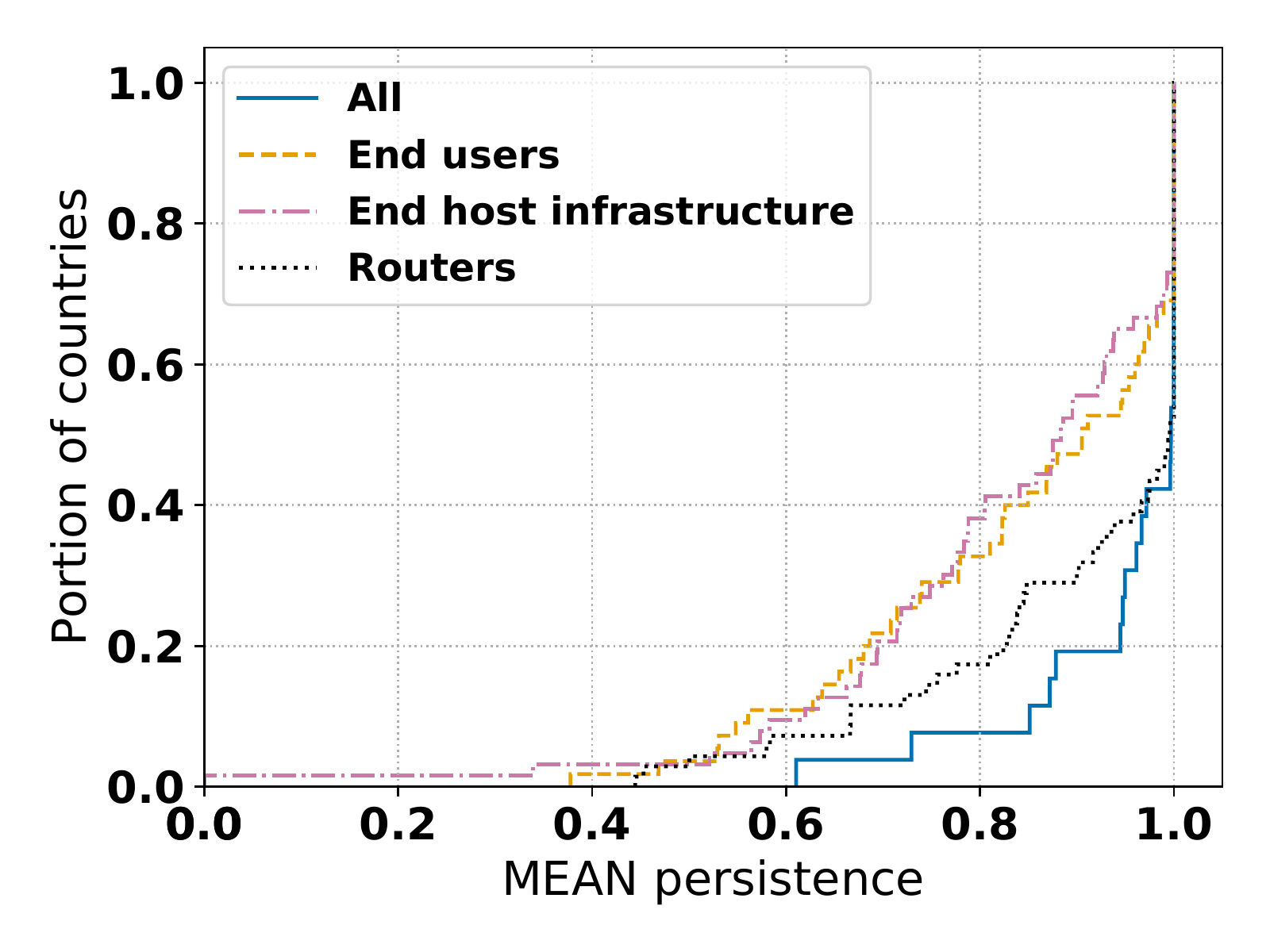}
  \caption{By country (Eq.~\ref{eq:persistence-location-average})}
  \label{fig:appendix:persistence-by-country-city}
\end{subfigure}
%

%
\begin{subfigure}{.24\textwidth}
\centering
  \includegraphics[width=\linewidth]{figures/metrics/persistence/country/persistence_snapshots_city_per_country_country_avg_year.pdf}
  \caption{All (Eq.~\ref{eq:persistence-location})}
  \label{fig:appendix:persistence-by-country-city-all}
\end{subfigure}%
\begin{subfigure}{.24\textwidth}
\centering
  \includegraphics[width=\linewidth]{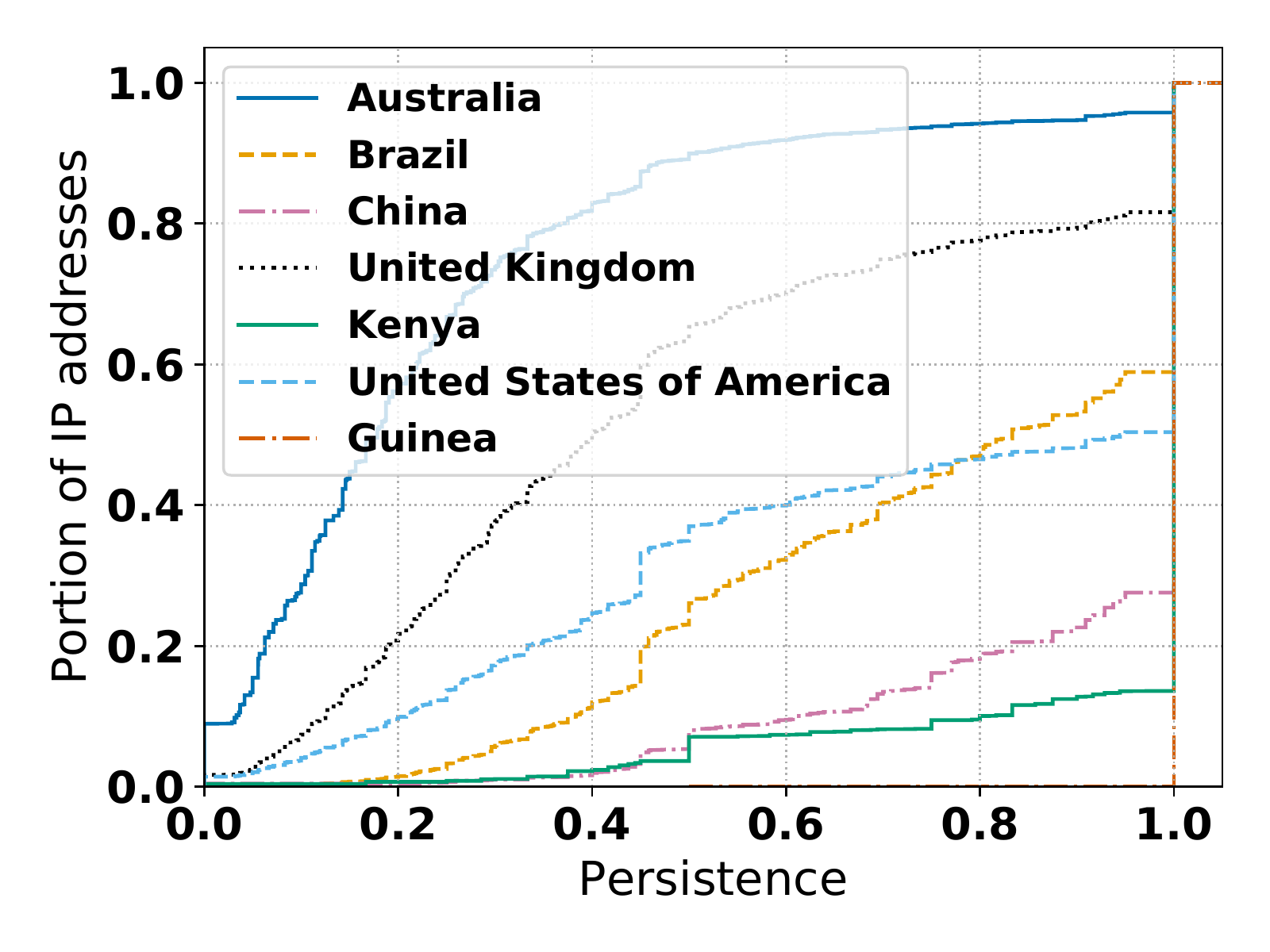}
  \caption{End users (Eq.~\ref{eq:persistence-location})}
  \label{fig:appendix:persistence-by-continent-city-end-users}
\end{subfigure}%
\begin{subfigure}{.24\textwidth}
\centering
  \includegraphics[width=\linewidth]{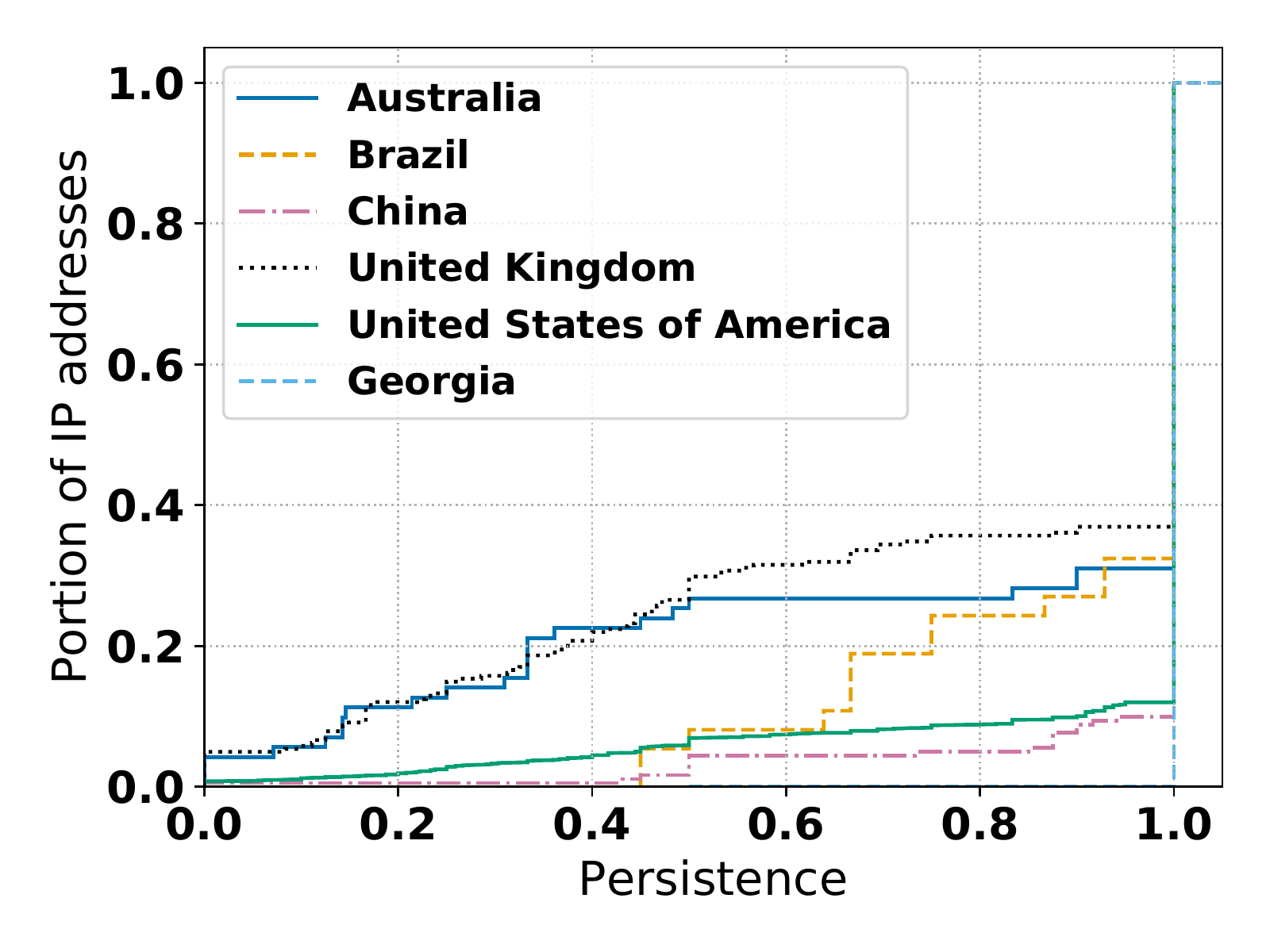}
  \caption{End host infra. (Eq.~\ref{eq:persistence-location})}
  \label{fig:appendix:persistence-by-continent-city-infrastructure}
\end{subfigure}%
\begin{subfigure}{.24\textwidth}
\centering
  \includegraphics[width=\linewidth]{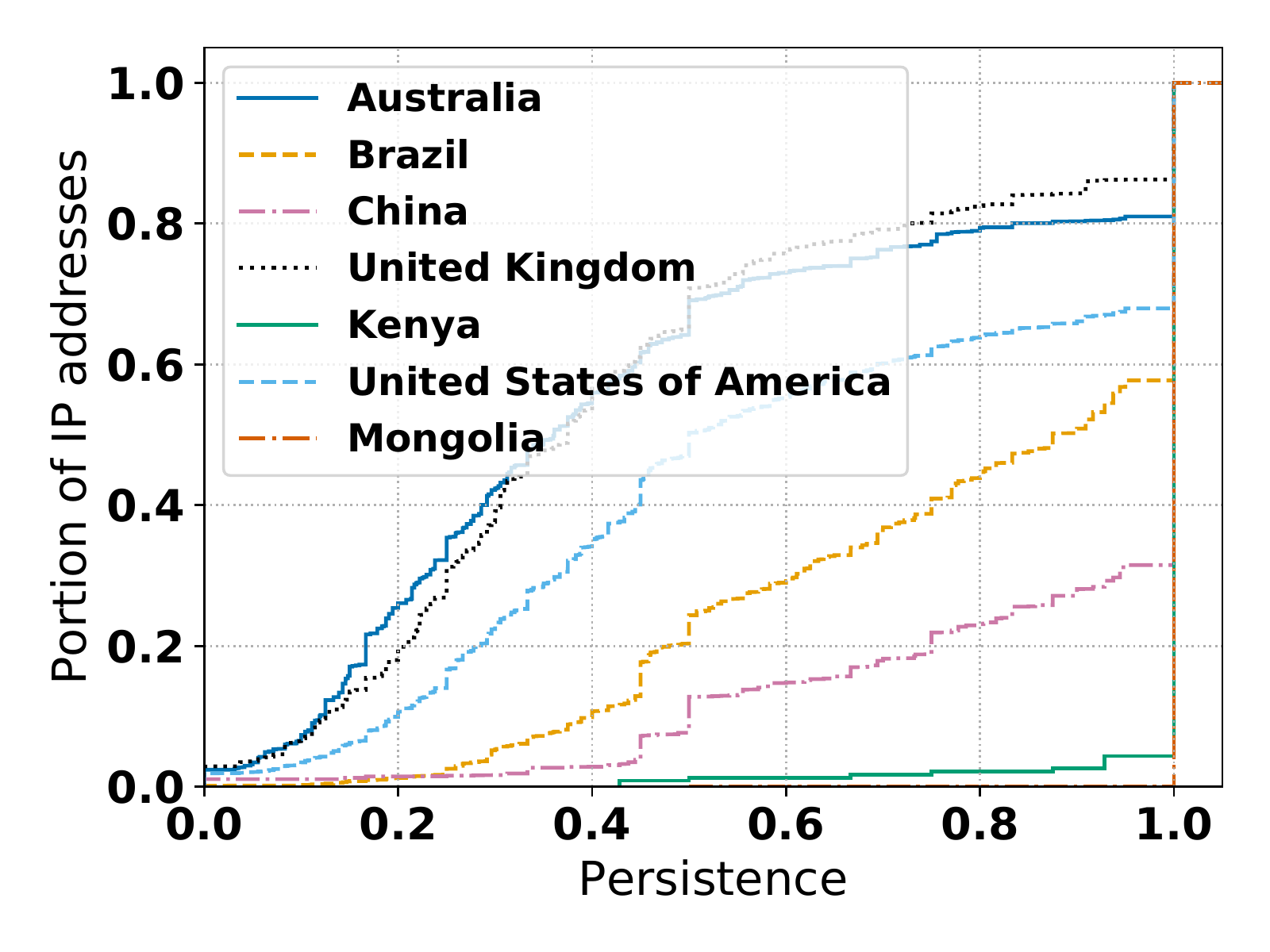}
  \caption{Routers (Eq.~\ref{eq:persistence-location})}
  \label{fig:appendix:persistence-by-continent-city-routers}
\end{subfigure}
 \caption{Persistence results on \maxmind city geolocation}
  \label{fig:appendix:persistence-city}
\end{figure*}

\subsection{Persistence on \maxmind city database}

\paragraph{Evolution of the persistence, 
Fig.~\ref{fig:appendix:persistence-over-time-city}, Eq.~\ref{eq:persistence-location-ip}}
Here is an interesting observation:
2018 and 2019 were the years where the maximum prevalence per IP address
were the lower. However, if 2019 is a year where the average persistence
is low compared to ohter years (except 2012), 2018 does not have a low
average persistence. This shows two things: an IP 
address can have a low maximum prevalence
but a not so low average persistence; city changes in 2018 were rather long term
chances compared to 2019.

\paragraph{Per type of IP addresses, 
Fig.~\ref{fig:appendix:persistence-by-ip-type-city}, Eq.~\ref{eq:persistence-location-ip}}
This graphs follows the one on prevalence, with end host infrastructure having
a higher average persistence than end users and routers.
\paragraph{Per geographic location, 
Fig.~\ref{fig:appendix:persistence-by-continent-city} and~\ref{fig:appendix:persistence-by-country-city},
Eq.~\ref{eq:persistence-location-average}}
Here we observe that the average persistence of a location 
highly depends on the continent and the country.
The range of average persistence goes from 0.30 to 0.90, while it goes from
0 to 1 for countries, depending on the type of IP addresses.
\paragraph{Detailed investigation on continents, second row of 
Fig.~\ref{fig:appendix:persistence-city}, Eq.~\ref{eq:persistence-location}}
These curves reveals that on all graphs, the curves of 
Oceania are
mostly above 
the other continents, meaning that the persistence is 
below the other continents. The order of the other continent depend on the type of 
IP addresses. Notice that Africa is among the higher persistence for the
three types of IP addresses.
\paragraph{Detailed investigation on country, third row of 
Fig.~\ref{fig:appendix:persistence-city}, Eq.~\ref{eq:persistence-location}}
Here again, we choose to show the countries having the higher average 
persistence.
The first statement that we make is that even for countries having 
the higher average persistence, they are not all one concentrated near 1:
some IP addresses have short city change, and some other have long.

We notice interesting differences with the graphs on prevalence:
For end users, all the countries, except United Kingdom, are 
countries in Africa, meaning that an IP address having a
city change in these countries is more likely
to last than in others. For routers, we also find totally difference countries
than the ones on prevalence; they are mostly European countries, with US and
Saudi Arabia added.

\begin{figure*}[h]

\begin{subfigure}{.24\textwidth}
\centering
  \includegraphics[width=\linewidth]{figures/metrics/distance/country/distance_snapshots_location_over_time_all_max_year.pdf}
  \caption{Over time (Eq.~\ref{eq:max-distance-metric})}
  \label{fig:appendix:distance-over-time}
\end{subfigure}%
\begin{subfigure}{.24\textwidth}
\centering
  \includegraphics[width=\linewidth]{figures/metrics/distance/country/distance_per_ip_type_location_database_all_max_year.pdf}
  \caption{By IP class (Eq.~\ref{eq:max-distance-metric})}
  \label{fig:appendix:distance-by-ip-type}
\end{subfigure}%
\begin{subfigure}{.24\textwidth}
\centering
  \includegraphics[width=\linewidth]{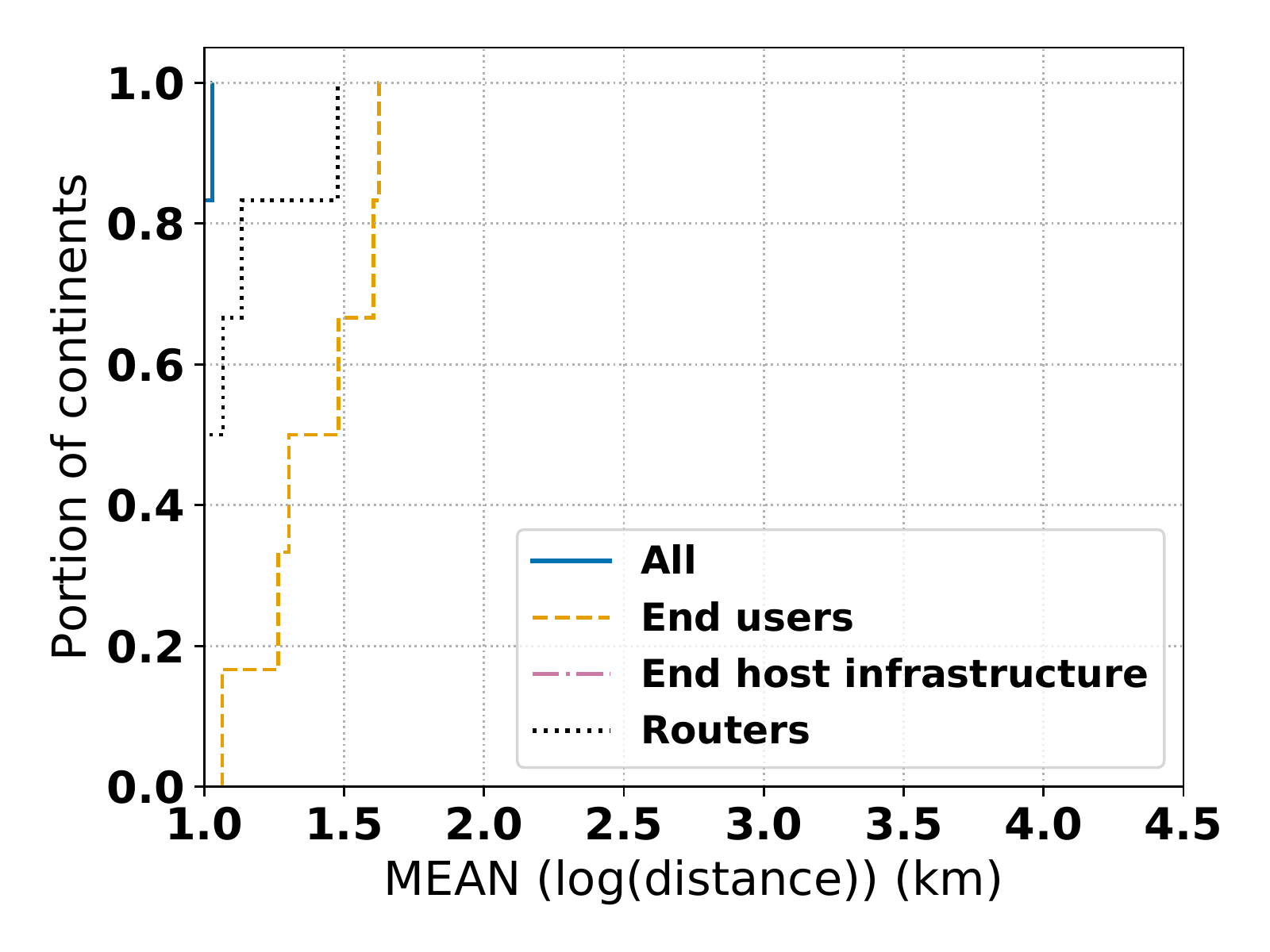}
  \caption{By continent (Eq.~\ref{eq:distance-metric-location-average})}
  \label{fig:appendix:distance-by-continent}
\end{subfigure}
\begin{subfigure}{.24\textwidth}
\centering
  \includegraphics[width=\linewidth]{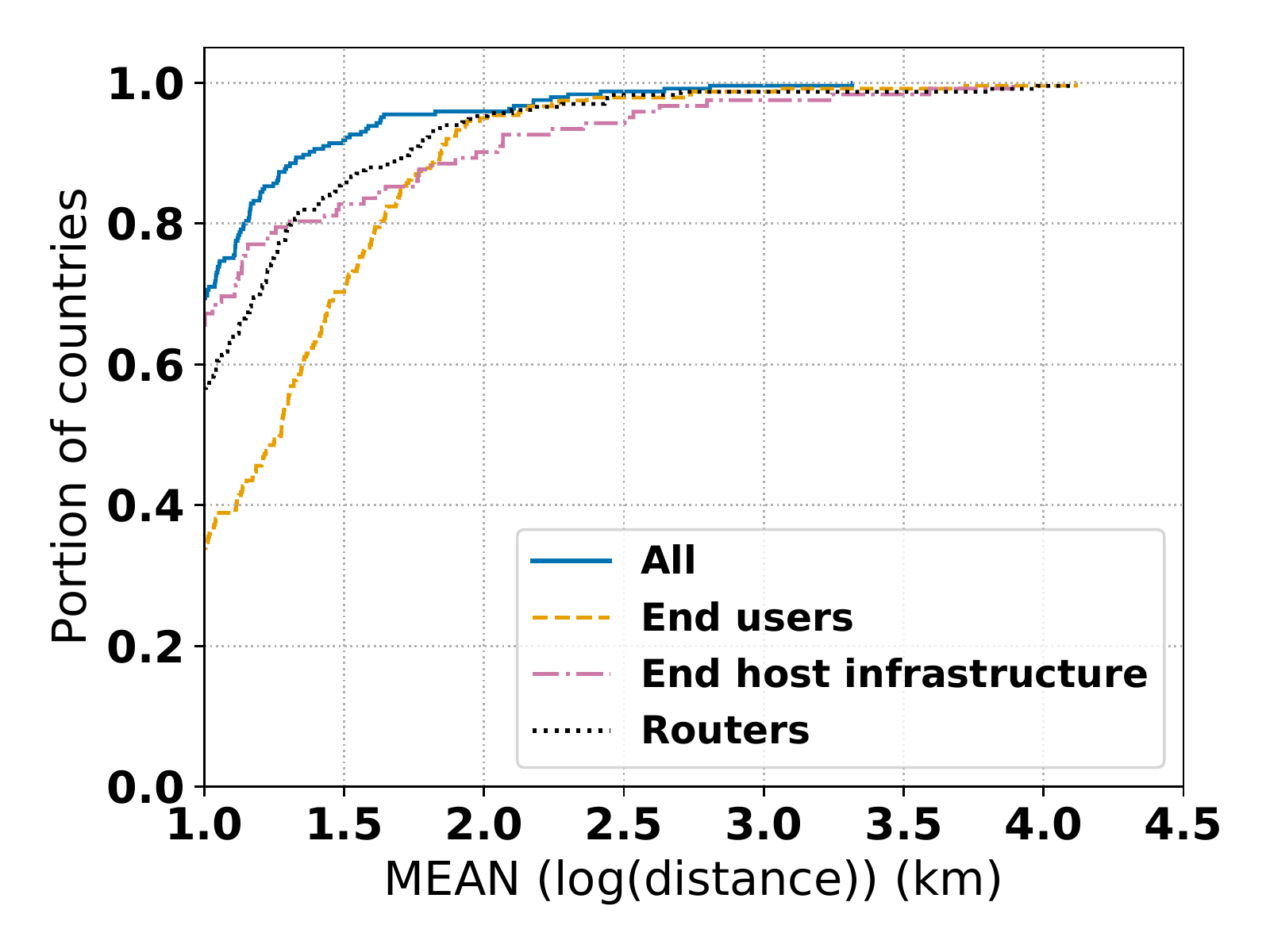}
  \caption{By country (Eq.~\ref{eq:distance-metric-location-average})}
  \label{fig:appendix:distance-by-country}
\end{subfigure}
%


\begin{subfigure}{.24\textwidth}
\centering
  \includegraphics[width=\linewidth]{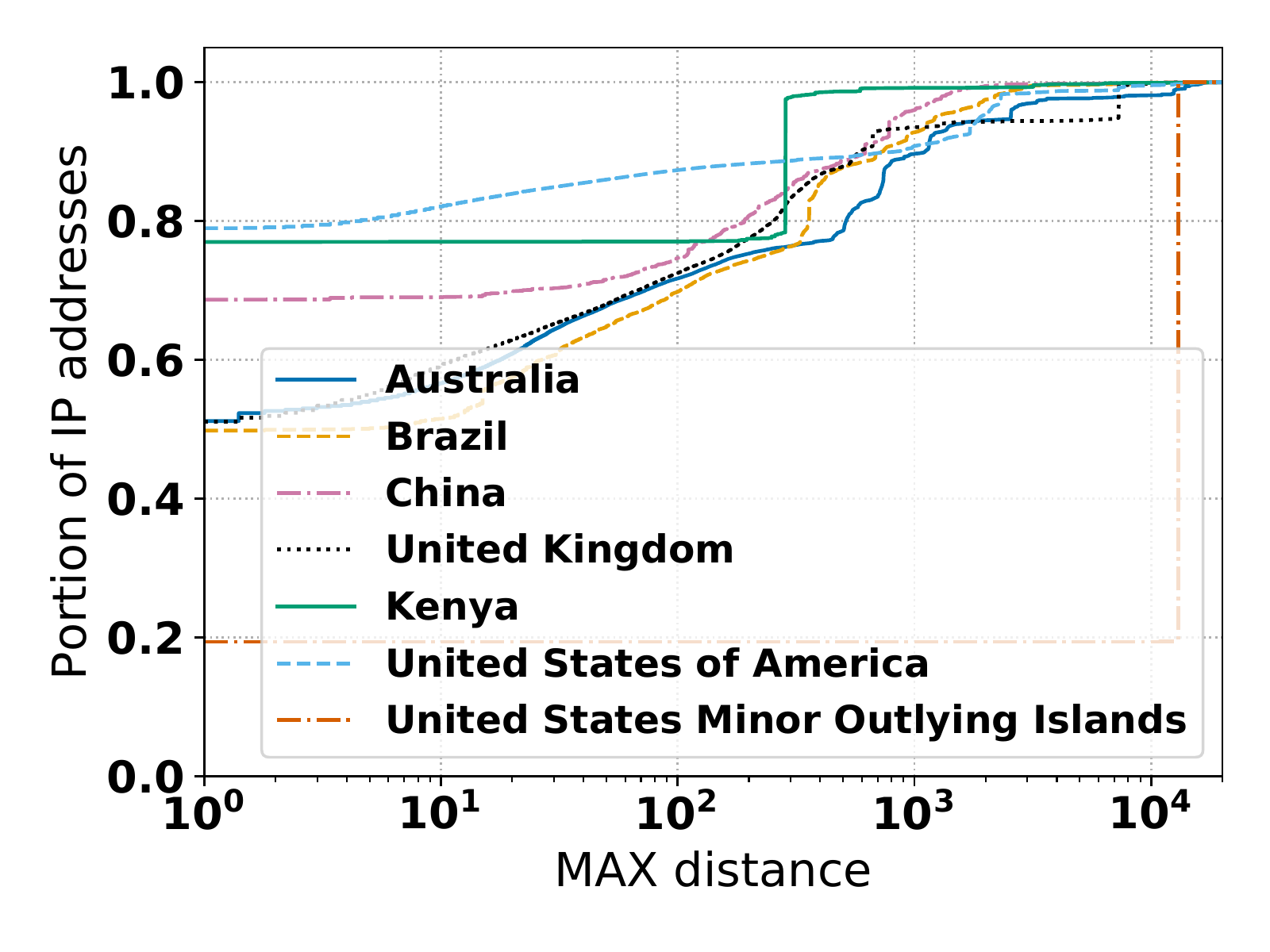}
  \caption{All (Eq.~\ref{eq:distance-metric-location})}
  \label{fig:appendix:distance-by-country-all}
\end{subfigure}%
\begin{subfigure}{.24\textwidth}
\centering
  \includegraphics[width=\linewidth]{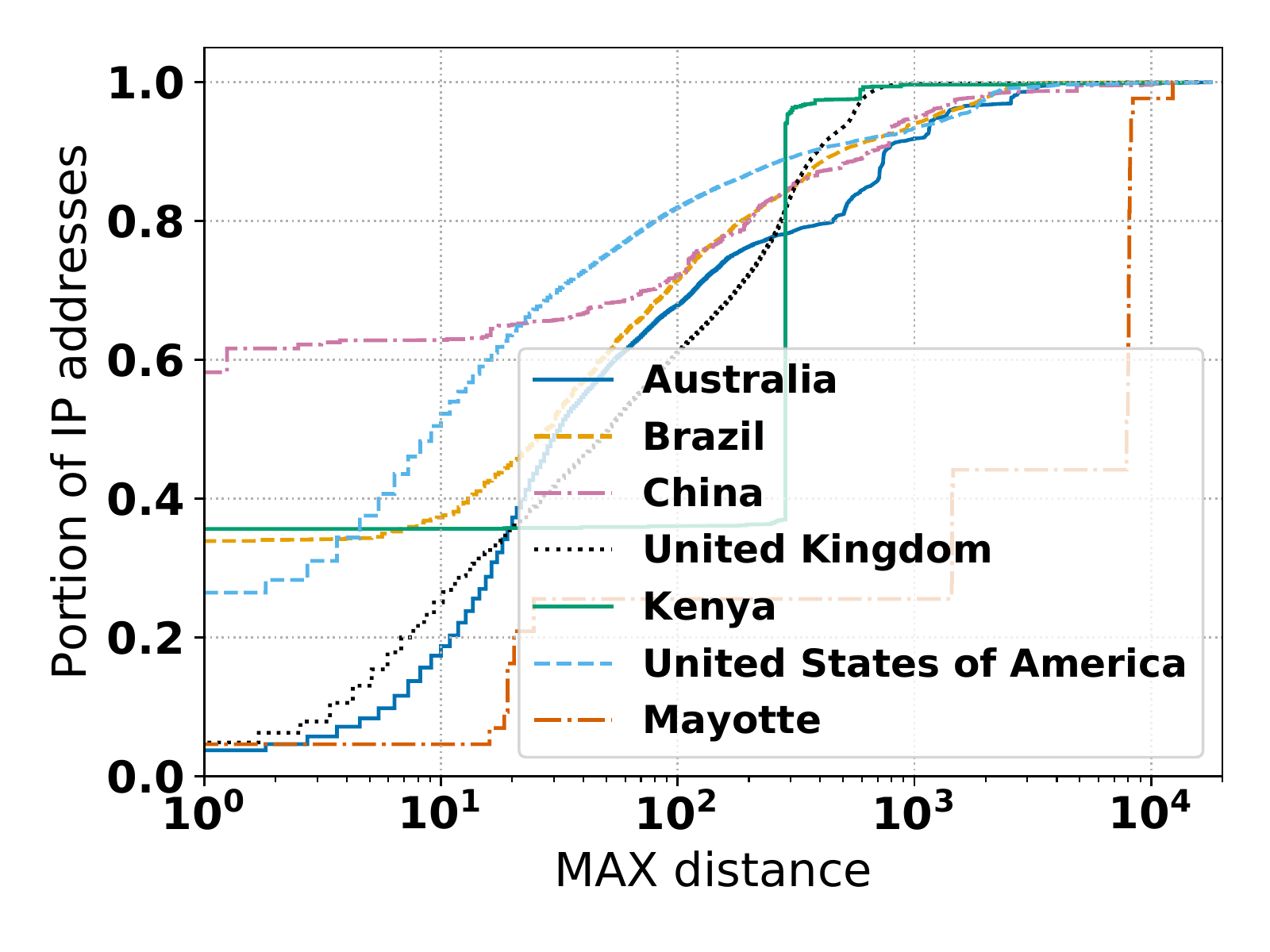}
  \caption{End users (Eq.~\ref{eq:distance-metric-location})}
  \label{fig:appendix:distance-by-country-end-users}
\end{subfigure}%
\begin{subfigure}{.24\textwidth}
\centering
  \includegraphics[width=\linewidth]{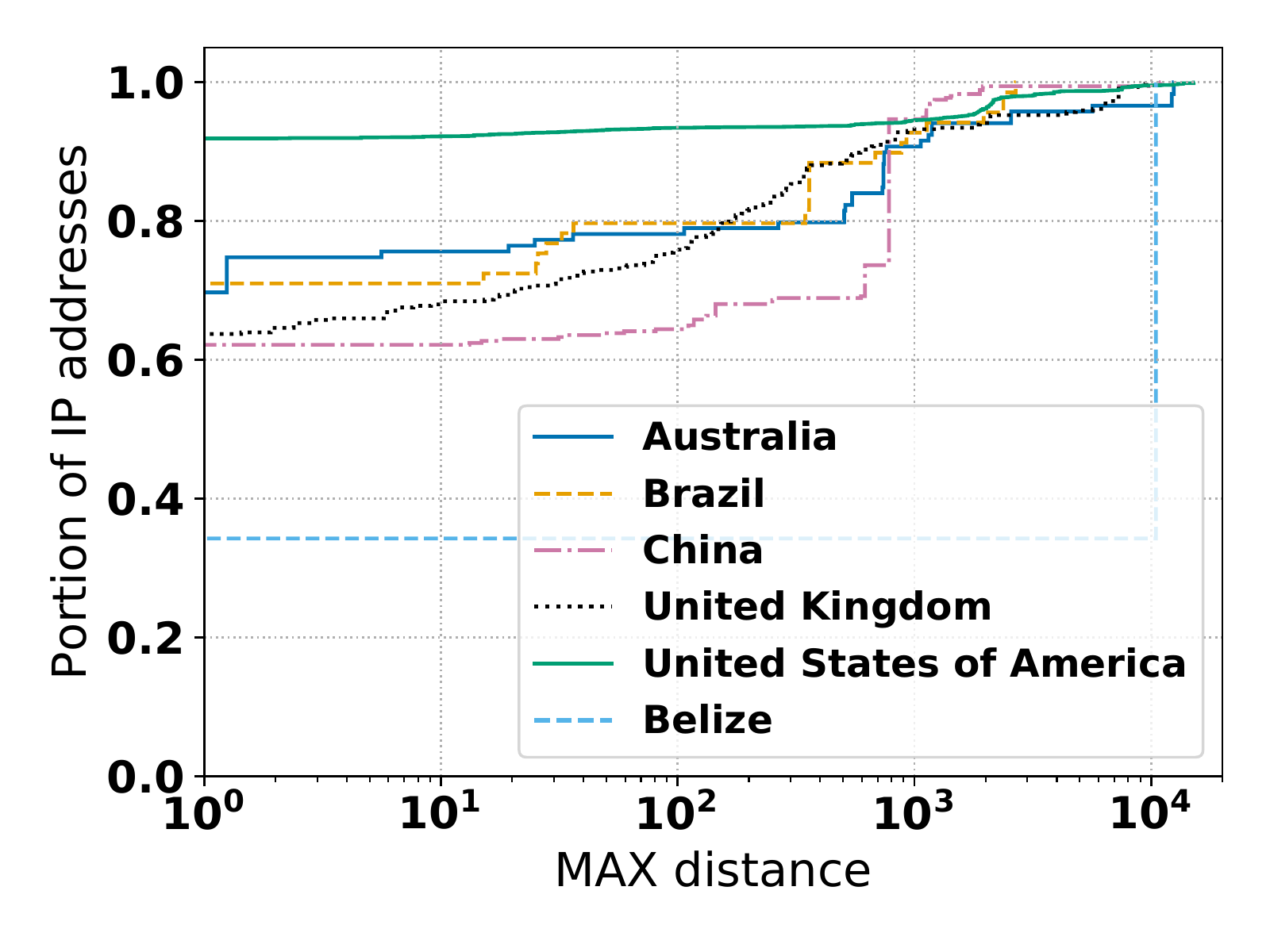}
  \caption{End host infra. (Eq.~\ref{eq:distance-metric-location})}
  \label{fig:appendix:distance-by-country-infrastructure}
\end{subfigure}%
\begin{subfigure}{.24\textwidth}
\centering
  \includegraphics[width=\linewidth]{figures/metrics/distance/country/distance_snapshots_location_router_per_country_country_max_year.pdf}
  \caption{Routers (Eq.~\ref{eq:distance-metric-location})}
  \label{fig:appendix:distance-by-country-routers}
\end{subfigure}
 \caption{Distance results on \maxmind coordinates}
  \label{fig:appendix:distance-coordinates}
\end{figure*}
\paragraph{Evolution over time (Fig.~\ref{fig:appendix:distance-over-time}, Eq.~\ref{eq:max-distance-metric})}
We can observe that there is a lot of variations across years. Indeed
there are from 10\% (in 2015) to 57\% of IP addresses (in 2012) that had a 
maximum distance of more than 100 KM.
\paragraph{Per type of IP addresses 
(Fig.~\ref{fig:appendix:distance-by-ip-type}, Eq.~\ref{eq:max-distance-metric})}
Again, the maximum distance depends on the type of IP address. 
End users have the greater portion of IP addresses experiencing
a maximum distance of more than 100 KM (35\%), whereas end 
host infrastructure have 10\%.
Notice also for all type of IP addresses, 
few IP addresses have a > 1000 KM maximum distance (6\% at most).
\paragraph{Per geographic location (Fig.~\ref{fig:appendix:distance-by-continent} 
and~\ref{fig:appendix:distance-by-country},
 Eq.~\ref{eq:distance-metric-location-average})}

Recall that our metric represents, per location, 
the average of the maximum logarithmic distance between 
two pairs of coordinates
of an IP address where at least one 
is in the continent or the country. 

First of all, we see that this metric for continents is rather small.
It goes from value near one (< 10 km), to a value of 1.6
(39 km). 
But on countries, like for prevalence, 
the distribution is more spread, going from less than one to more than 4, which
correspond to 10,000 km.
We detail this distribution in the next paragraphs.

\paragraph{Detailed investigation on continents (second row 
of Fig.~\ref{fig:appendix:distance-coordinates}, Eq.~\ref{eq:distance-metric-location})}

In general, on all four graphs, whatever the continent, there are
a significant portion of IP addresses that have a maximum distance between two
of their locations of more than 100 KM.
We can also observe that the maximum distance in continents depend on the 
type of IP address; North America is have statistically fewer distance for end 
users and end host infrastructure, whereas it is Asia for routers. 

\paragraph{Detailed investigation on countries (third row 
of Fig.~\ref{fig:appendix:distance-coordinates}, Eq.~\ref{eq:distance-metric-location})}
In these graphs, we show the ten countries that have the highest value for 
the metric defined in Eq.~\ref{eq:distance-metric-location-average}, i.e., the
countries where the distances are the higher. 
Again, like for prevalence and persistence, the countries depend on 
the type of IP address. For end users, they are mainly located in small countries 
in islands. For end host infrastructure, these are either medium or big 
countries in all continents (e.g., India, Bulgaria, Argentina). For routers, 
we find countries in small islands and countries in Africa.